\documentclass{aa}  

\usepackage{graphicx}
\usepackage{txfonts}
\usepackage{verbatim} 
\usepackage[normalem]{ulem}

\usepackage[dvipsnames]{xcolor}
\usepackage{soul}
\usepackage{bm}

\usepackage[toc,page]{appendix}

\newcommand{\mlpd}[2]{\bgroup\markoverwith{\textcolor{red}{\rule[0.5ex]{2pt}{0.4pt}}}\ULon{#1} \textcolor{blue} {#2}}

\def\co2{CO$_2$}
\def\ch4{CH$_4$}
\def\h2{H$_{2}$}
\def\h2o{H$_2$O}
\def\nh3{NH$_3$}
\def\hdu18{HD~189733\,b}
\def\hd20{HD~209458\,b}

\begin{document} 

\title{Discriminating between hazy and clear hot-Jupiter atmospheres with CARMENES}
%
%
\titlerunning{Discriminating between hazy and clear hot Jupiter atmospheres with CARMENES}
\author{A. S\'anchez-L\'opez\inst{1}, M. L\'opez-Puertas\inst{2}, I. A. G. Snellen\inst{1}, E. Nagel\inst{3}, F. F. Bauer\inst{2}, E. Pall\'e\inst{4,5}, L. Tal-Or\inst{6,9}, \\
P. J. Amado\inst{2}, J. A. Caballero\inst{7}, S. Czesla\inst{8}, L. Nortmann\inst{9}, A. Reiners\inst{9}, I. Ribas\inst{10,11}, A. Quirrenbach\inst{12}, J. Aceituno\inst{13}, V.~J.~S.~B{\'e}jar\inst{4,5}, N.\,Casasayas-Barris\inst{4,5}, Th. Henning\inst{14}, K. Molaverdikhani\inst{14}, D. Montes\inst{15}, M. Stangret\inst{4,5}, M.\,R.\,Zapatero~Osorio\inst{16}, and M.\,Zechmeister\inst{9}}
\institute{Leiden Observatory, Leiden University, Postbus 9513, 2300 RA, Leiden, The Netherlands\\
\email{alexsl@strw.leidenuniv.nl}
\and
Instituto de Astrof{\'i}sica de Andaluc{\'i}a (IAA-CSIC), Glorieta de la Astronom{\'i}a s/n, 18008 Granada, Spain
\and
Th{\"u}ringer Landessternwarte Tautenburg, Sternwarte 5, 07778 Tautenburg, Germany
\and
Instituto de Astrof{\'i}sica de Canarias (IAC), Calle V{\'i}a L{\'a}ctea s/n, 38200 La Laguna, Tenerife, Spain
\and
Departamento de Astrof{\'i}sica, Universidad de La Laguna, 38026  La Laguna, Tenerife, Spain
\and
Department of Physics, Ariel University, Ariel 40700, Israel
\and
Centro de Astrobiolog{\'i}a (CSIC-INTA), ESAC, Camino bajo del castillo s/n, 28692 Villanueva de la Ca{\~n}ada, Madrid, Spain
\and
Hamburger Sternwarte, Universit{\"a}t Hamburg, Gojenbergsweg 112, 21029 Hamburg, Germany
\and
Institut f{\"u}r Astrophysik, Georg-August-Universit{\"a}t, Friedrich-Hund-Platz 1, 37077 G{\"o}ttingen, Germany
\and
Institut de Ci\`encies de l'Espai (CSIC-IEEC), Campus UAB, c/ de Can Magrans s/n, 08193 Bellaterra, Barcelona, Spain
\and
Institut d'Estudis Espacials de Catalunya (IEEC), 08034 Barcelona, Spain
\and
Landessternwarte, Zentrum f\"ur Astronomie der Universit\"at Heidelberg, K\"onigstuhl 12, 69117 Heidelberg, Germany
\and
Observatorio de Calar Alto, Sierra de los Filabres, 04550 G\'ergal, Almer\'{\i}a, Spain
\and
Max-Planck-Institut f{\"u}r Astronomie, K{\"o}nigstuhl 17, 69117 Heidelberg, Germany
\and
Departamento de F{\'i}sica de la Tierra y Astrof{\'i}sica \& IPARCOS-UCM (Instituto de F{\'i}sica de Part{\'i}culas y del Cosmos de la UCM), Facultad de Ciencias F{\'i}sicas, Universidad Complutense de Madrid,  28040 Madrid, Spain
\and
Centro de Astrobiolog{\'i}a (CSIC-INTA), Carretera de Ajalvir km 4,
28850 Torrej{\'o}n de Ardoz, Madrid, Spain
}
\authorrunning{A. S\'anchez-L\'opez et al.}

\date{}

 
\abstract
{Relatively large radii of some hot Jupiters observed in the ultraviolet (UV) and blue-optical are generally interpreted to be due to Rayleigh scattering by high-altitude haze particles. However, the haze composition and its production mechanisms are not fully understood, and observational information is still limited.}
{We aim to study the presence of hazes in the atmospheres of \hd20 and \hdu18 with high spectral resolution spectra by analysing the strength of water vapour cross-correlation signals across the red optical and near-infrared wavelength ranges.}
{A total of seven transits of the two planets were observed with the CARMENES spectrograph at the 3.5\,m Calar Alto telescope. 
Their Doppler-shifted signals were disentangled from the telluric and stellar contributions using the detrending algorithm {\tt SYSREM}. The residual spectra were subsequently cross-correlated with water vapour templates at 0.70--0.96\,$\mu$m to measure the strength of the water vapour absorption bands.}
{The optical water vapour bands were detected at $5.2 \sigma$ in \hd20 in one transit, whereas no evidence of them was found in four transits of \hdu18. Therefore, the relative strength of the optical water bands compared to those in the near-infrared were found to be larger in \hd20 than in \hdu18.}
{We interpret the non-detection of optical water bands in the transmission spectra of \hdu18, compared to the detection in \hd20, to be due to the presence of high-altitude hazes in the former planet, which are largely absent in the latter. This is consistent with previous measurements with the \textit{Hubble Space Telescope}. We show that currently available CARMENES observations of hot Jupiters can be used to investigate the presence of haze extinction in their atmospheres.}

   \keywords{planets and satellites: atmospheres -- planets and satellites: individual: HD\,209458\,b, HD\,189733\,b -- techniques: spectroscopic -- infrared: planetary systems}

   \maketitle
%

\section{Introduction}
\label{introduction}

\begin{figure*}[ht]
\centering
\includegraphics[angle=0, scale=0.23]{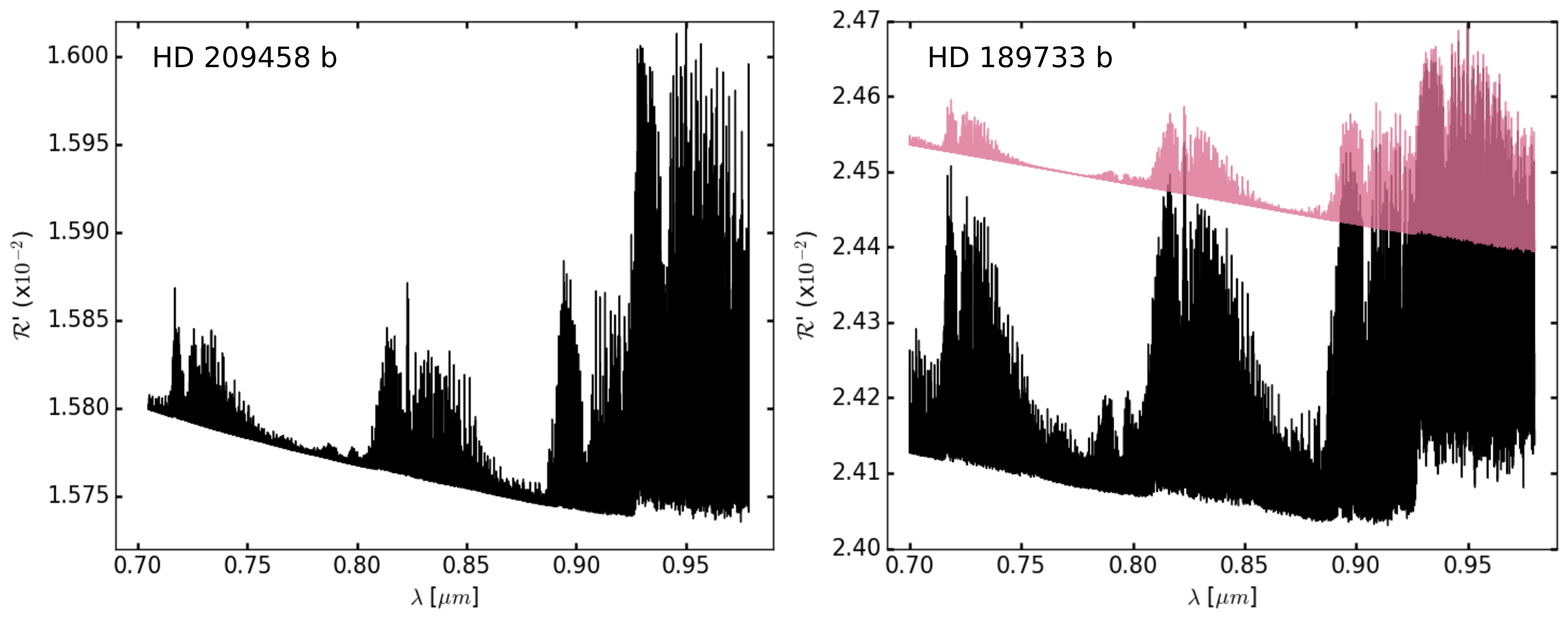}
\caption{Model transmission spectra of \hd20 (\textit{left}) and \hdu18 (\textit{right}) using the pressure-temperature profiles from \cite{Brogi17} and \cite{Brogi18} and the water vapour volume-mixing ratios of 10$^{-5}$ and 10$^{-4}$, respectively. The same model for \hdu18 with an additional haze contribution at an arbitrary level is shown in magenta to illustrate the reduction of the water signal. All models were computed at the CARMENES spectral resolution in the VIS channel ($\mathcal{R} = 94,600$) and they included the contributions from Rayleigh scattering and collision-induced absorption.} 
\label{model_spec}
\end{figure*}

Ground-based spectroscopic observations at high resolution ($\mathcal{R}$\,$>$\,40,000) have flourished as a very powerful tool for the unprecedented characterisation of exoplanet atmospheres  \citep[e.g.][]{Birkby18, Ehrenreich2020}. In recent years, a wide variety of atomic and molecular species, such as Fe~{\sc i}, Fe~{\sc ii}, TiO, CO, \h2o, CH$_4$, HCN, and \nh3, among others, have been successfully identified in the atmospheres of hot Jupiters using cross-correlation techniques \citep{Snellen10, Birkby13, deKok13, Nugroho17, Brogi17, Hawker2018, Yan18, Casasayas18, Brogi18, Hoeijmakers18, Brogi19, Alonso19, Sanchezlopez2019, Guilluy2019, Casasayas19}. 
These detections provide unique information on the physical and chemical processes in hot planet atmospheres, with the aim to constrain their composition and ultimately their formation and evolution scenarios. 

Much effort has been made to the characterisation of the two most studied hot Jupiters: \hd20 \citep{Charbonneau00, Henry00} and \hdu18 \citep{Bouchy05}. In particular, space observations of the former are better understood when considering partial cloud coverage \citep{Barstow17, MacDonald17}, whereas the latter presents a rather steep Rayleigh scattering slope produced by a strong haze extinction \citep{Sing16, Barstow17}. In this context, the study of \h2o cross-correlation signals, using ground-based high resolution spectrographs across different spectral intervals, can provide strong constraints on the presence of these types of atmospheric aerosols \citep{Pino18}. This is because the strength of the \h2o rovibrational bands as well as the intensity of the haze extinction change with wavelength. Specifically, the strength of the water vapour bands decreases towards shorter wavelengths, whereas the extinction due to hazes increases (e.g. see right panel in Fig.\,\ref{model_spec}).

Earlier studies have investigated the presence of \h2o in \hdu18 and \hd20 from the ground at near-infrared wavelengths, finding robust detections \citep{Brogi18, Hawker2018, Cabot19, Alonso19, Sanchezlopez2019}. In previous work by our group, we compared the water vapour detections using the CARMENES spectrograph in \hdu18 \citep{Alonso19} and \hd20 at 0.96--1.71\,$\mu$m \citep{Sanchezlopez2019}. We found a strong \h2o signal using the band at $\sim$1.0\,$\mu$m in \hd20, while only weak signatures from this band were obtained for \hdu18. In the latter planet, however, we obtained strong \h2o signals using the $\sim$1.15\,$\mu$m and $\sim$1.40\,$\mu$m bands. This result already hinted at a possible muting of the shorter-wavelength signals in \hdu18 by a strong haze extinction. However, different telluric conditions in the two nights and the variability of the telluric water vapour reported in \citet{Sanchezlopez2019} prevented drawing further conclusions.

In this work, we aim to probe the atmospheres of \hdu18 and \hd20 in the red optical by observing several transits of these hot Jupiters with CARMENES. We focus our attention on the 0.70--0.96\,$\mu$m spectral region, where the strongest \h2o absorption in the optical occurs. Previous attempts at detecting water vapour in \hdu18 in the optical were reported in \citet{Allart17}, who studied the $\sim$0.65\,$\mu$m spectral band with the HARPS spectrograph \citep{Mayor2003}. They reported a non-detection and a $5 \sigma$ upper limit of 100\,ppm on this band's strength. Similar efforts to detect \h2o in the optical were undertaken in \citet{Esteves17}, targeting the super-Earth 55~Cancri\,e. The authors reported a non-detection of water vapour in the optical after combining ESPaDOns \citep[506--795\,nm,][]{Donati03} and Subaru HDS \citep[524--789\,nm,][]{Noguchi02} data. However, no similar studies have been reported for \hd20 in the optical region so far. 

Here, we report a detection of water vapour in \hd20 in the optical (0.70--0.96\,$\mu$m) from one transit. In contrast, we do not detect \h2o in \hdu18 in this spectral region even after combining four transits, which is in line with previous studies. In Sect. \ref{observations} we describe the observations and first steps of the analyses. In Sect.\,\ref{data_analysis}, we describe the telluric and stellar signal removal using the {\tt SYSREM} algorithm and present the cross-correlation technique to extract planet atmospheric signals. In Sect.\,\ref{results_discussion} we discuss the results and their implications followed by the conclusions in Sect. \ref{conclusions}.

\section{Observations and data reduction} 
\label{observations}

\begin{figure*}[ht]
\includegraphics[angle=0, width=1.0\columnwidth]{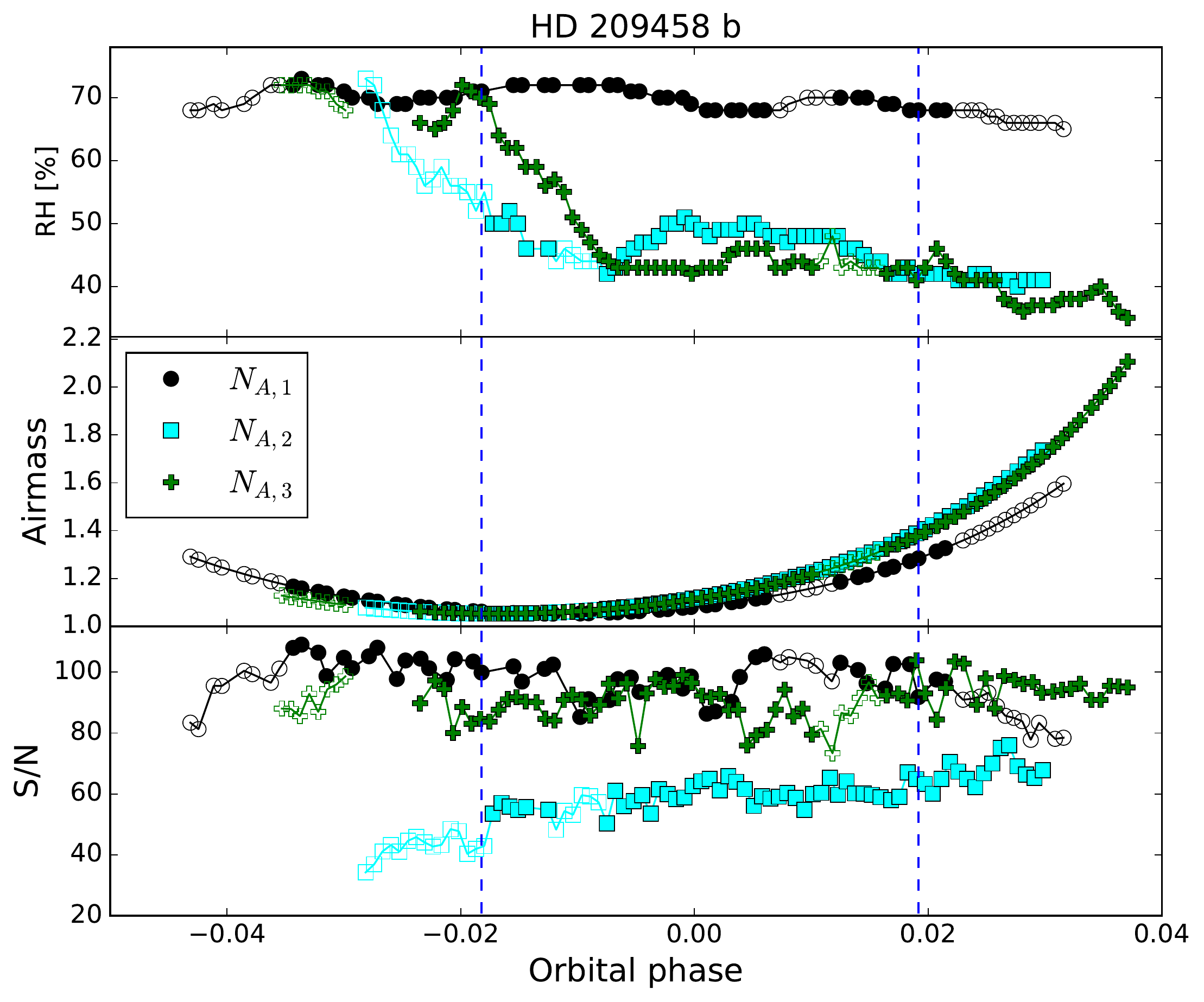}\includegraphics[angle=0, width=1.0\columnwidth]{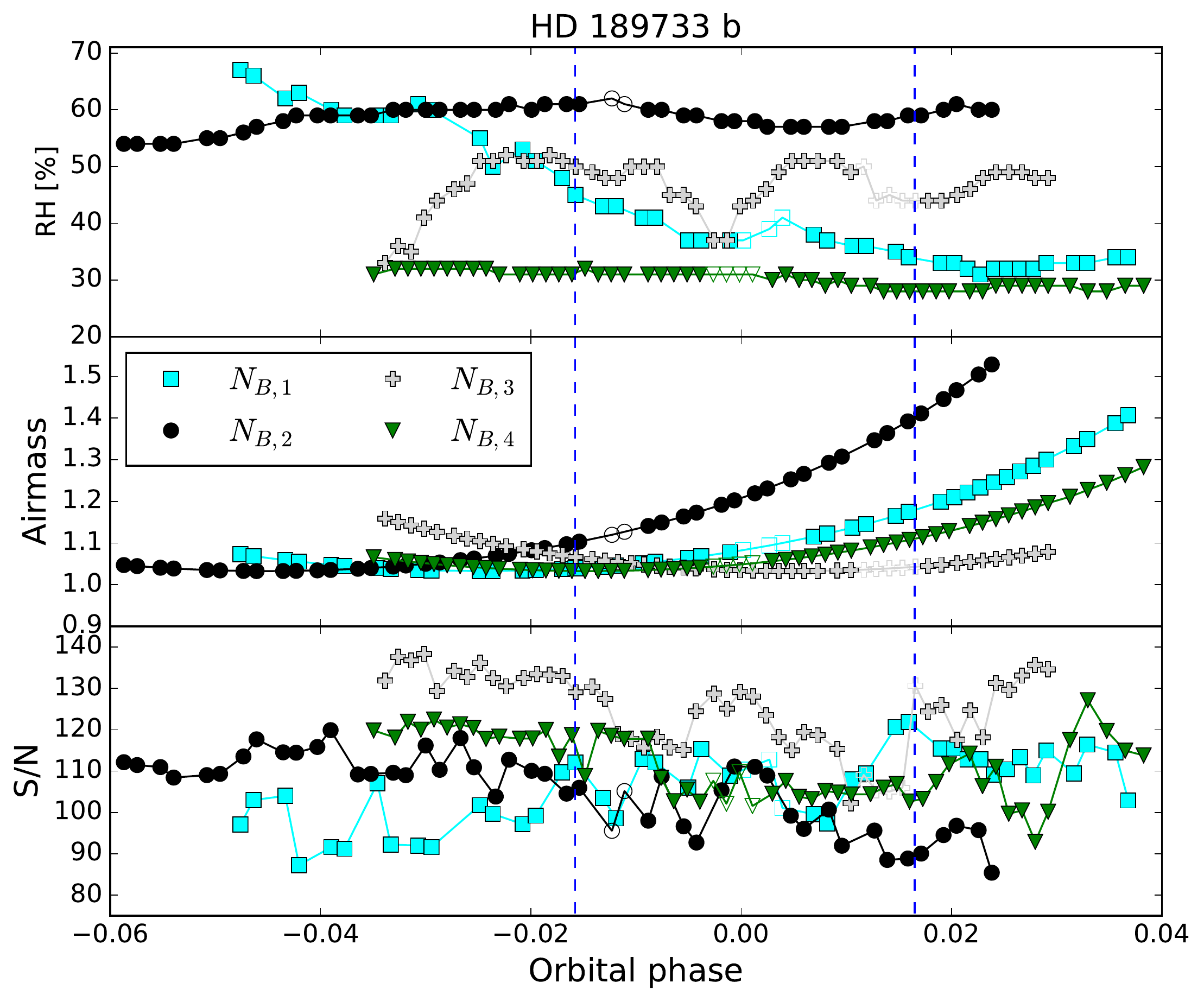}
\caption{Relative humidity (\textit{top panel}), air mass (\textit{middle panel}), and mean S/N per spectra (\textit{bottom panel}) during the transit observations of \hd20 (\textit{left}) and \hdu18 (\textit{right}). The transits occur between the vertical dashed lines. Out-of-transit open symbols represent the spectra that are not included in the analyses because of their low S/N. Open symbols during transit represent spectra that are not included due to the overlap of telluric and planet lines when the latter has a near-zero velocity in the observer's frame.} 
\label{obs_conditions}
\end{figure*}

We used the CARMENES spectrograph \citep{Quirrenbach16, Quirrenbach18} to observe three transits of \hd20 (target A) during the nights of 16 September 2016 (hereafter $N_{A,\,1}$), 8 November 2016 (hereafter $N_{A,\,2}$), and 6 September 2018 (hereafter $N_{A,\,3}$), and four transits of \hdu18 (target B) on 8 August 2016 (hereafter $N_{B,\,1}$), 16 September 2016 (hereafter $N_{B,\,2}$), 30 June 2019 (hereafter $N_{B,\,3}$), and 9 August 2019 (hereafter $N_{B,\,4}$). CARMENES is mounted at the 3.5\,m telescope at the Calar Alto Observatory (Almer\'ia, Spain). We analysed the observations of its visible channel in the 0.52--0.96\,$\mu$m spectral region at a resolving power of $\mathcal{R}=96,000$. We restricted the analyses to the spectral orders covering the 0.70--0.96\,$\mu$m interval, where the strongest \h2o absorption occurs (see Fig.\,\ref{model_spec}). We did not include the weak \h2o band at $\sim$0.65\,$\mu$m, because of its expected small contribution. 

The relative humidity, air mass, and mean signal-to-noise ratio (S/N) per spectrum for each night are shown in Fig.\,\ref{obs_conditions}. In each case, the raw spectra were processed using the CARMENES data reduction pipeline {\tt caracal} v2.20 \citep{Zechmeister2014, Caballero2016} and, subsequently, analysed using custom Python subroutines. For $N_{A,\,1}$, eight spectra observed at the beginning of the night and another 12 recorded at the end were discarded due to their decreasing S/N, which hampered the telluric correction procedure discussed below (see out-of-transit open symbols in Fig.\,\ref{obs_conditions}). Regarding $N_{A,\,2}$, we discarded the first 15 pre-transit spectra to avoid low (<\,50) S/N observations. The significantly lower S/N of the observations during this night was due to worse seeing and extinction conditions. Unfortunately, the observational epoch in all nights caused some in-transit spectra to be recorded at times when the velocity of the exoplanet with respect to the Earth ranged from --2.6 to +2.6\,km\,s$^{-1}$ (two wavelength steps around zero). This yielded a poor Doppler separation of the exo-atmospheric lines from the telluric contribution, which would result in a strong telluric contamination in the final signal. Therefore, these spectra were also discarded from the analysis (see open symbols during transit in Fig.\,\ref{obs_conditions}).

We normalised each spectral order by using second-order polynomial fits for their pseudo-continuum. This step provided a self-calibration that allowed us to compare flux variations between spectra at small spectral scales. Next, we masked the spectral regions where the telluric absorption was larger than 80\% of the flux continuum. Unfortunately, this necessary step also prevented us from obtaining information from the pixels with the largest possible \h2o signal from the planet atmosphere. In addition, we used the methods described in \citet{Alonso19} and \citet{Sanchezlopez2019} to mask the outliers present in the data, which are likely produced by cosmic rays, and telluric emission lines.

\section{Data analysis}
\label{data_analysis}
\subsection{Detrending using {\tt SYSREM}} 
\label{sysrem}

With the objective of recovering the weak exo-atmospheric signal, we removed the telluric and stellar contributions that dominated the spectral matrix for each night. We used {\tt SYSREM} \citep{Tamuz05, Mazeh07}, which is a principal component analysis algorithm that has been widely used and tested in the past for exo-atmospheric studies \citep{Birkby13, Birkby17, Nugroho17, Hawker2018, Alonso19, Sanchezlopez2019, Gibson2020, Stangret2020}. In total, we ran 15 {\tt SYSREM} iterations in each spectral order, night, and target, and we stored the resulting 2520 residual matrices for further analyses. The individual treatment of the orders allowed us to take into account their different contamination levels, caused by different telluric or stellar contributions, their different detector efficiencies, or changing instrumental responses during the night. An imperfect removal of these contributions is unavoidable regardless of the number of iterations performed because {\tt SYSREM} fits the spectral matrix by working with a set of two linear and independent coefficients that change with time and wavelength, respectively. The residual spectral matrix is obtained after subtracting the best least square fit from the original spectral matrix, which presents a non-linear behaviour by itself. Nevertheless, the aforementioned studies extensively showed the suitability of this algorithm for removing telluric and stellar contributions even in the presence of high water vapour levels above the observatory.

After a certain number of {\tt SYSREM} iterations, which could be different for each spectral order, the algorithm also identifies and removes a possible signal from the exo-atmosphere. In order to minimise this effect, we injected a synthetic \h2o signal at the expected velocities for the planets early in the procedure (i.e. before the telluric correction) and analysed its behaviour. We used the cross-correlation technique (see Sect.\,\ref{signalretrieval}) to maximise the recovery of the injected signal \textbf{in each spectral order and} for each of the seven nights (see Figs.\,\ref{sn_evol_hd20} and \ref{sn_evol_hd18}). Since the water vapour bands in the 0.70--0.96\,$\mu$m spectral range are rather weak, we injected the signal at five times ($5\times$) the expected strength to ensure that it was clearly recovered above the noise. The spectral orders for which the S/N of the injected signal peak was lower than three (below the dashed horizontal lines in both figures) were discarded due to the very small actual signal that we would expect to retrieve from them. For the remaining orders and for the rest of the analysis, we used the iteration that allowed us to maximise the recovery of the injected signal (marked with star symbols in Figs.\,\ref{sn_evol_hd20} and \ref{sn_evol_hd18}).

Additionally, we visually inspected the cross-correlation results for each spectral order and each night individually in search for possible residuals that might not be detected in the previous step. For instance, a strong injected signal might still be retrieved above a strong residual, but the latter would mask a weaker, real peak. We found that some spectral orders presented noise structures around the expected planet velocities or telluric residuals in the form of large cross-correlation values around 0\,km\,s$^{-1}$ (i.e. the Earth's rest-frame). Hence, these spectral orders were left out of the analyses (grey curves in the figures of Appendix\,\ref{appendix:a}). 

In total, the discarded spectral orders due to both the weak recovery of injections and the presence of telluric residuals represented 39\%, 46\%, and 50\% of the total data of \hd20 on $N_{A,\,1}$, $N_{A,\,2}$, and $N_{A,\,3}$, respectively. For \hdu18, 50\%, 25\%, 32\%, and 26\% of the total data were excluded on $N_{B,\,1}$, $N_{B,\,2}$, $N_{B,\,3}$, and $N_{B,\,4}$, respectively. The highest fractions of lost spectral orders seem to correlate with the nights presenting a higher variability of the relative humidity (see Fig.\,\ref{obs_conditions}). Therefore, a large telluric variability can preclude a successful telluric removal with {\tt SYSREM} in some spectral orders, as is discussed by \citet{Sanchezlopez2019}.

\subsection{Planet atmospheric signal extraction}
\label{signalretrieval}

In order to probe the planet atmospheric signal, we followed the same procedure as \citet{Sanchezlopez2019}. We cross-correlated the residual spectra resulting from {\tt SYSREM} with high-resolution templates of the \h2o absorption computed at the CARMENES spectral resolution with {\tt KOPRA} \citep{Stiller2002}. In order to compare our results directly with our previous findings, we used the same atmospheric parameters for \hd20 and \hdu18 as \citet{Sanchezlopez2019} and \citet{Alonso19}, respectively (see Fig.\,\ref{model_spec}). 

Due to the largely unknown pressure level at which the local continuum around the water vapour bands is produced, these observations, similar to most transit measurements, result in a large intrinsic uncertainty in absolute abundances \citep[e.g. Fig.\,9 in][]{Alonso19}. Therefore, we did not test other combinations of temperatures or \h2o abundances for these atmospheres. 

The cross-correlations were performed for each night and spectral order separately in a velocity interval from --200 to +200\,km\,s$^{-1}$  (with respect to the Earth's rest-frame) in steps of 1.3\,km\,s$^{-1}$, which corresponds to the mean pixel size of the instrument in the red optical range. At this stage, no planet atmospheric features could be observed in the orders individually. With the objective of enhancing a possible signal, the cross-correlations matrices of all orders were co-added to form one total cross-correlation matrix for each night. Consecutively, we shifted the matrices to the exoplanet rest-frame and co-added all in-transit spectra over time, which allowed us to obtain a 1D cross-correlation function (CCF) per night. We tested a range of planet orbital velocities ($K_p$), creating a grid from --280 to +280\,km\,s$^{-1}$  (i.e. 561 CCFs). This allowed us to inspect a wide velocity space to check for spurious signals, indications of telluric or stellar residuals at low $K_p$ values, and possibly strong correlated noise sources appearing in the (unphysical) negative $K_p$ space \citep{Birkby17}.

In order to assess the significance of the observed signals, we calculated the S/N of each CCF at each $K_p$ by dividing each cross-correlation value by the standard deviation obtained from the rest of the CCF velocity interval, excluding a $\pm$15.6\,km\,s$^{-1}$  region around it. The selection of different velocity intervals can impact the assessment of the noise and, hence, we chose a wide interval (i.e. $\pm$\,200\,km\,s$^{-1}$)   to obtain an accurate measurement of each CCF noise.

\section{Results and discussion}
\label{results_discussion}

\subsection{\hd20}
\label{hd20_results}

\begin{figure*}
\centering
\includegraphics[angle=0, width=0.62\columnwidth]{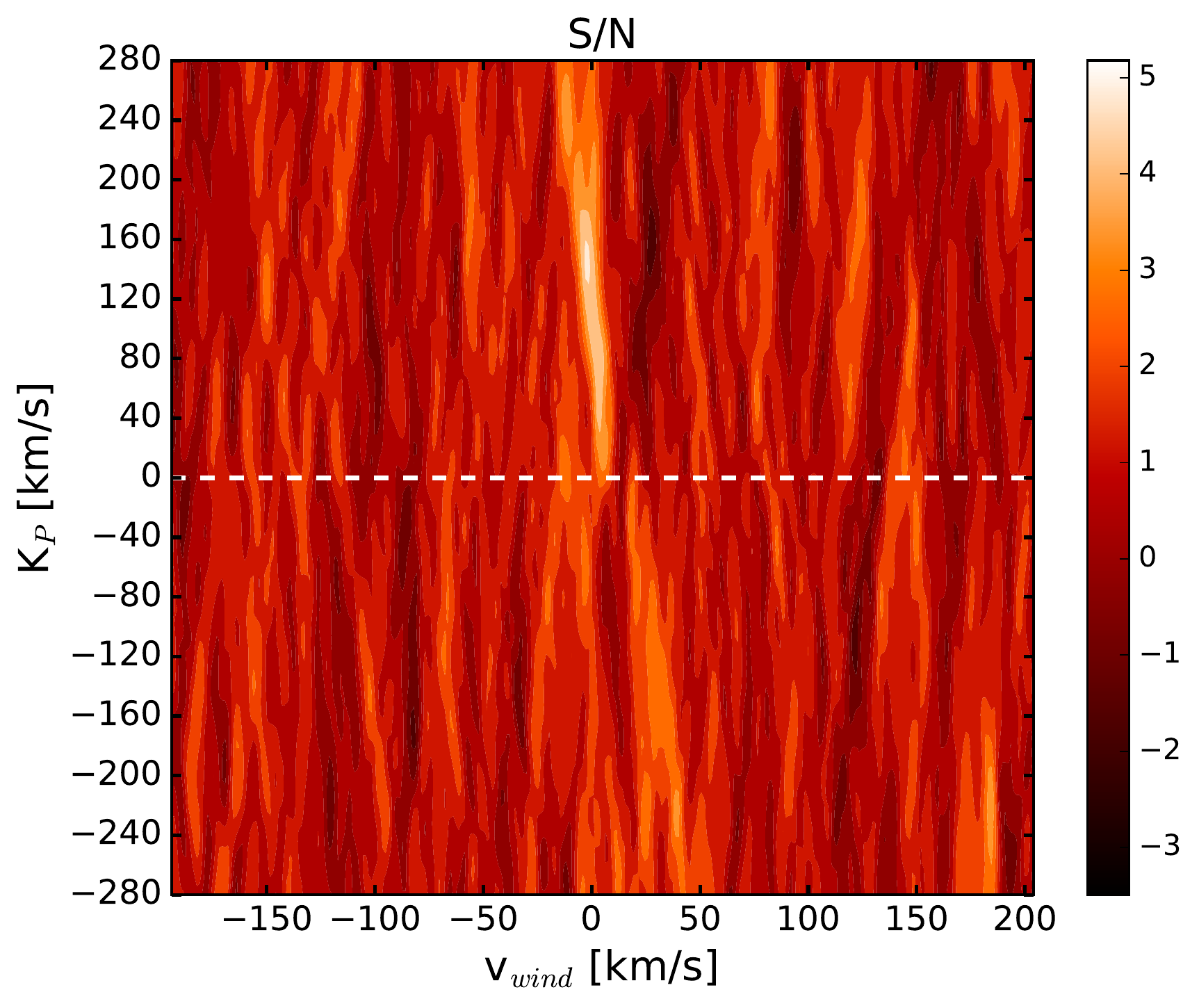}\includegraphics[angle=0, width=0.65\columnwidth]{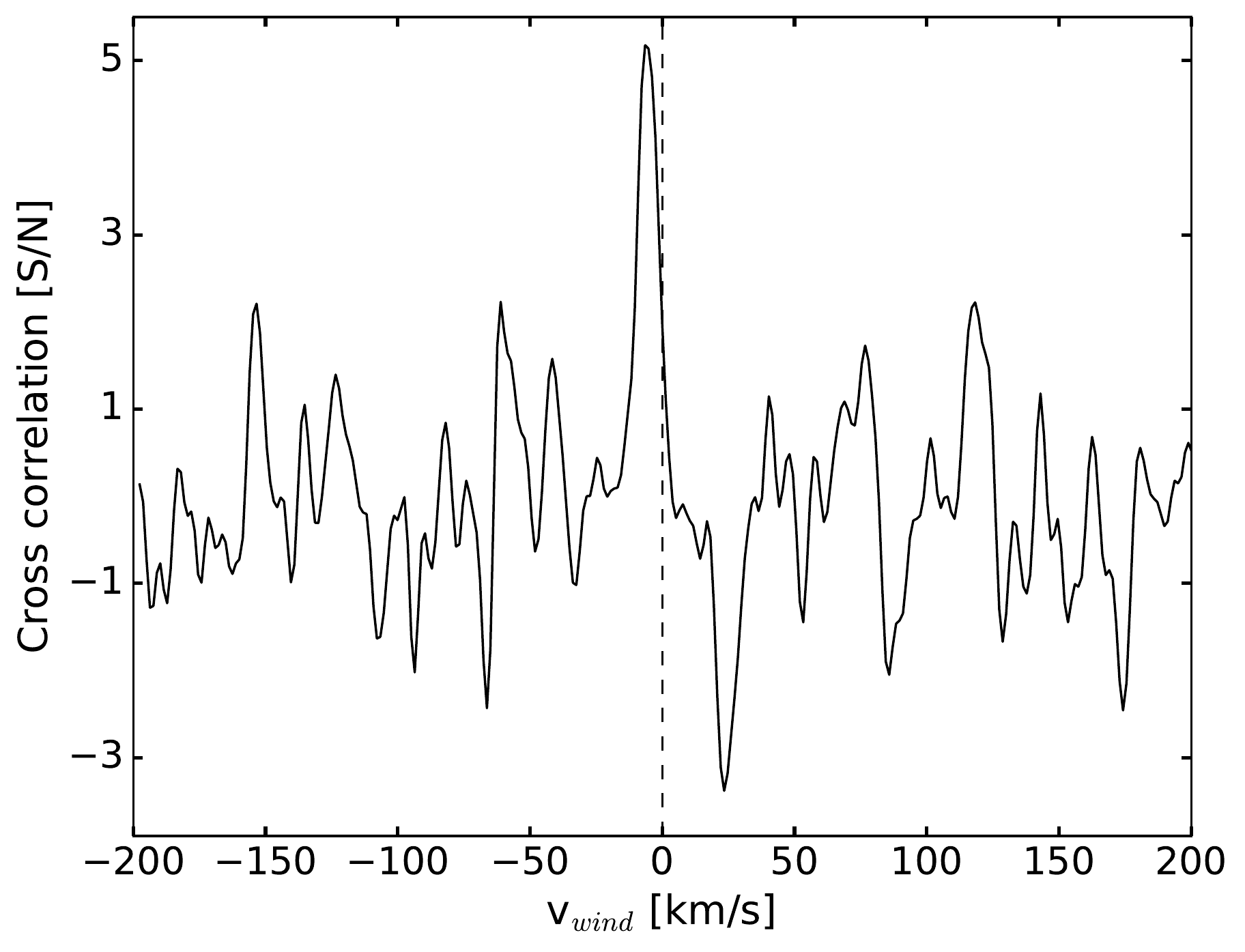}\includegraphics[angle=0, width=0.76\columnwidth]{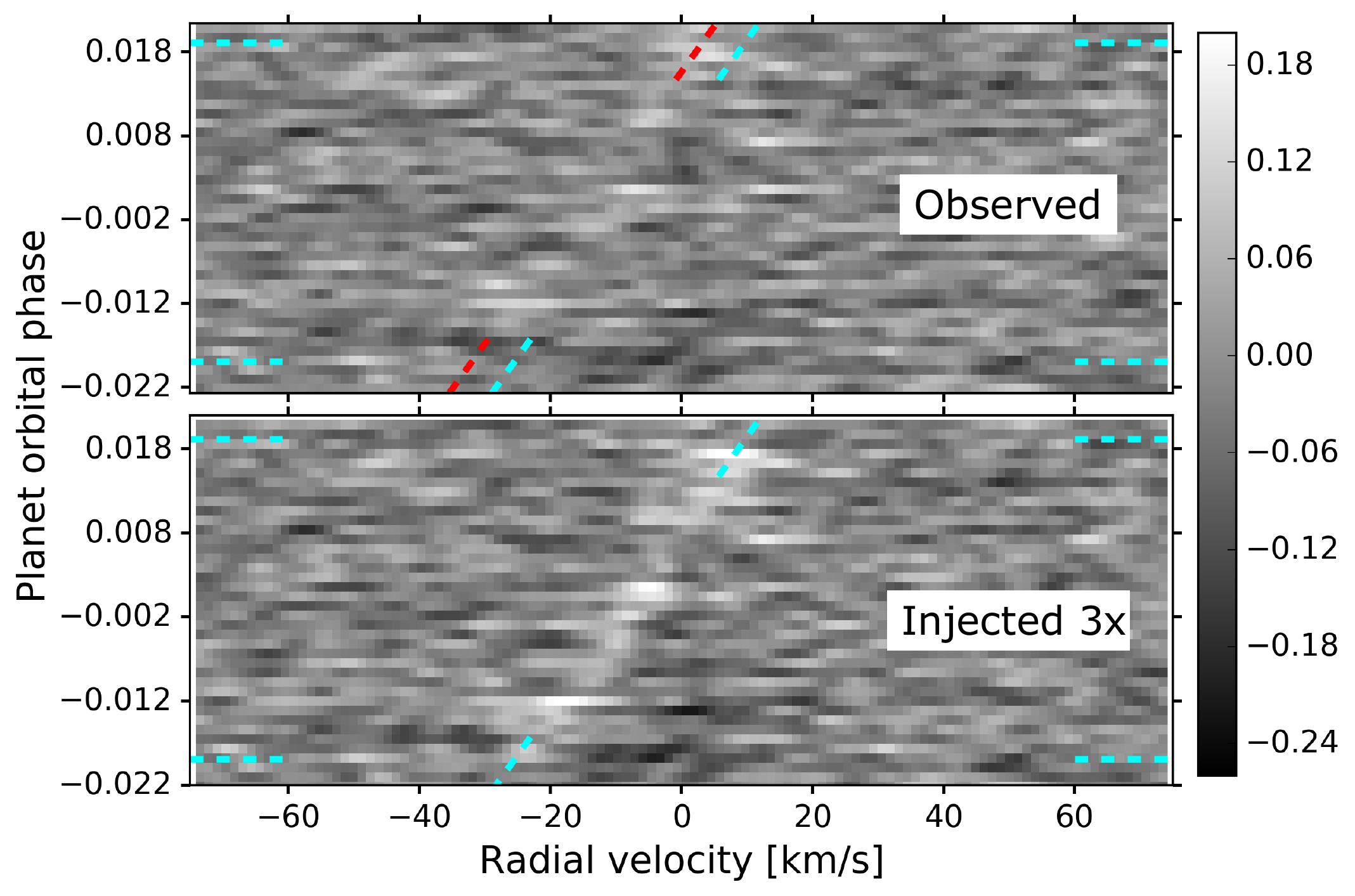}\\
\includegraphics[angle=0, width=0.63\columnwidth]{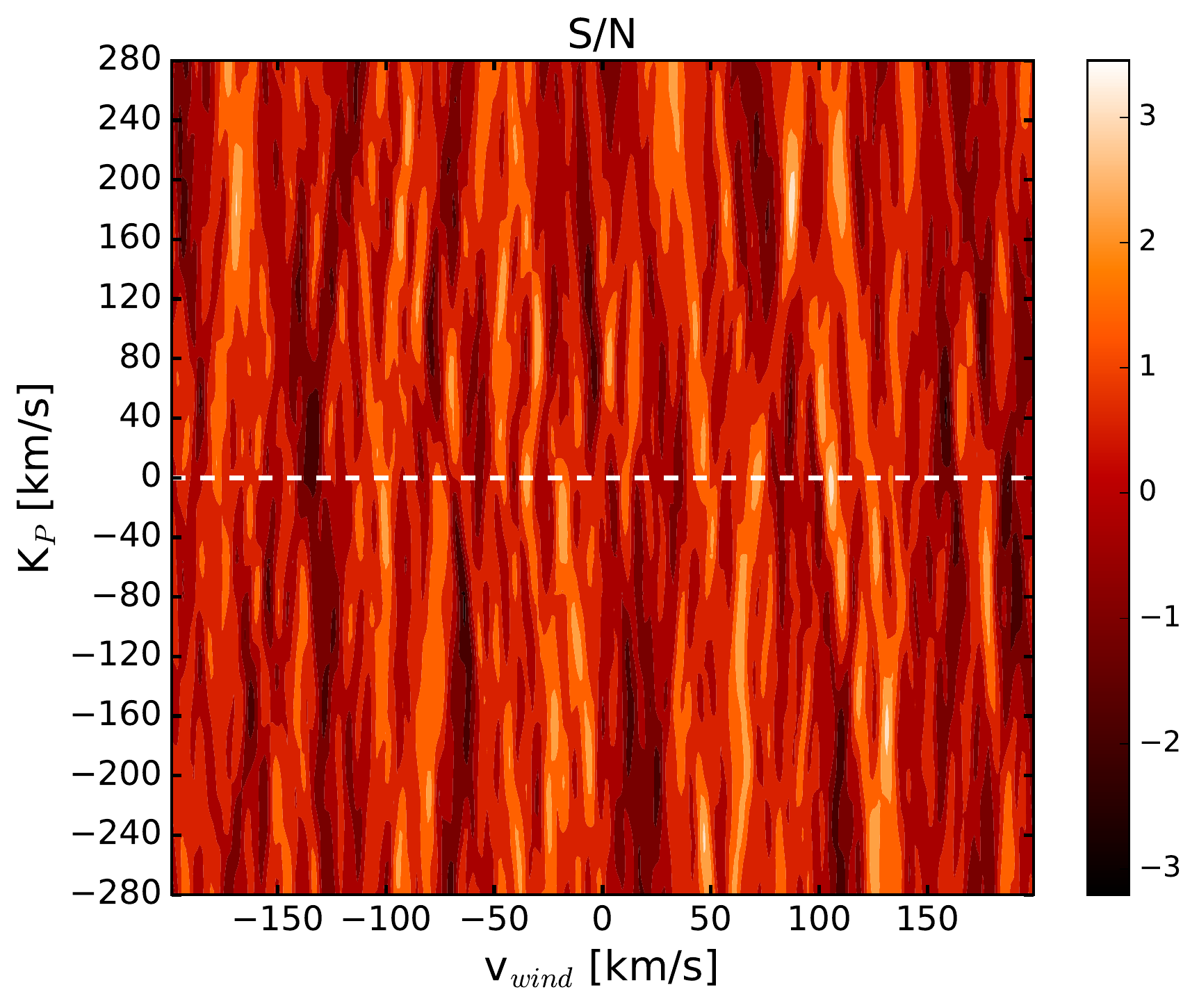}\includegraphics[angle=0, width=0.65\columnwidth]{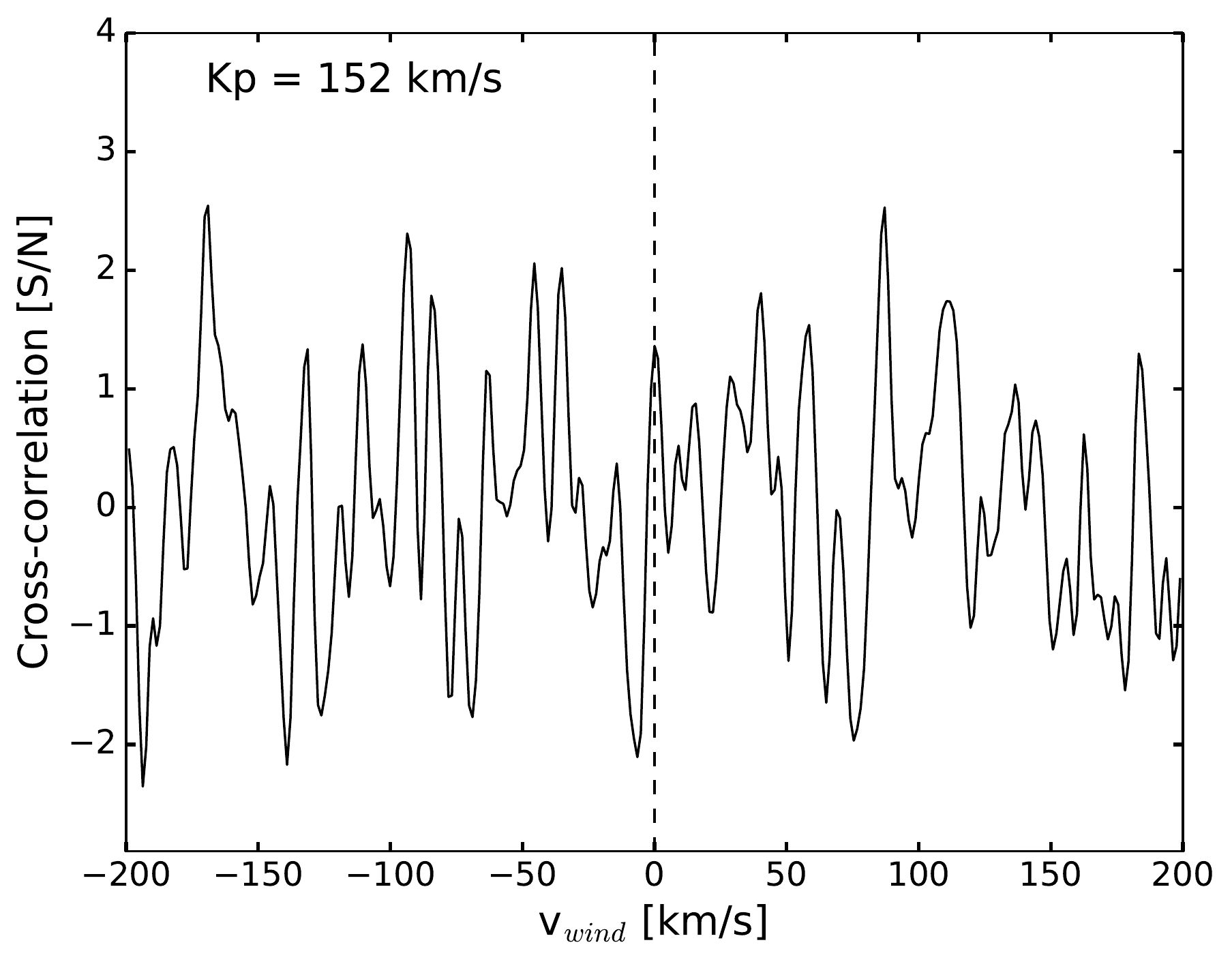}\includegraphics[angle=0, width=0.76\columnwidth]{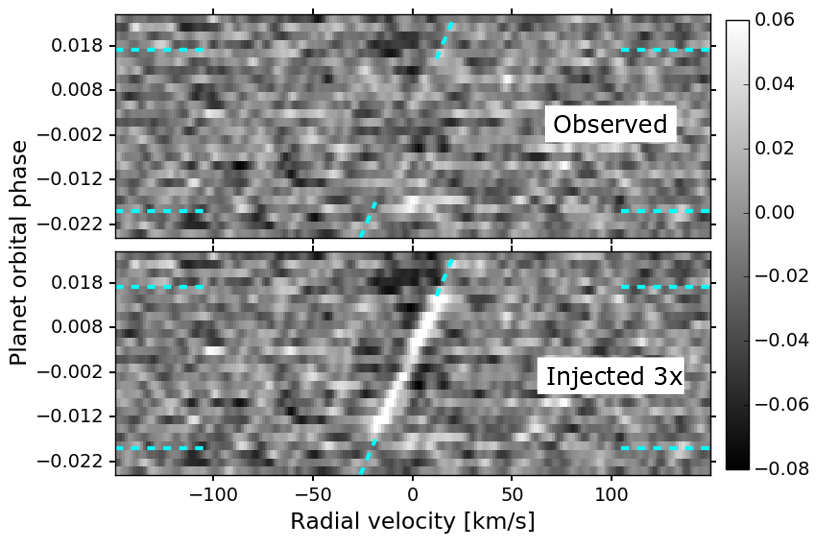}
\caption{Cross-correlation results obtained on the night  N$_{A,\,1}$ for \hd20 (\textit{top row}) and after combining four transits of \hdu18 (\textit{bottom row}). \textit{Left column}: S/N maps of potential water vapour signals with respect to the exoplanet rest-frame (horizontal axis) and K$_p$ (vertical axis). \textit{Middle column}: Slice through the left panels at the $K_p$ with the largest significance peak for \hd20 (145\,km\,s$^{-1}$) and at the expected $K_p$ of \hdu18. \textit{Right column}: Cross-correlation matrices in the Earth rest-frame for the observed data and after injecting a signal at 3$\times$ the expected strength. The transit occurs between the cyan horizontal dashed lines. The cyan tilted dashed lines mark the expected velocities of the exoplanets during the observations. The tilted red dashed line in the case of \hd20 marks the position of the most significant signal observed.}
\label{results}
\end{figure*}

We measured a water vapour signal on the night $N_{A,\,1}$ in \hd20 with an S/N\,=\,5.2 at $K_p$\,=\,145$\pm31$\,km\,s$^{-1}$, which is a planet orbital velocity that is in good agreement with the widely accepted value of 145.9\,km\,s$^{-1}$  \citep{Brogi17} and previous results \citep{Snellen10, Hawker2018, Brogi19, Sanchezlopez2019}. 
The top row of Fig.\,\ref{results} shows the S/N map with respect to the planet rest frame as well as a function of $K_p$ (left panel) and a slice at $K_p$\,=\,145\,km\,s$^{-1}$, showing the CCF peak with maximum significance (middle panel). The obtained signal follows the expected trail of the exoplanet during the observations as depicted in the cross-correlation matrix in the Earth's rest-frame (right panel). 
However, it shows a significant blueshift of $-6.5^{+3.9}_{-2.6}$\,km\,s$^{-1}$, which is consistent with what was measured in the near-infrared from the long-wavelength wing of the \h2o band at $\sim$1\,$\mu$m \citep[$-6.5^{+2.6}_{-1.3}$\,km\,s$^{-1}$, ][]{Sanchezlopez2019} and with model predictions \citep{Rauscher2012, Showman13, Amundsen2016}. This blueshift could be caused by global high-altitude winds blowing at the terminator from the day to the night side hemisphere of \hd20.

With the objective of studying the robustness of the observed \h2o signal, we repeated the cross-correlation analysis of this night by using the same number of {\tt SYSREM} iterations in all the useful spectral orders, which minimises model-dependencies at the expense of a possible under- or over-correction in some of them \citep{Alonso19, Stangret2020}. We were able to recover a very similar $5 \sigma$ blueshifted signal at the same planet $K_p$ after applying five {\tt SYSREM} iterations to all orders (see Fig.\,\ref{sn_map_5_its}). For a larger number of iterations, the signal was increasingly removed by the algorithm (Fig.\,\ref{sysrem_evol_freeze}).

Furthermore, we investigated the individual contributions from the three water vapour bands probed in the 0.70--0.96\,$\mu$m spectral region. The resulting S/N maps are shown in Fig.\,\ref{multiband_hd20}. We did not find significant signals around the expected $K_p$\,=\,145.9\,km\,s$^{-1}$ for any of the bands individually. The strongest contributions were observed at that $K_p$ for the 0.82\,$\mu$m and 0.95\,$\mu$m bands, but these signals were not significant enough to claim single-band detections. Also, residual signals were observed at other velocities, making them somewhat uncertain. Therefore, the total signal discussed above did not arise from a small sample of spectral points, but rather from using a wide spectral interval that included several \h2o bands.
In particular, the 0.95\,$\mu$m band was already detected with the CARMENES data from its infrared channel (\citet{Sanchezlopez2019} by using the long-wavelength wing of the water feature. The optical data presented here mainly covers the short-wavelength wing, showing a weak signal with an S/N\,=\,3.2 at the expected $K_p$. We note that the efficiency of the optical arm of CARMENES around these long wavelengths decreases, reducing the S/N of the observations of this band. 

The same procedure was applied for the nights $N_{A,\,2}$ and $N_{A,\,3}$. However, we did not find any evidence of planetary water absorption in these nights (see Fig.\,\ref{sn_cc_hd20_mask_02}). In particular, in $N_{A,\,2}$, the systematic residuals at 0\,km\,s$^{-1}$  (Fig.\,\ref{sn_cc_hd20_mask_02}, top right) caused the telluric signal at $K_p$\,=\,0\,km\,s$^{-1}$  to dominate the S/N map (Fig.\,\ref{sn_cc_hd20_mask_02}, top left). For $N_{A,\,3}$, a strong contamination at an orbital phase $\sim$\,0.1 (Fig.\,\ref{sn_cc_hd20_mask_02}, bottom right) produced spurious signals that peaked in the negative $K_p$ space (Fig.\,\ref{sn_cc_hd20_mask_02}, bottom left).

\begin{figure}[htb!]
\centering
\includegraphics[angle=0, width=0.98\columnwidth]{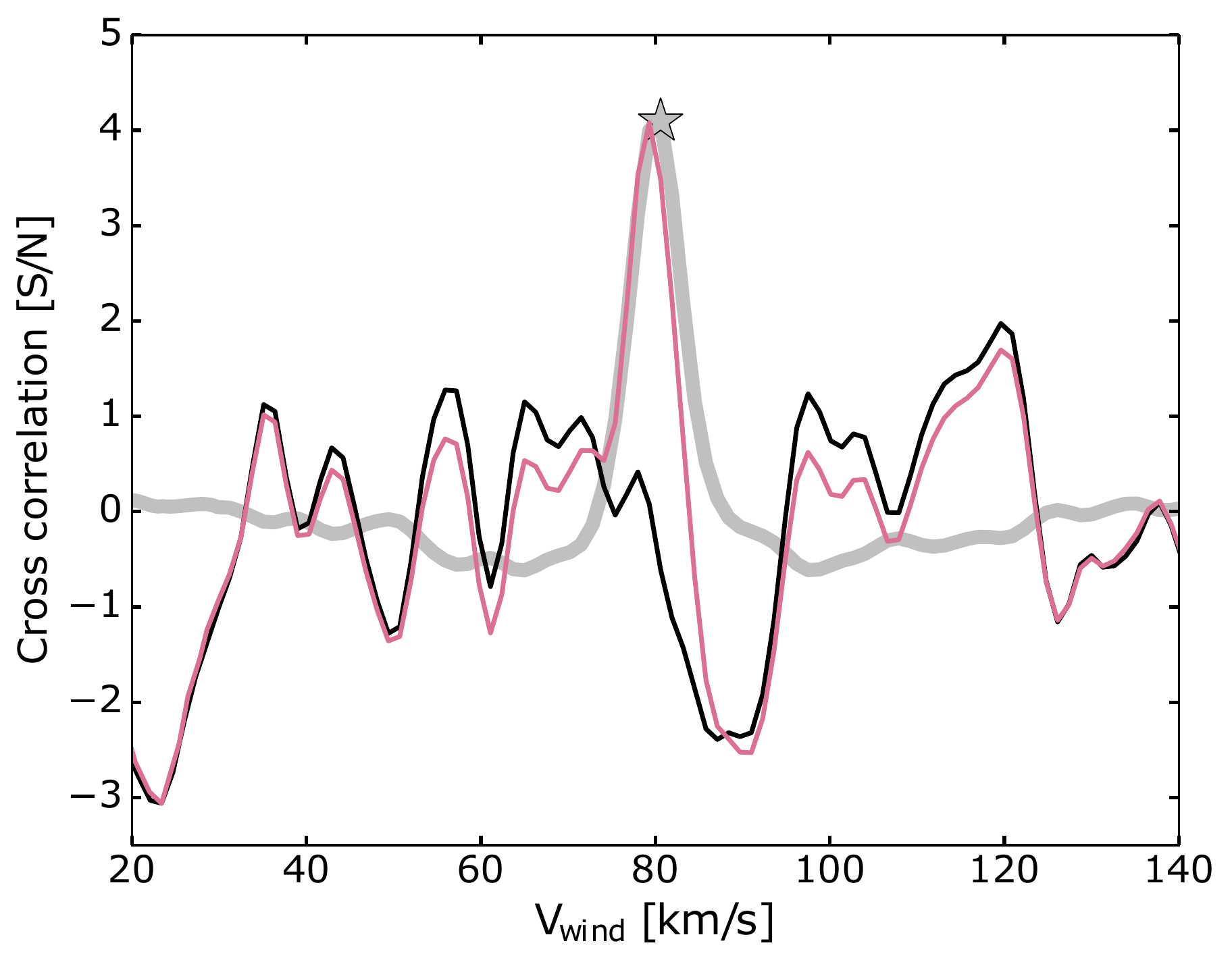}
\includegraphics[angle=0, width=0.93\columnwidth]{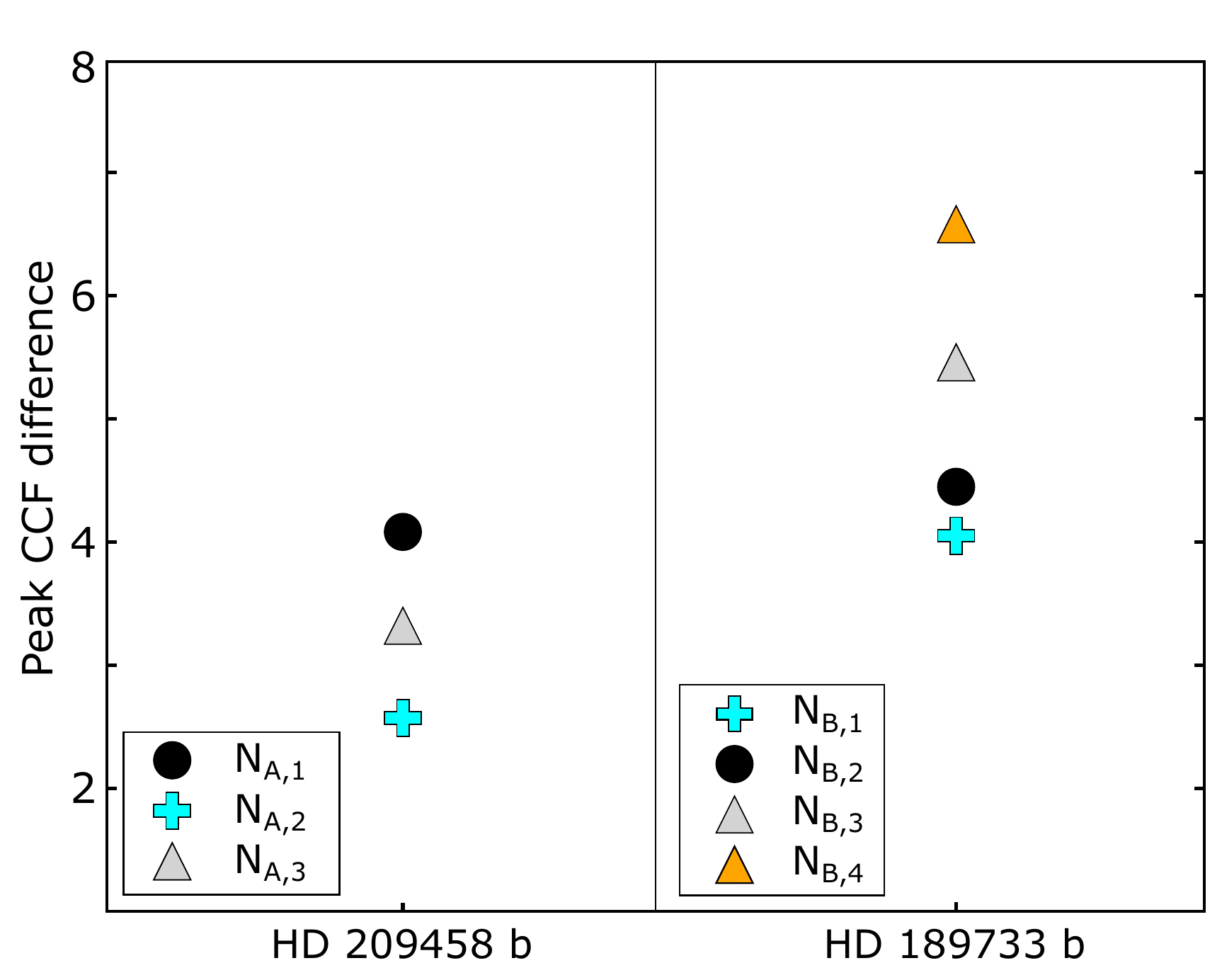}
\caption{\textit{Top}: Representative CCF obtained with an injected signal at the expected level (magenta) and without an injection (black) at a $K_p$ of 180\,km\,s$^{-1}$ and $\varv_{\text{wind}}$\,=\,80.6\,km\,s$^{-1}$. The difference between both CCFs is shown in grey and its peak is marked with a star symbol. The illustration corresponds to the results of $N_{A,\,1}$. The CCFs for the other nights have similar shapes, but different peak values. \textit{Bottom:} Peak differences between the CCFs with and without an injection for all \hd20 (\textit{left}) and \hdu18 (\textit{right}) data sets.} 
\label{injection_tests}
\end{figure}

Since we only observed a significant \h2o CCF peak in one of the three transits of \hd20, we assessed the capabilities of our methodology for observing this planet's atmosphere on the three nights. We injected signals at the expected level (1$\times$) of \h2o absorption and investigated their recovery (see top panel of Fig.\,\ref{injection_tests}). These types of signals can be rather weak in individual orders, especially if they cover a region with low \h2o absorption. Thus, instead of using the peak value of the CCF with injection directly, we studied the difference between the CCFs with an injection and without one. This metric is robust against the contribution of potential noise sources at the velocities of the injection, which are subtracted. Furthermore, in order to avoid the influence of any potential real planetary signals or telluric residuals, we injected the model at very different velocities from those expected for the planet (i.e. $K_p$\,=\,180\,km\,s$^{-1}$,  $\varv_{\text{wind}}$\,=\,80.6\,km\,s$^{-1}$).

We found that the injected signal was significantly better recovered on $N_{A,\,1}$ with respect to $N_{A,\,2}$ and $N_{A,\,3}$, in which the planet signal was at the noise level (see bottom left panel of Fig.\,\ref{injection_tests}). In other words, if there was \h2o optical absorption in \hd20, we would most likely only be able to detect it with this methodology on
$N_{A,\,1}$, which is what we observed. The reasons behind the poorer recovery of injections on $N_{A,\,2}$ and $N_{A,\,3}$ are unclear, although they were likely related to the poor S/N of the spectra on $N_{A,\,2}$ and the unsuccessful telluric removal on both nights. The latter could be caused by a high telluric variability, inducing additional non-linear trends that hinder the performance of {\tt SYSREM}.
Indications of this type of variability can be inferred from the observed changes in the relative humidity during these nights (see top panel on the left side of Fig.\,\ref{obs_conditions}). In particular, the rather rapid variability of the telluric water vapour content on $N_{A,\,3}$ during the transit could also be the cause of the unsuccessful telluric removal in the reddest orders of the CARMENES near-infrared channel presented by \citet{Sanchezlopez2019}. Finding the exact reasons behind the different injection recoveries would require an extensive study of the telluric conditions for the different nights. However, we did not perform this type of analysis since the injection recoveries discussed above already provided us with a metric of the goodness of each data set for finding \h2o signals using the CCF technique.

\subsection{\hdu18}
\label{hd18_results}

\begin{figure}[ht]
\centering
\includegraphics[angle=0, width=1\columnwidth]{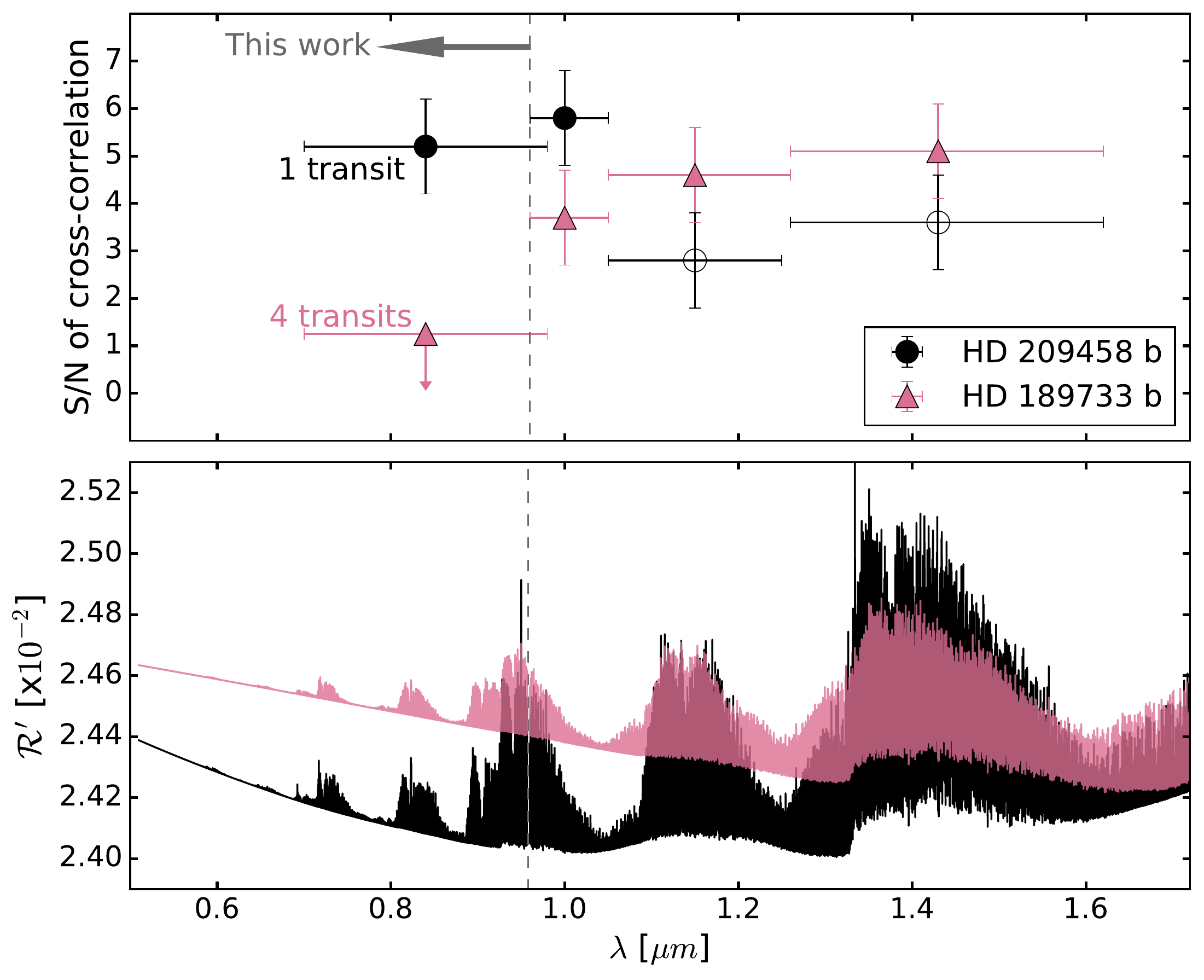}
\caption{\textit{Top}: S/N of the \h2o cross-correlation signal obtained for different wavelength intervals (indicated by horizontal bars) for \hd20 (black and open circles) and \hdu18 (magenta triangles) with CARMENES. The $1 \sigma$ uncertainties are indicated by the vertical bars. The results in the 0.96--1.71\,$\mu$m spectral region were obtained with one transit for \hd20 and \hdu18 in \citet{Sanchezlopez2019}. Open symbols show spectral regions for which the measurements of \hd20 are uncertain due to a problematic telluric removal. The larger signal towards shorter wavelengths in \hd20 compared to \hdu18 is indicative of the higher atmospheric transparency of the former planet. \textit{Bottom}: Model \h2o transmission spectra for \hd20 assuming a clear atmosphere (black) and for \hdu18 including a stronger haze extinction at an arbitrary level (magenta). The spectrum of 
\hd20 is offset by $10^{-2}$ for illustration purposes.} 
\label{comparison}
\end{figure}

For \hdu18, we did not find significant water vapour signatures in any of the four transits analysed (see Fig.\,\ref{sn_cc_hd18_mask_02}). On $N_{B,\,1}$ and $N_{B,\,3}$, the S/N maps were dominated by telluric residuals at a $K_p$ of 0\,km\,s$^{-1}$  (see first and third rows in Fig.\,\ref{sn_cc_hd18_mask_02}). The best recovery of injections was found on $N_{B,\,3}$ and $N_{B,\,4}$ (see bottom right panel in Fig.\,\ref{injection_tests}), for which we did not observe any evidence of \h2o optical absorption. The maximum significance CCF at the expected $K_p$ of 152.5\,km\,s$^{-1}$ was found for $N_{B,\,2}$ with an S/N of 3.1. However, this night presented a significantly weaker recovery of injections, suggesting that this low-significance CCF peak was likely produced by noise. Given the similar S/N of the four transit measurements, the different observational conditions were likely the reason for the different strength of the injection recoveries.

In order to improve the detectability of a potential signal from \hdu18, we combined the cross-correlation matrices from the four transits weighted by their respective strength of injection recoveries (see bottom row in Fig.\,\ref{results}). 
However, it did not allow us to identify any significant signals either. We repeated the analysis by combining only $N_{B,\,2}$ and $N_{B,\,4}$, which presented fewer telluric residuals in their S/N maps and also had the most stable relative humidity conditions (see Fig.\,\ref{obs_conditions}), but we found similar results (i.e. non-detection of optical \h2o absorption). However, the injected models of \hdu18 at the expected strength were well recovered in all data sets, which suggests that it would be possible to detect water vapour if the relative depth of the planetary \h2o lines were similar to those of our cloud-free model. In assuming the $\sim$1$\sigma$ CCF peak that we observed at the expected velocities (i.e. $K_p$\,=\,152.5\,km\,s$^{-1}$  and $\varv_{\text{wind}}$\,=\,0\,km\,s$^{-1}$) was a real planet atmospheric signal, we estimated that between 60 and 70 transits, observed under similar conditions to those presented here, would be required to detect the water vapour bands in the red optical at $4 \sigma$ with CARMENES.

\subsection{Comparison of the optical \h2o analysis in both planets}
\label{planet_comparison}

The results obtained for both planets when applying the same \h2o cross-correlation analyses allowed us to compare the two different atmospheres. We were able to detect \h2o in \hd20 from the red optical bands, whereas we did not find any significant signals in four transit data sets of \hdu18 over the same spectral range. However, previous studies have reported the detection of a similarly strong \h2o absorption in both planets in the near-infrared (see top panel in Fig.\,\ref{comparison}). In this context, we interpret the non-detection of \h2o optical signals in \hdu18 to be due to the presence of atmospheric hazes, which cause a strong extinction and mute most of the water vapour molecular features (see bottom panel in Fig.\,\ref{comparison}). In contrast, we found that \hd20 presents a clearer atmosphere, which facilitates the identification of the weak \h2o optical signatures. Furthermore, these results are in line with space observations of these objects, which found a steep Rayleigh scattering slope in \hdu18, but a more gentle one in \hd20 \citep{Sing16, Barstow17}. However, we were not able to discard the presence of other opacity contributors in the red-optical, such as TiO and VO, which might also present a contribution. This was because {\tt KOPRA} does not yet include the opacities of these molecules and, in the case of the publicly available {\tt petitRADTRANS} \citep{Molliere2019}, the linelists used are those from \citet{Plez1998}, which might not be accurate enough for high-resolution CCF studies \citep{Merritt2020}.

\section{Conclusions} \label{conclusions}

We have presented the analysis of water vapour atmospheric signals in \hd20 and \hdu18 in the 0.70--0.96\,$\mu$m spectral interval, applying the cross-correlation technique to high-resolution CARMENES data of seven transits. We detected \h2o in the atmosphere of \hd20 using one transit data set, but we were not able to reveal \h2o with confidence in \hdu18 after combining four transits.
Since the latter planet exhibits a similar or even somewhat stronger \h2o signal in the near-infrared bands, we attribute the relative weakness of the optical water bands in \hdu18 to the presence of haze extinction, as already proposed for this planet from the strong UV and blue optical Rayleigh scattering signal \citep{Sing16, Barstow17}. This study shows that a distinct level of aerosol extinction in different exoplanets can be inferred in a similar way as proposed by \citet{Pino18} from currently available observations of high-resolution spectrographs on 4\,m-class ground-based telescopes. However, single band detections and comparisons are very challenging with our data sets due to the presence of telluric residuals, systematics, and other noise sources. This is especially important in the red optical, where the information contained in three \h2o bands had to be combined to detect water vapour in one transit of \hd20. Moreover, our tests using injected signals highlight the importance of telluric stability when analysing several transits of the same planet. Having similar observational conditions at Calar Alto on the different nights is a key ingredient in our capacity to disentangle planet atmospheric signals with CARMENES using transit data and cross-correlation. 

High-resolution observations at higher S/N and with better observational conditions than those we analysed could allow the exploration of the aerosol content in hot-Jupiter atmospheres using the methods proposed by \citet{Pino18}. In addition, this could be attempted by performing retrievals \citep[e.g.][]{Brogi17, Brogi19, Fisher2020}, including the strength of the wavelength-dependent haze extinction as an additional parameter. For instance, a simple model for small aerosol particles could be implemented as in \citet{Sing16} in the calculation of the templates. By using a wide spectral interval, the different strengths of \h2o bands should favour atmospheric models that include a larger opacity due to hazes, hence helping us constrain their general contribution. Thus, these analyses can complement the current knowledge obtained from space observations by using a different method.

\begin{acknowledgements}
A.S.L. and I.S. acknowledge funding from the European Research Council (ERC) under the European Union's Horizon 2020 research and innovation program under grant agreement No 694513. 
CARMENES is an instrument for the Centro Astron\'omico Hispano-Alem\'an de Calar Alto (CAHA, Almer\'ia, Spain). CARMENES is funded by the German Max-Planck-Gesellschaft (MPG), the Spanish Consejo Superior de Investigaciones Cient\'ificas (CSIC), the European Union through FEDER/ERF FICTS-2011-02 funds, and the members of the CARMENES Consortium (Max-Planck-Institut f\"{u}r Astronomie, Instituto de Astrof\'isica de Andaluc\'ia, Landessternwarte Königstuhl, Institut de Ci\`encies de l'Espai, Insitut f\"{u}r Astrophysik G\"{o}ttingen, Universidad Complutense de Madrid, Th\"{u}ringer Landessternwarte Tautenburg, Instituto de Astrof\'isica de Canarias, Hamburger Sternwarte, Centro de Astrobiolog\'ia and Centro Astron\'omico Hispano-Alem\'an), with additional contributions by the the Spanish Ministerios de Ciencia e Innovaci\'on
and of Econom\'ia y Competitividad, the Fondo Europeo de Desarrollo
Regional (FEDER/ERF), the Agencia estatal de investigaci\'on, the Fondo
Social Europeo under grants AYA2011-30 147-C03-01, -02, and -03, AYA2012-
39612-C03-01, ESP2013-48391-C4-1-R, ESP2014–54062–R, ESP 2016–76076–
R, ESP2016-80435-C2-2-R, ESP2017-87143-R, PGC2018-098153-B-C31, BES–2015–073500, and BES–
2015–074542, the German Science Foundation through the Major Research Instrumentation Programme and DFG Research Unit FOR2544 “Blue Planets around Red Stars”, the Klaus Tschira Stiftung, the states of Baden-W\"{u}rttemberg and Niedersachsen, and by the Junta de Andaluc\'ia. IAA authors acknowledge financial support from the State Agency for Research of the Spanish MCIU through the ``Center of Excellence Severo Ochoa" award SEV-2017-0709.
Based on observations collected at the Centro Astronómico Hispano-Alem\'an (CAHA) at Calar Alto, operated jointly by Junta de Andaluc\'ia and Consejo Superior de Investigaciones Cientí\'ificas (IAA-CSIC). We thank the anonymous referee for the very useful comments.
\end{acknowledgements}


\onecolumn
\appendix
\section{Supplementary figures}
\label{appendix:a}

\begin{figure*}[htb!]
\centering
\includegraphics[angle=0, width=0.34\columnwidth]{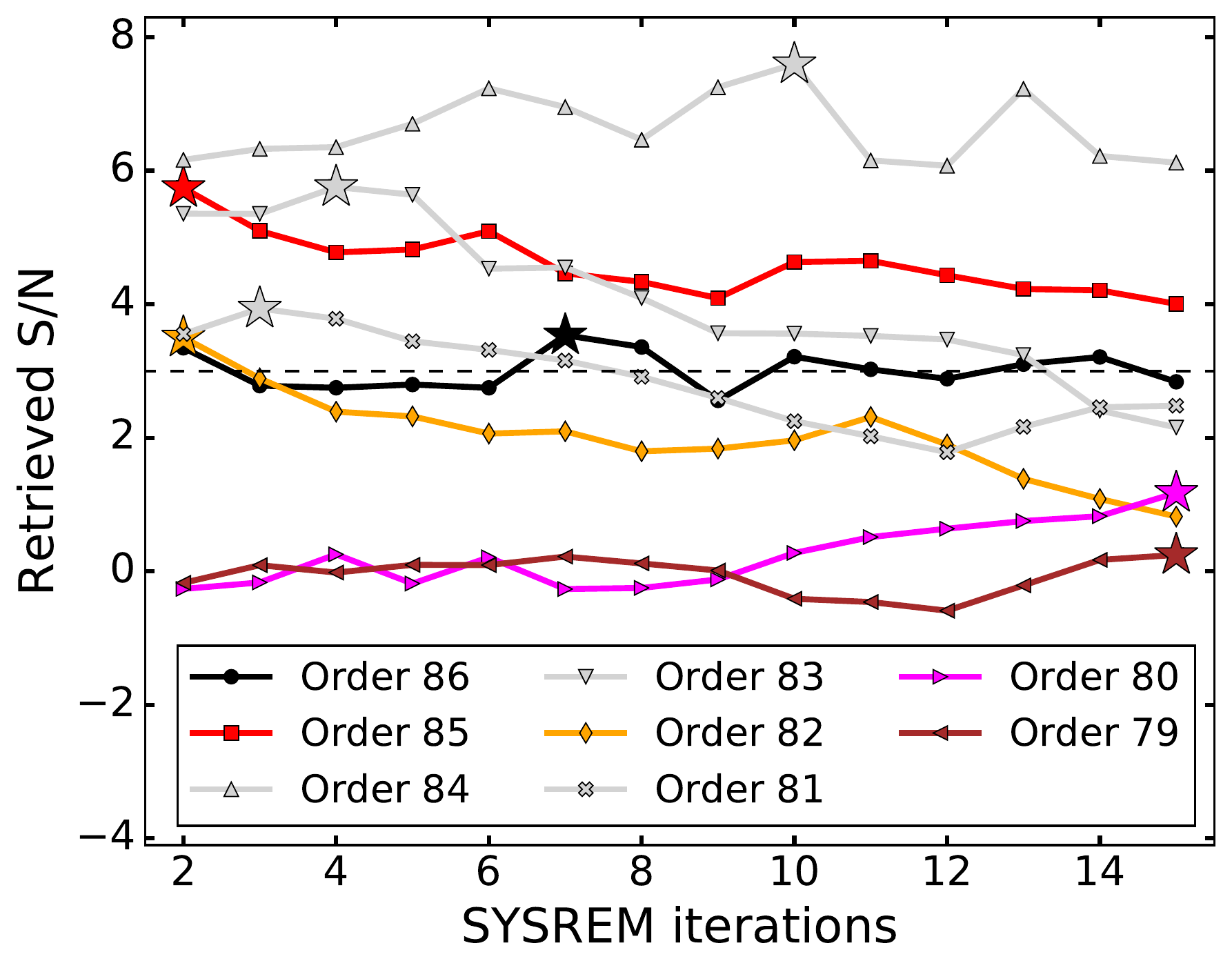}\includegraphics[angle=0, width=0.34\columnwidth]{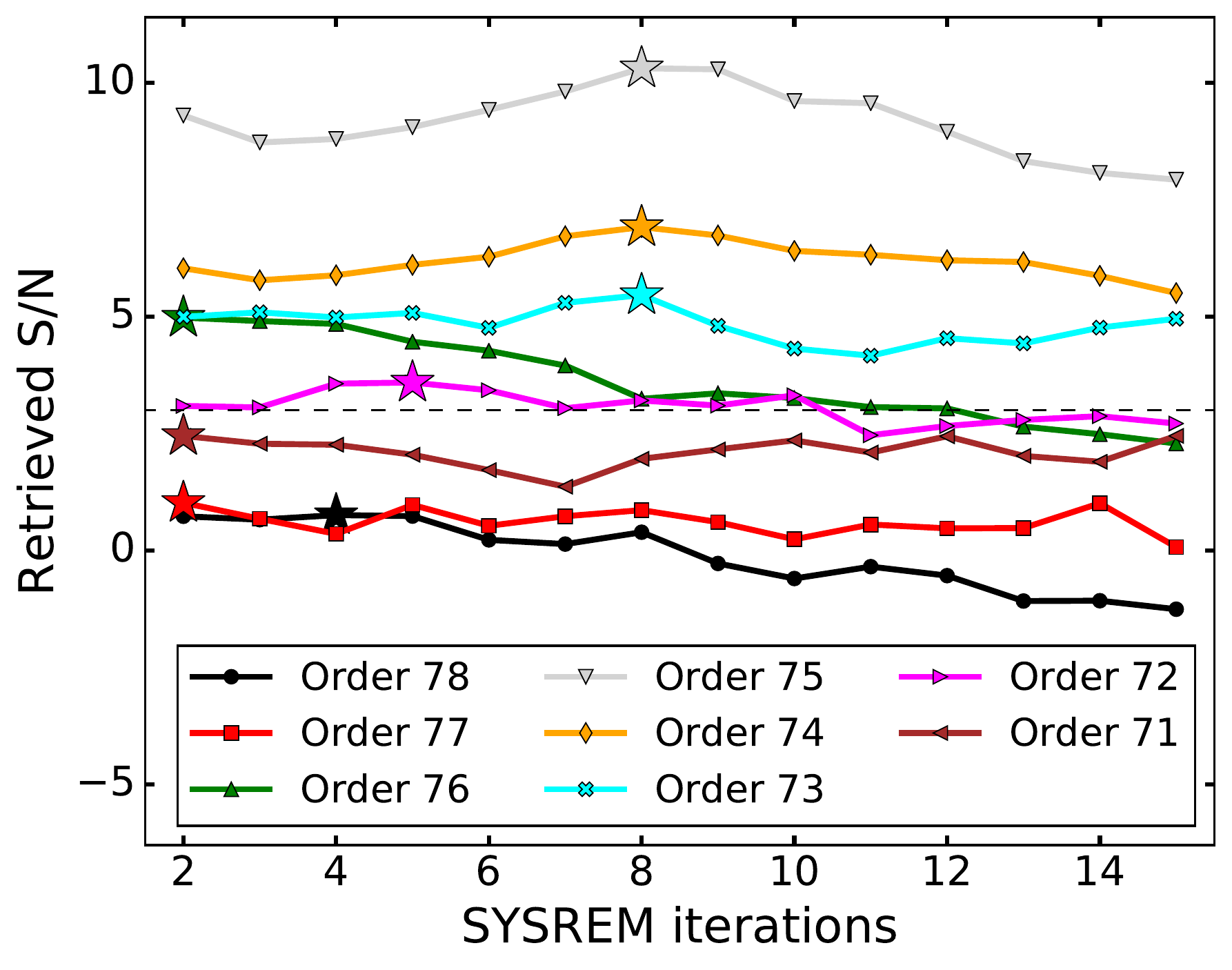}\includegraphics[angle=0, width=0.34\columnwidth]{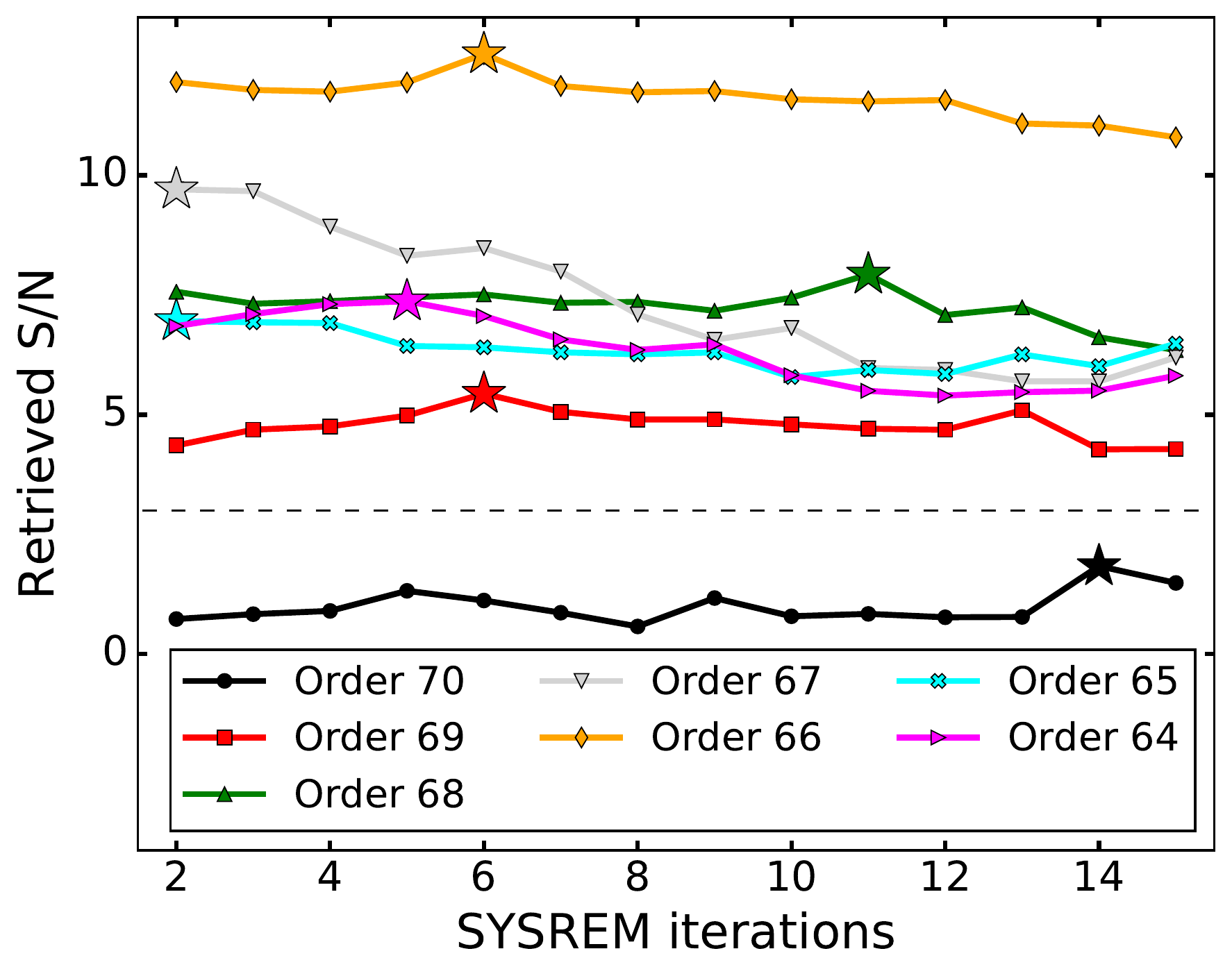}
\includegraphics[angle=0, width=0.34\columnwidth]{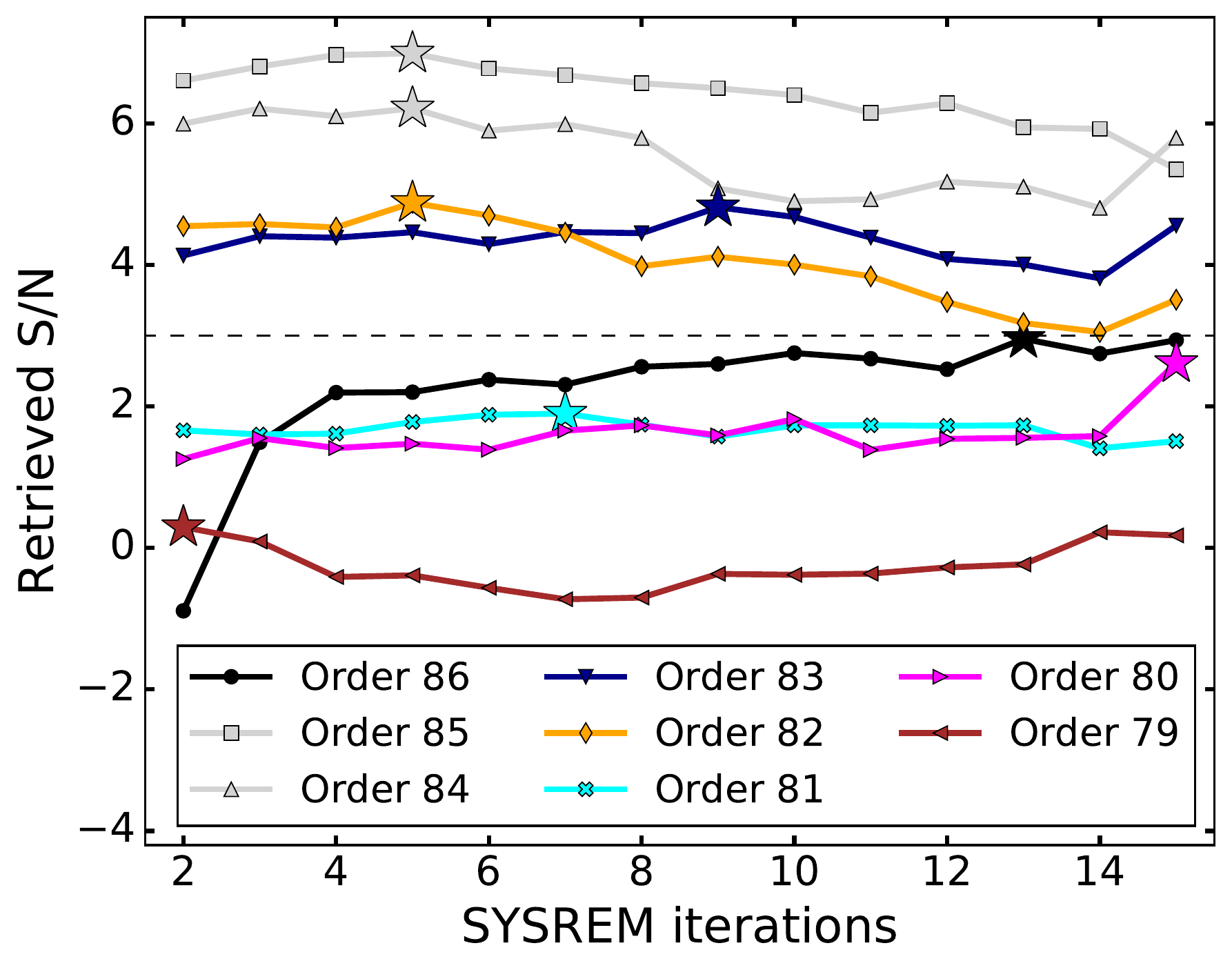}\includegraphics[angle=0, width=0.34\columnwidth]{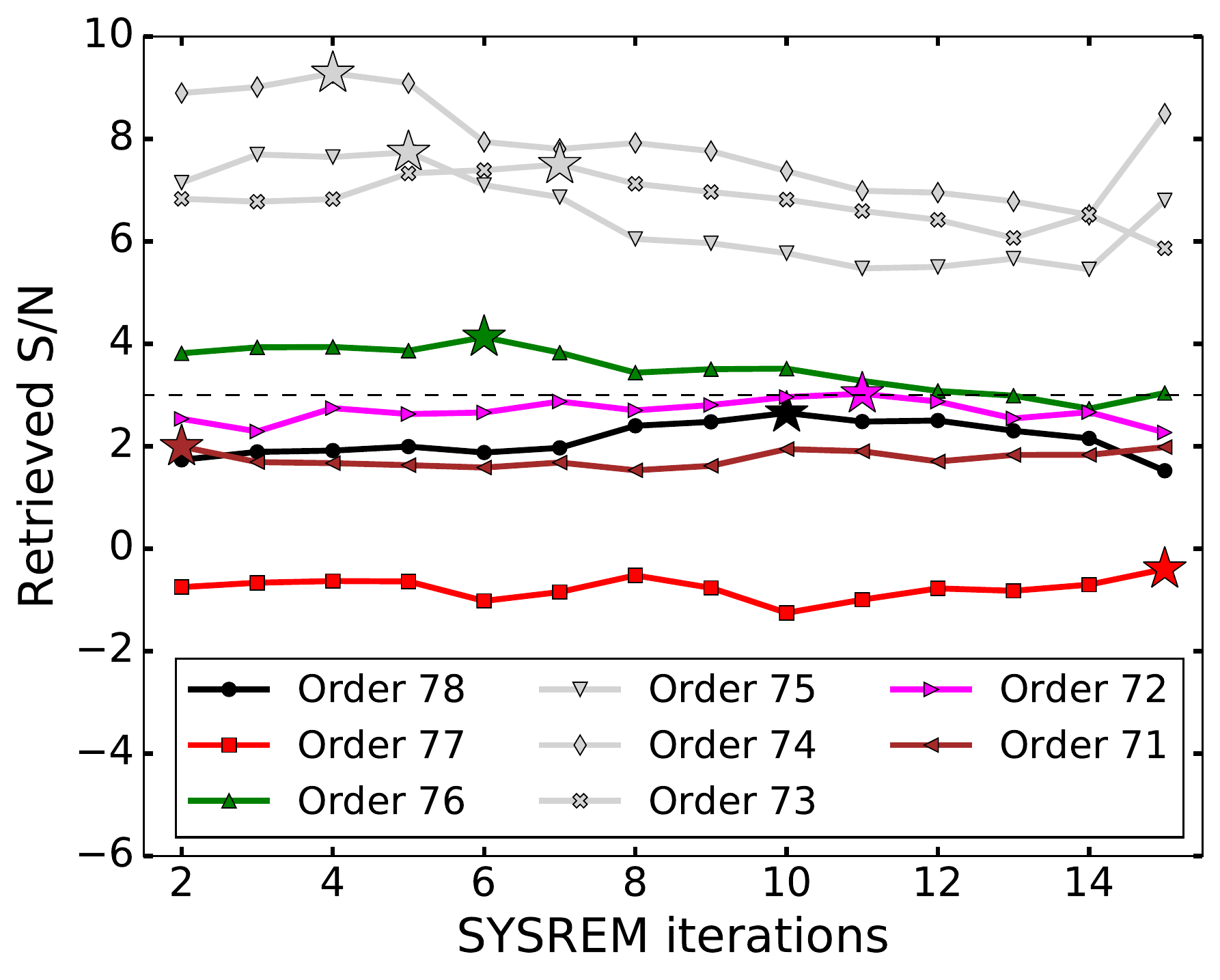}\includegraphics[angle=0, width=0.34\columnwidth]{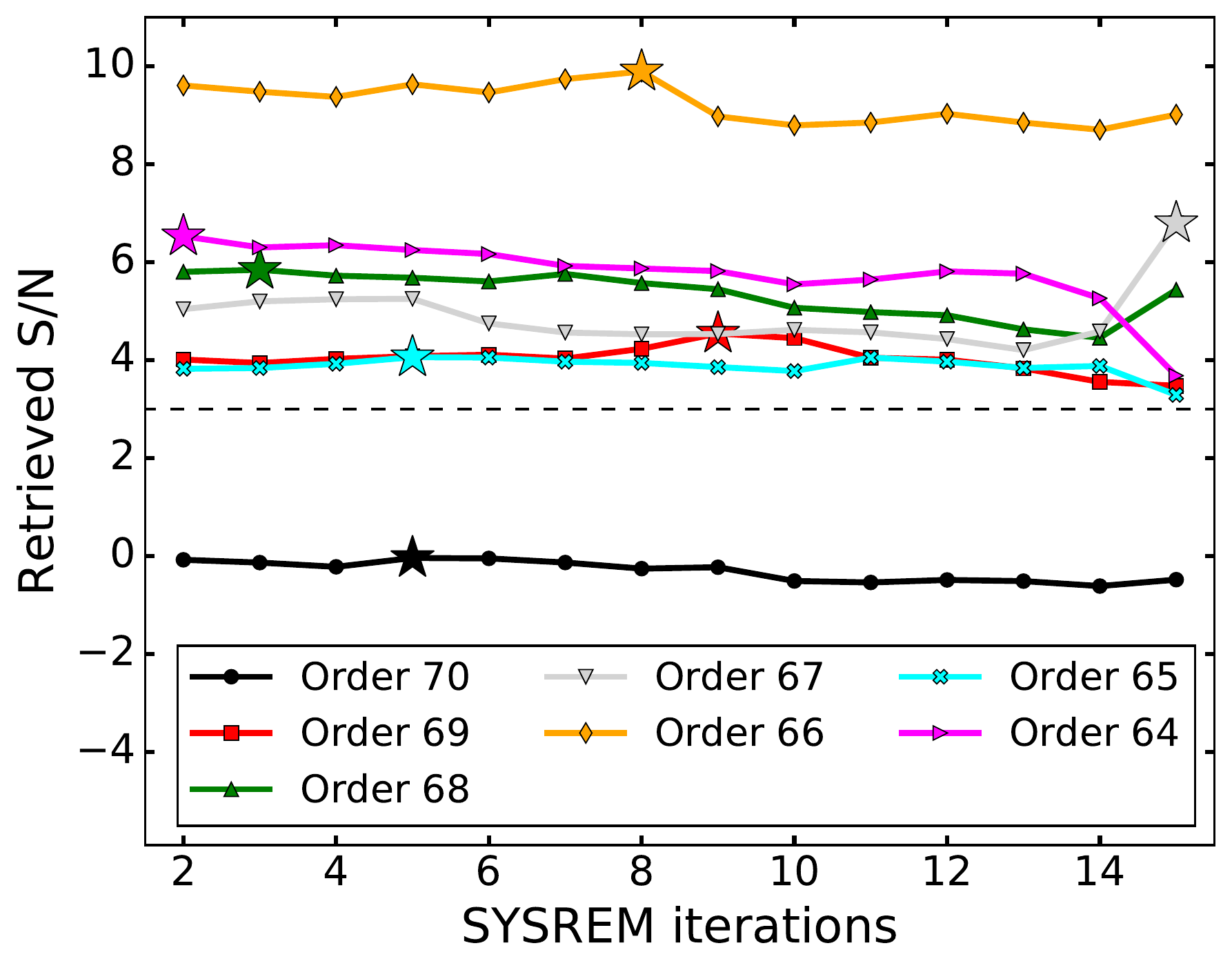}
\includegraphics[angle=0, width=0.34\columnwidth]{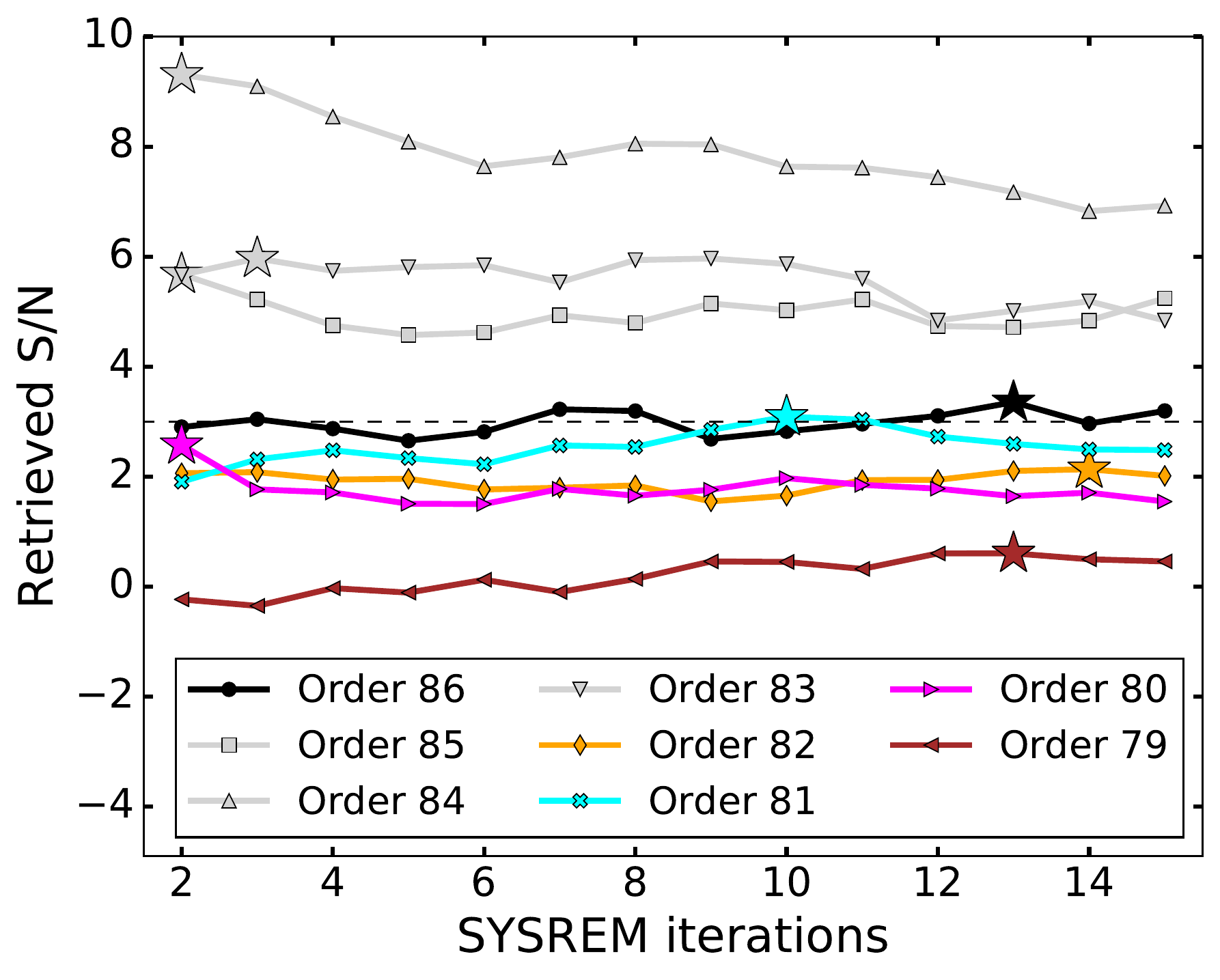}\includegraphics[angle=0, width=0.34\columnwidth]{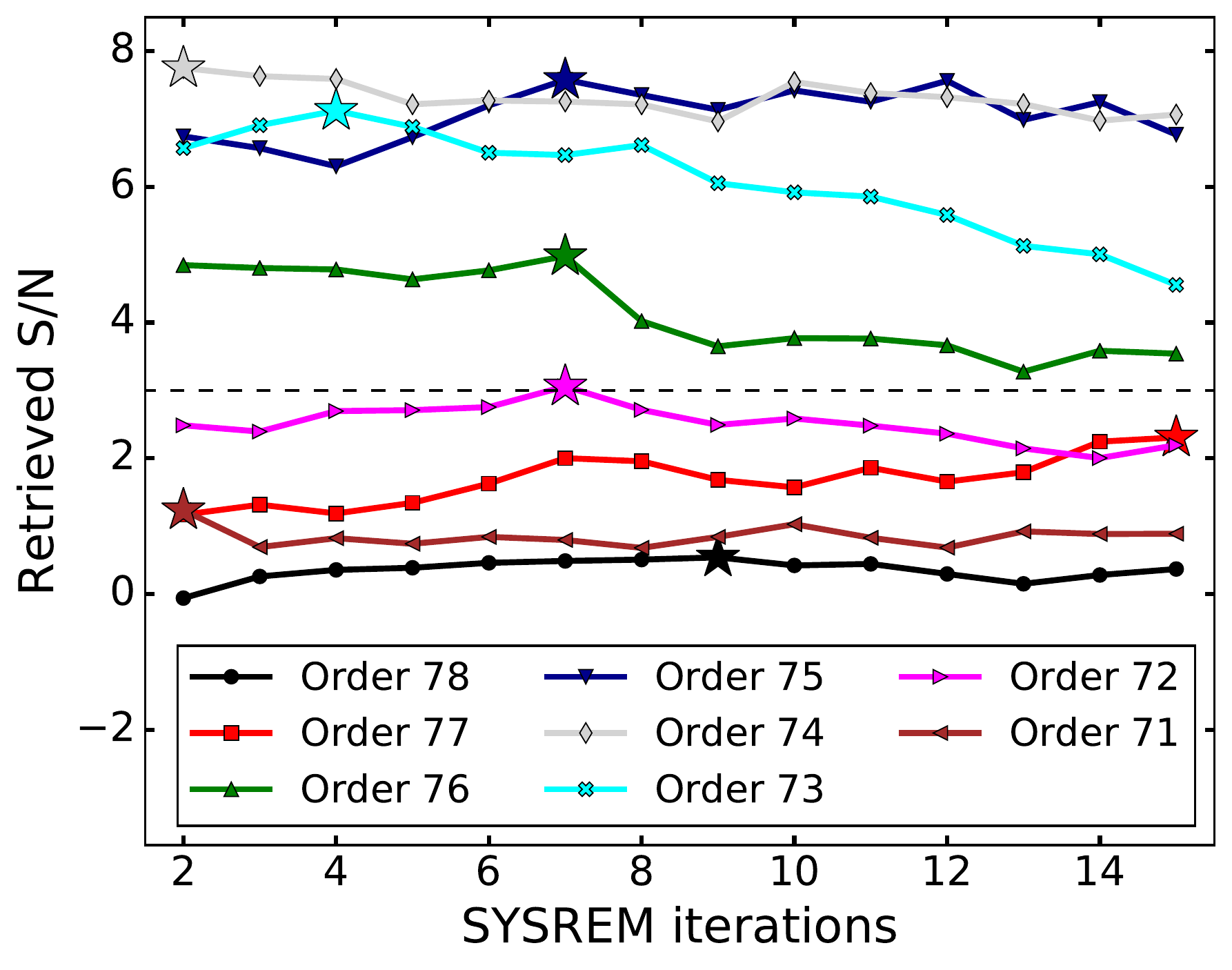}\includegraphics[angle=0, width=0.34\columnwidth]{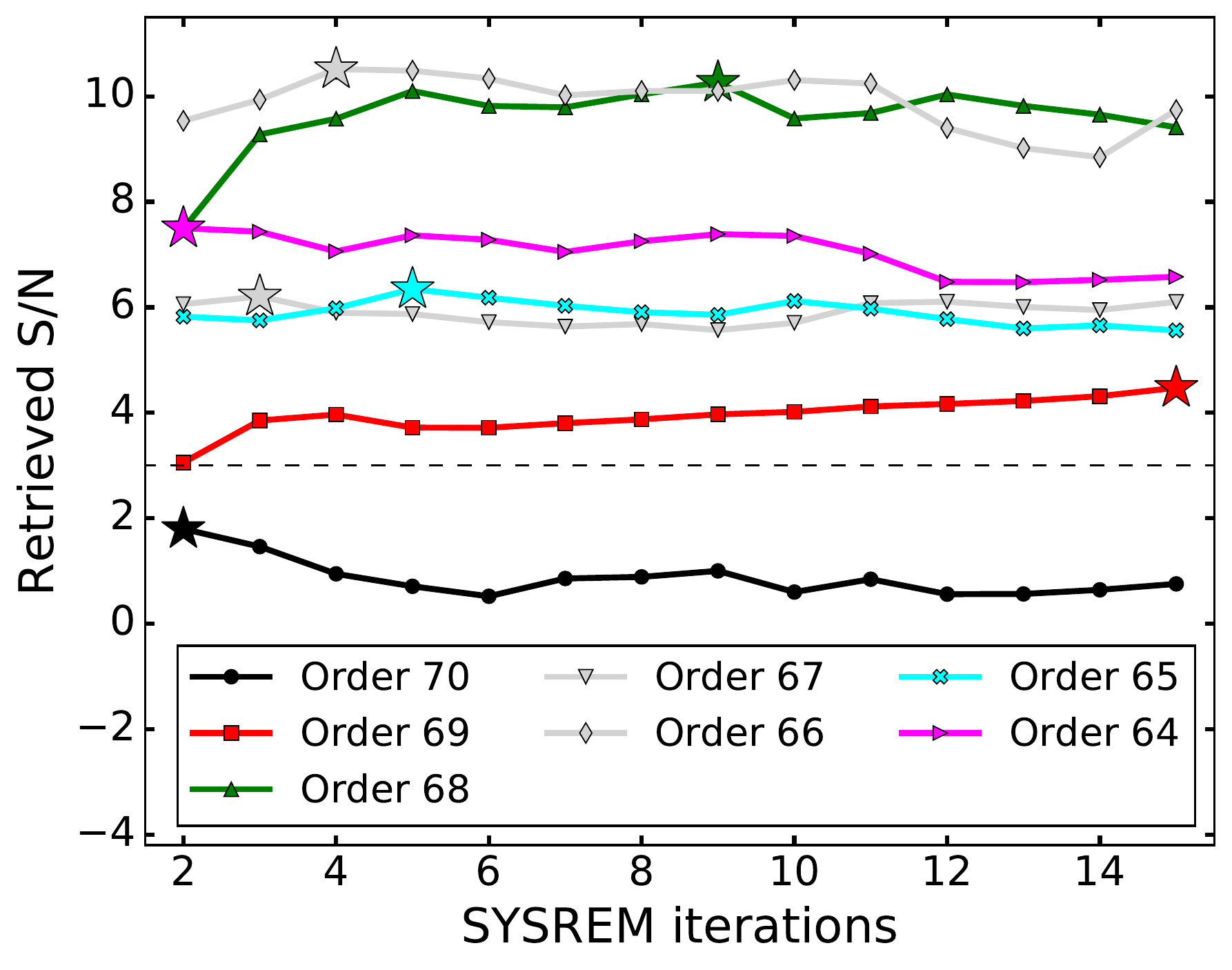}
\caption{Evolution of the S/N of the retrieved injected signal with subsequent {\tt SYSREM} iterations for \hd20 on the nights of $N_{A,\,1}$ (\textit{top}), $N_{A,\,2}$ (\textit{middle}), and $N_{A,\,3}$ (\textit{bottom}) for the \h2o bands at $\sim$0.72\,$\mu$m (\textit{left}), $\sim$0.82\,$\mu$m (\textit{central}), and $\sim$0.95\,$\mu$m (\textit{right}). The model signals were injected at 5$\times$ the expected strength. The horizontal dashed lines mark the $\text{S/N}=3$ level. The star symbols mark the iteration in which the recovery of the injected signal is maximised.} 
\label{sn_evol_hd20}
\end{figure*}

\begin{figure*}[htb!]
\centering
\includegraphics[angle=0, width=0.34\columnwidth]{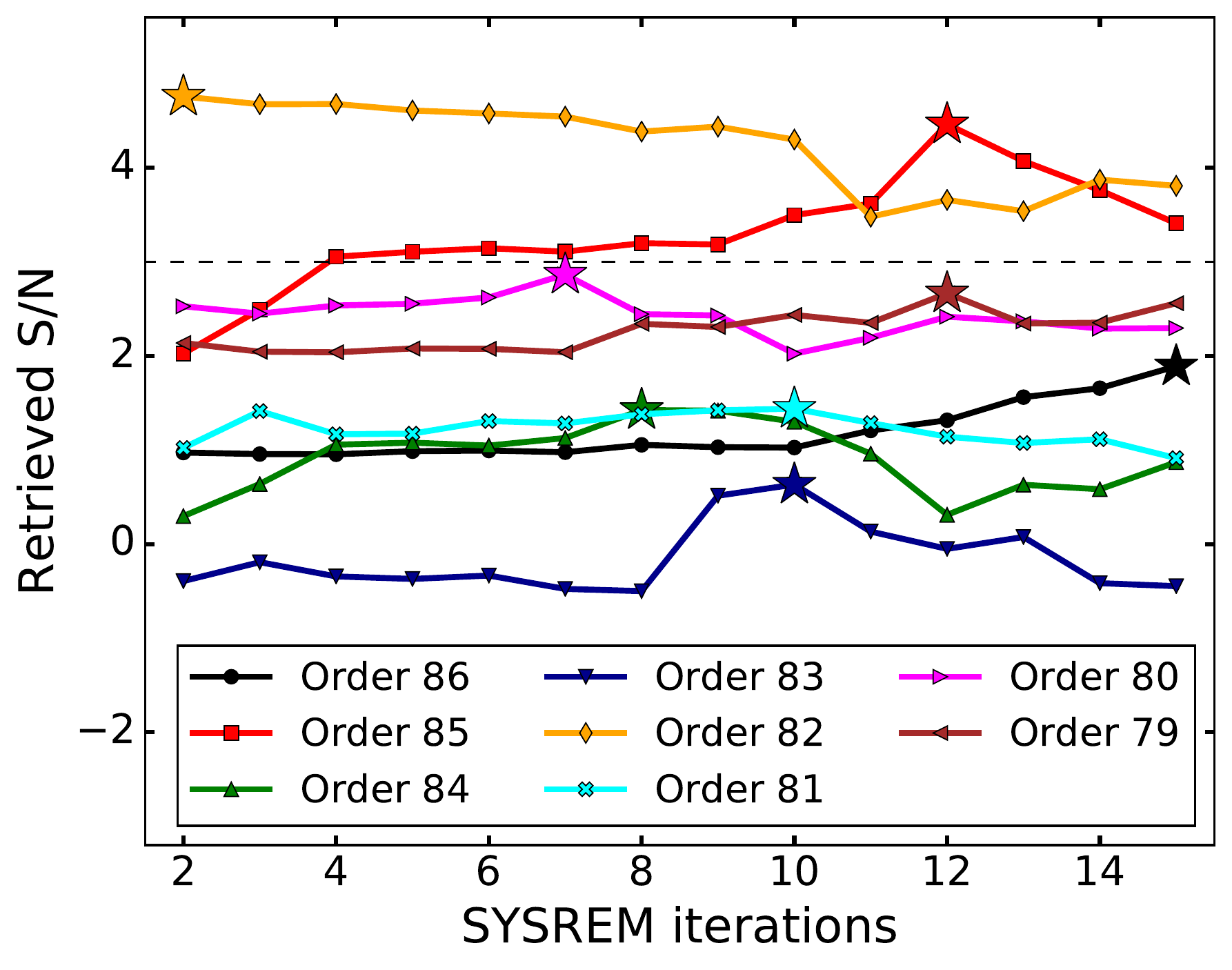}\includegraphics[angle=0, width=0.34\columnwidth]{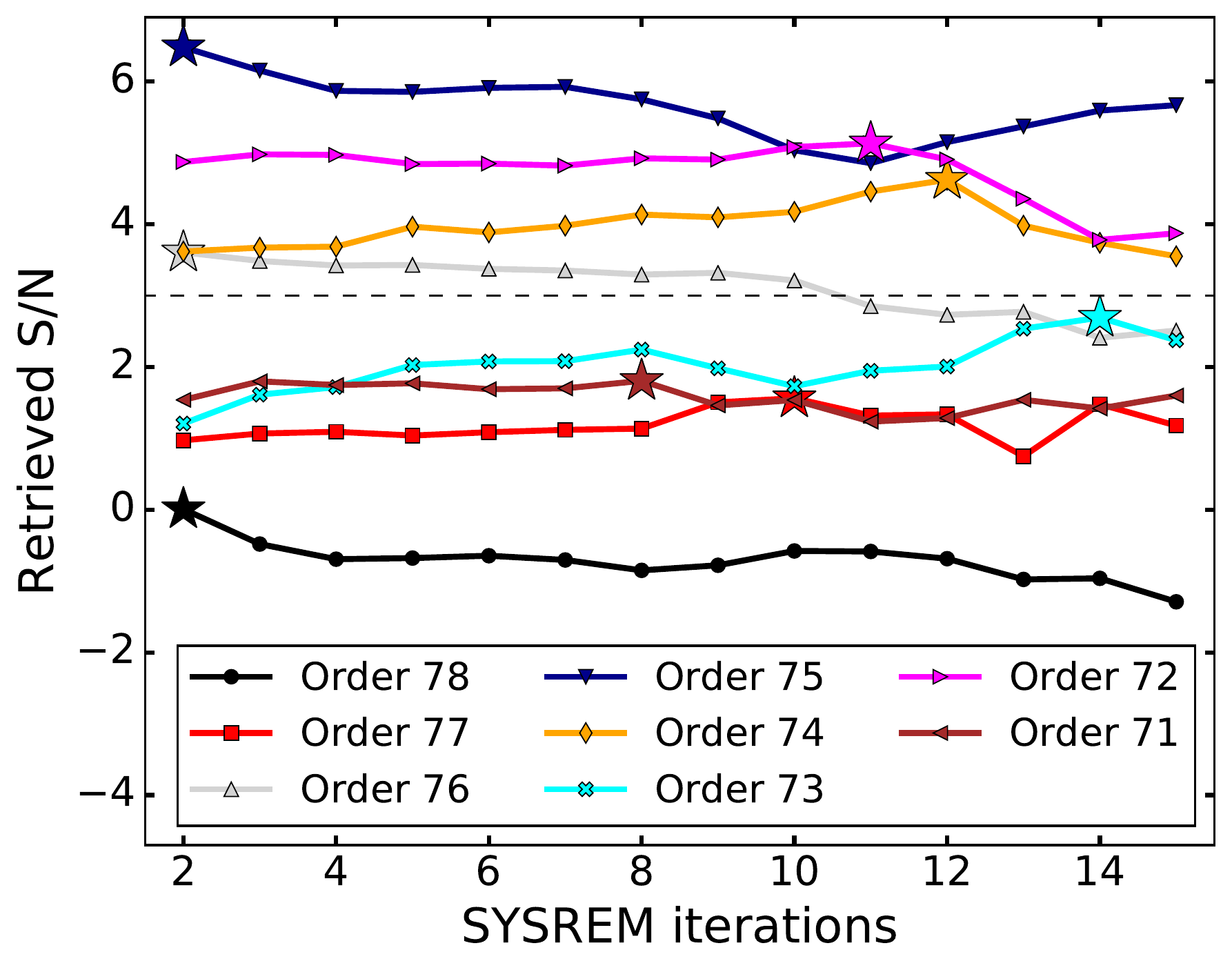}\includegraphics[angle=0, width=0.34\columnwidth]{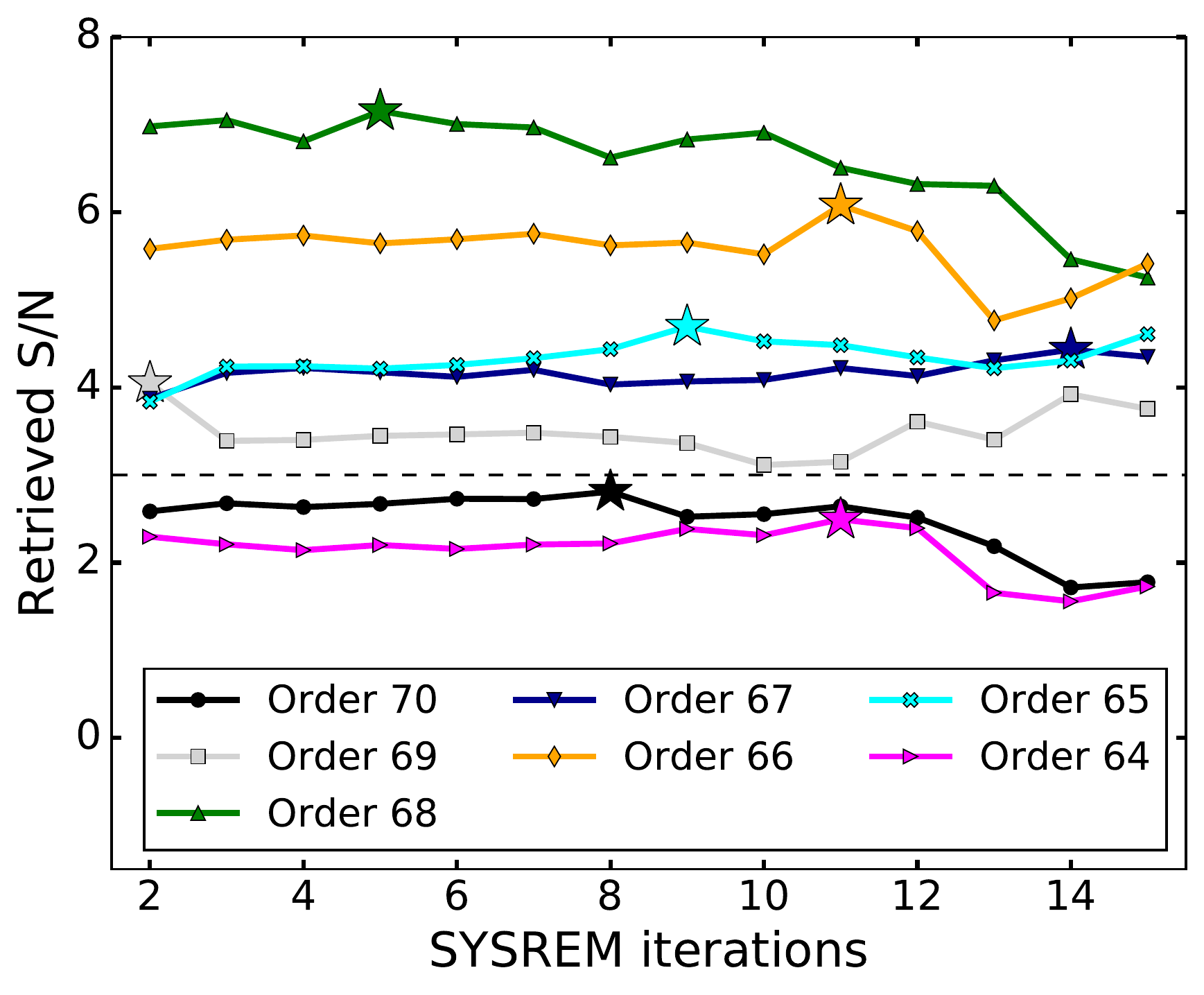}
\includegraphics[angle=0, width=0.34\columnwidth]{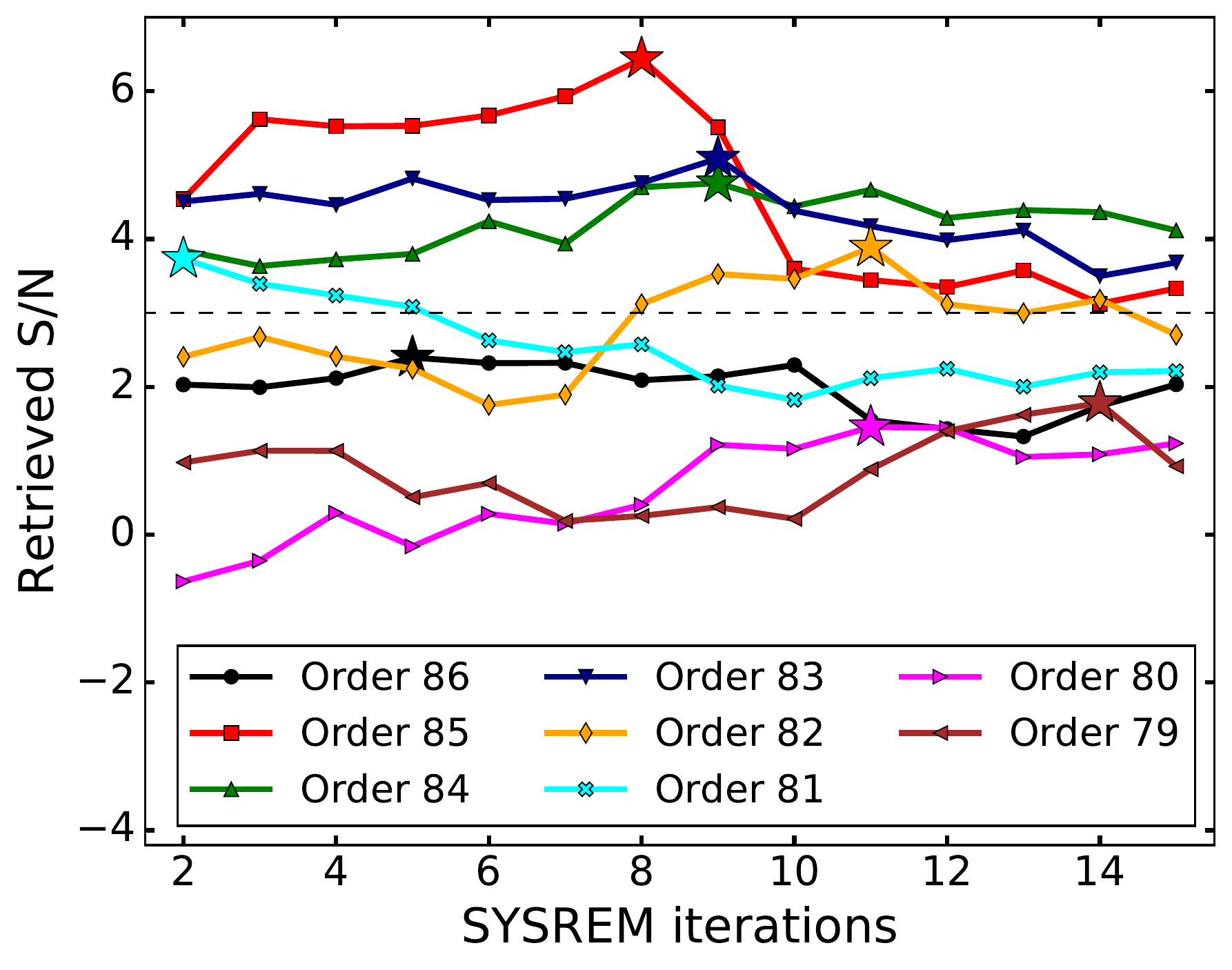}\includegraphics[angle=0, width=0.34\columnwidth]{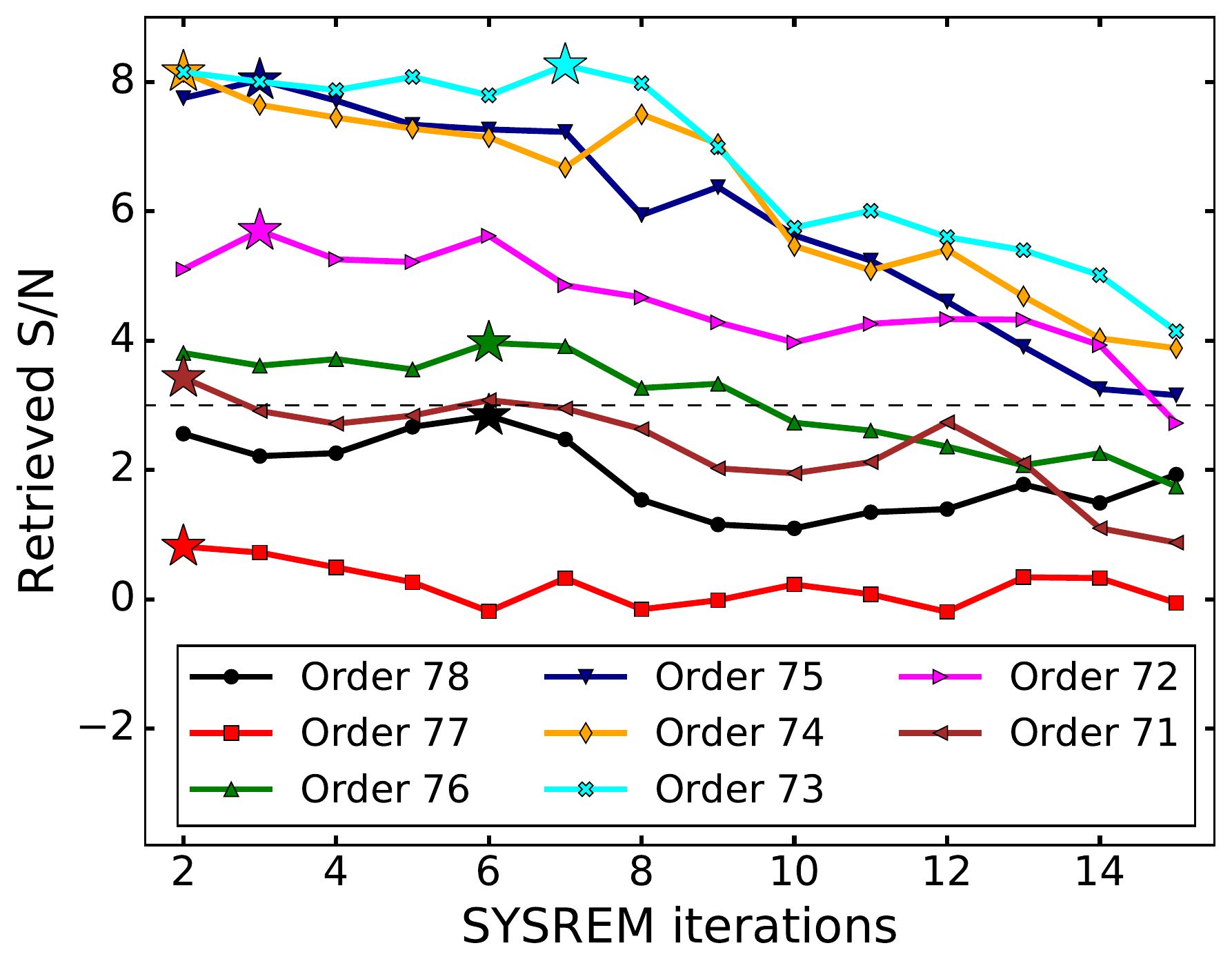}\includegraphics[angle=0, width=0.34\columnwidth]{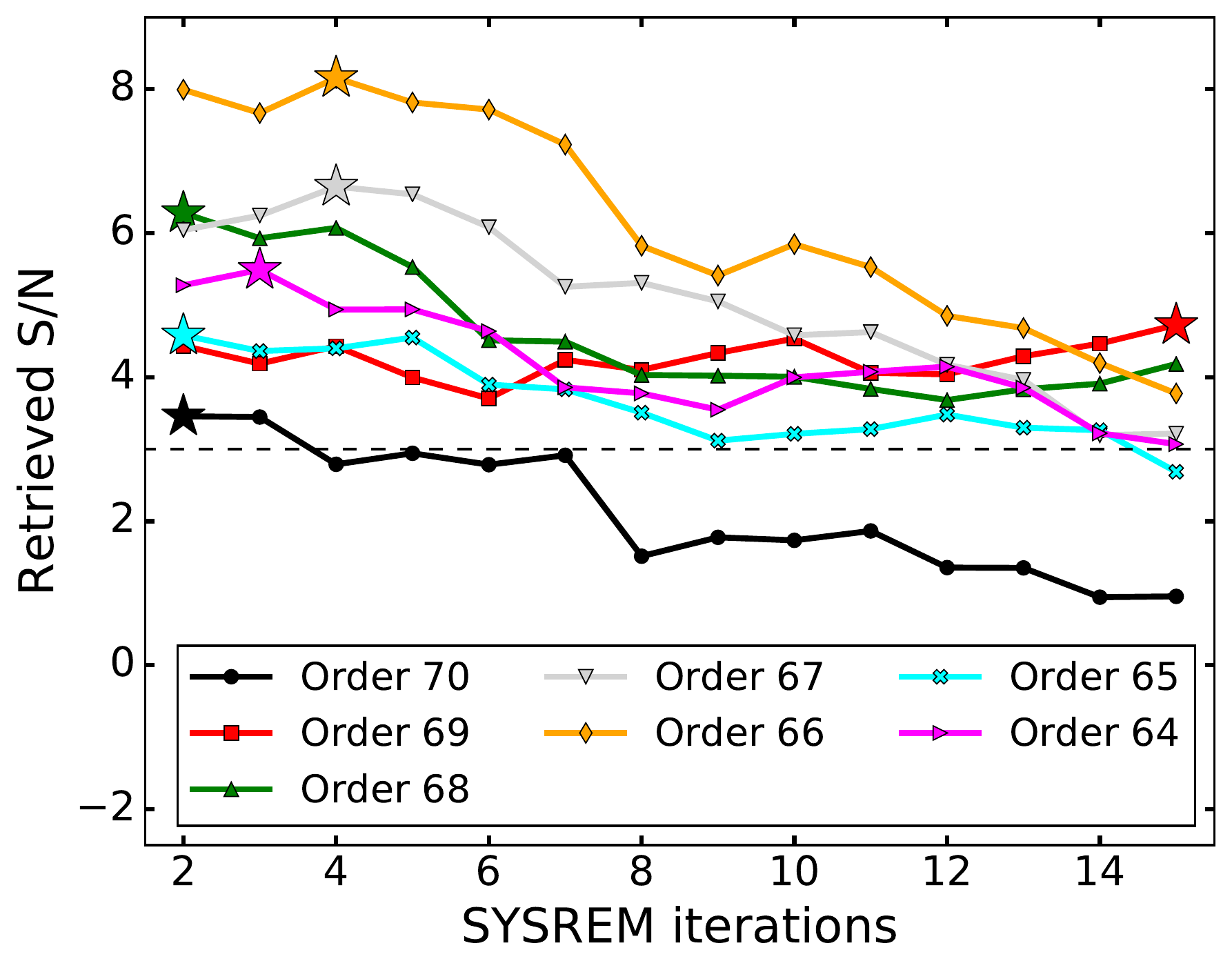}
\includegraphics[angle=0, width=0.34\columnwidth]{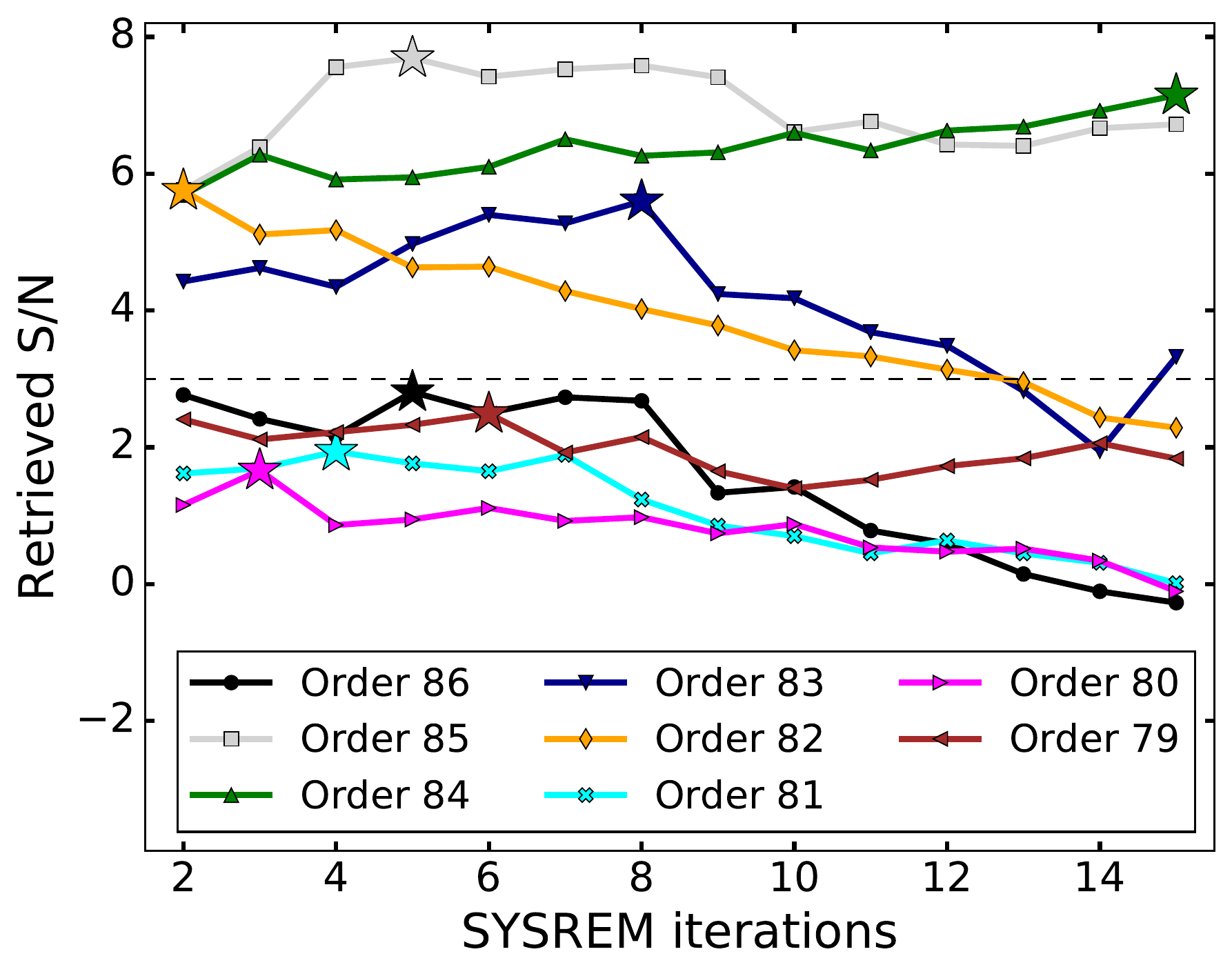}\includegraphics[angle=0, width=0.34\columnwidth]{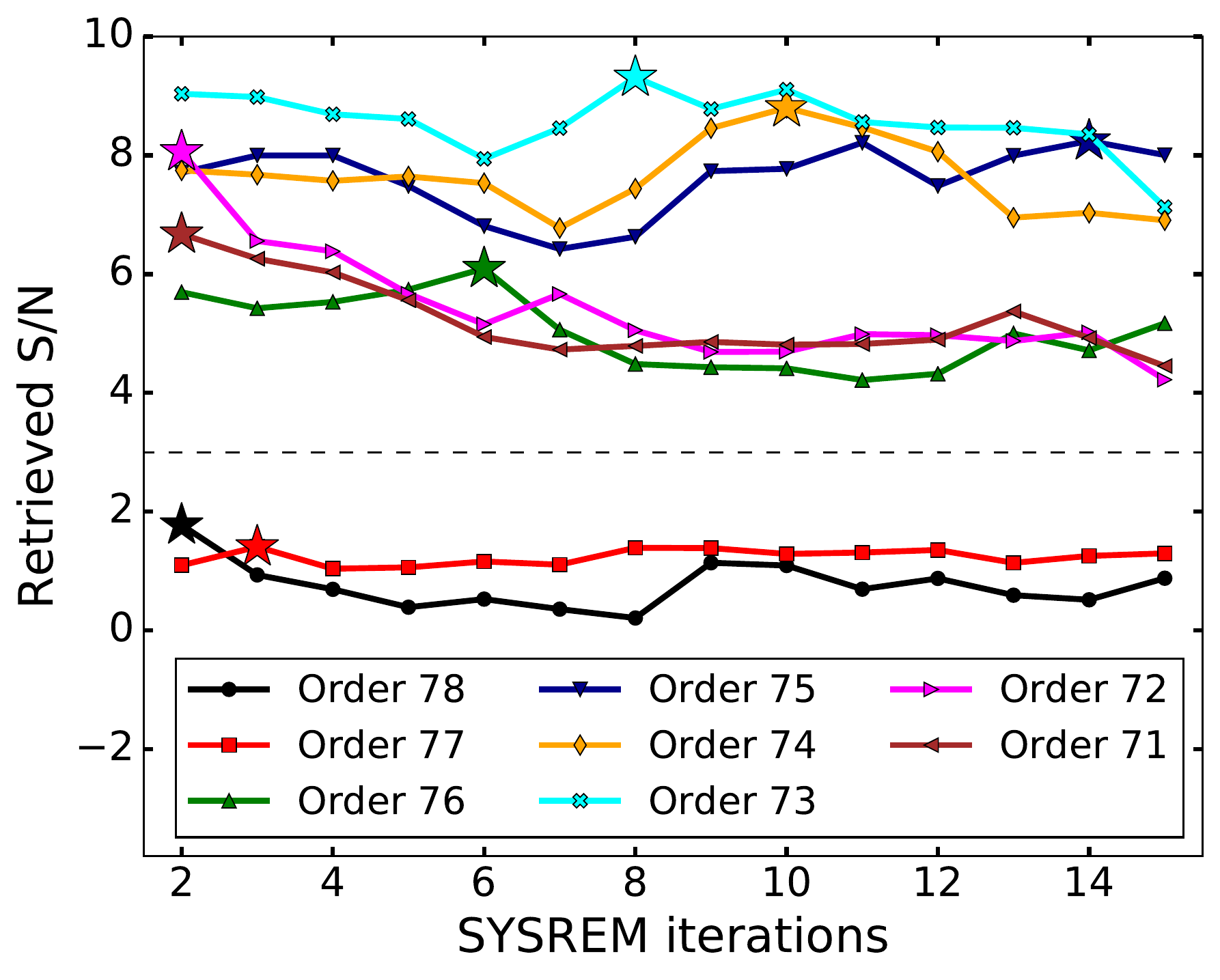}\includegraphics[angle=0, width=0.34\columnwidth]{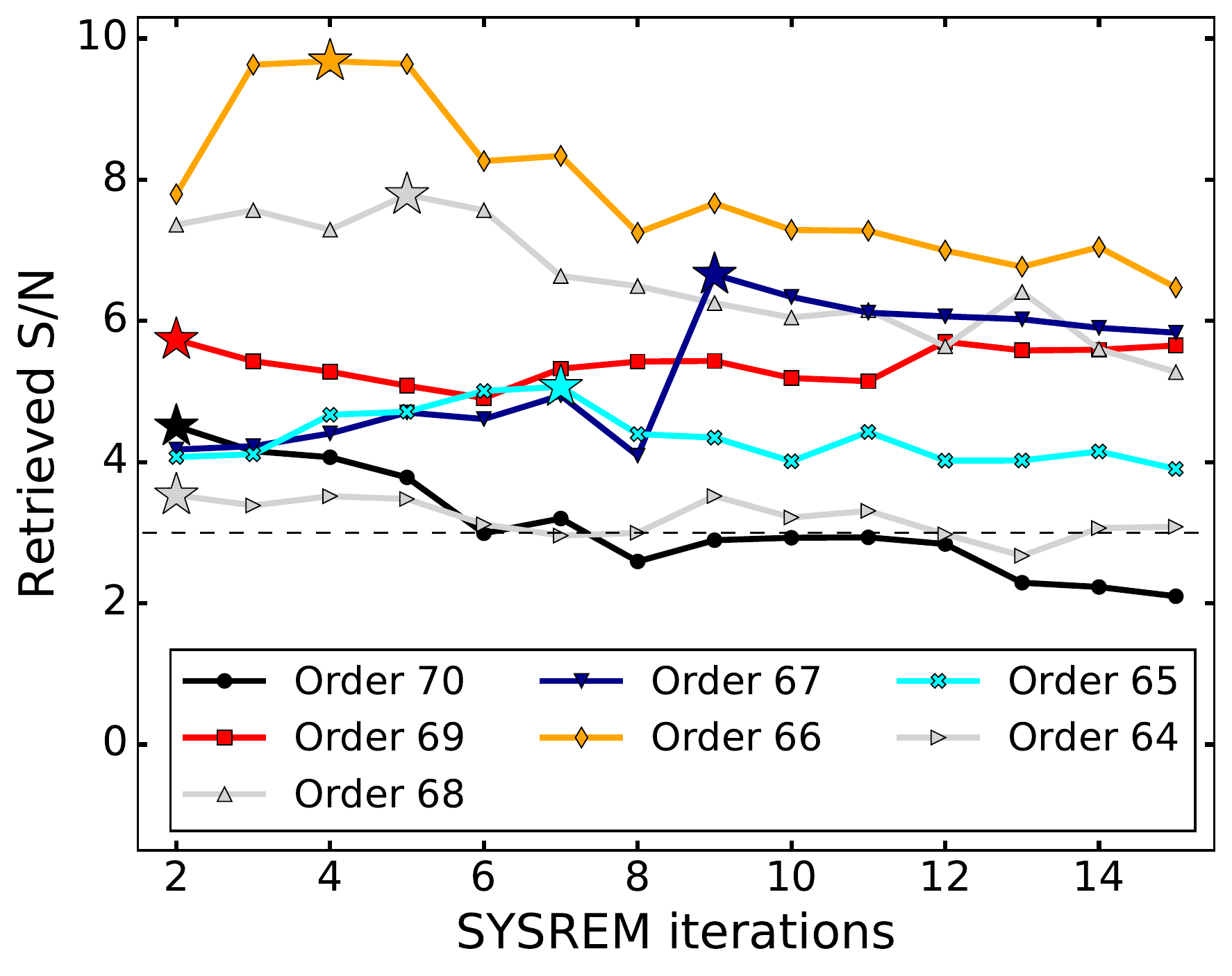}
\includegraphics[angle=0, width=0.34\columnwidth]{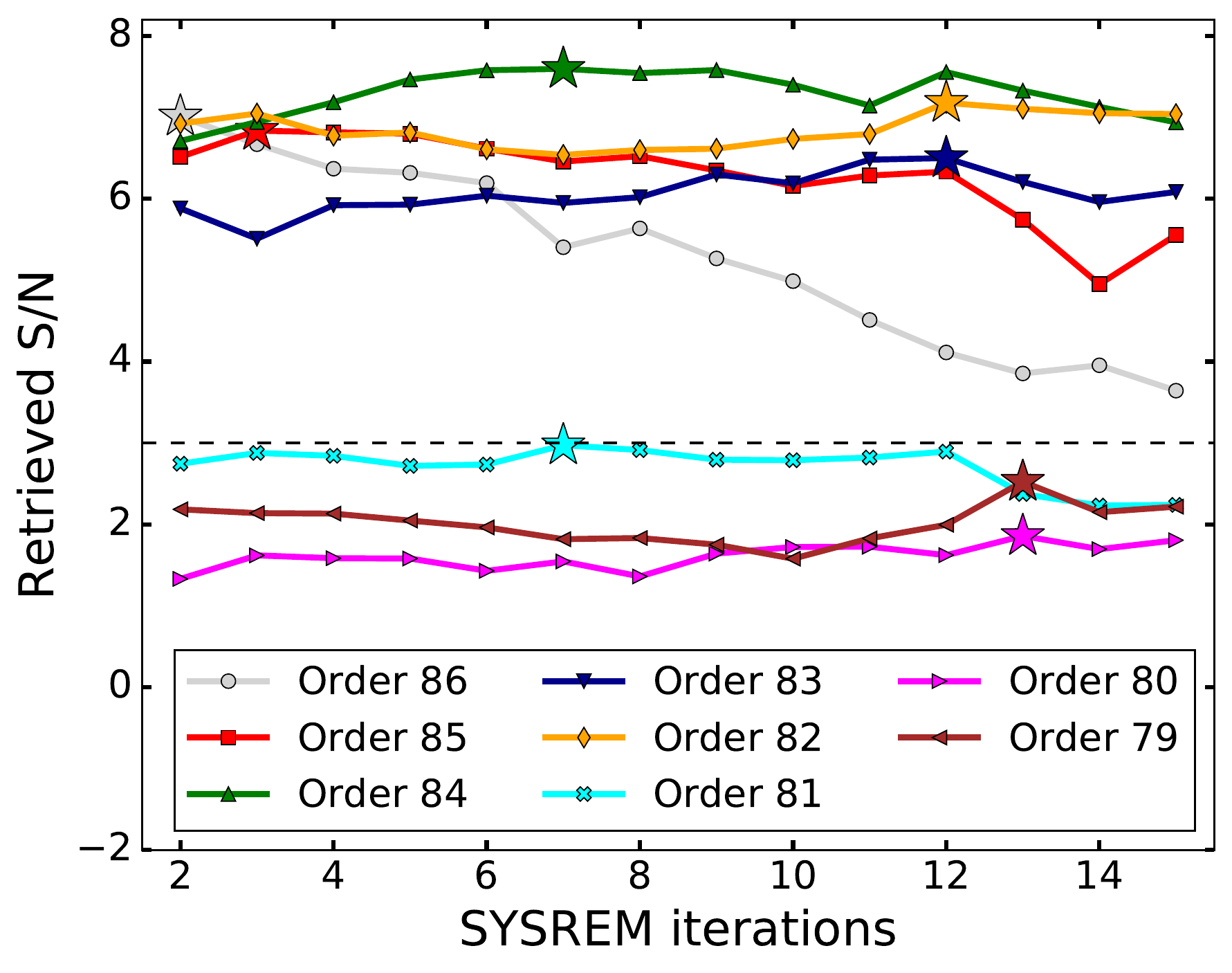}\includegraphics[angle=0, width=0.34\columnwidth]{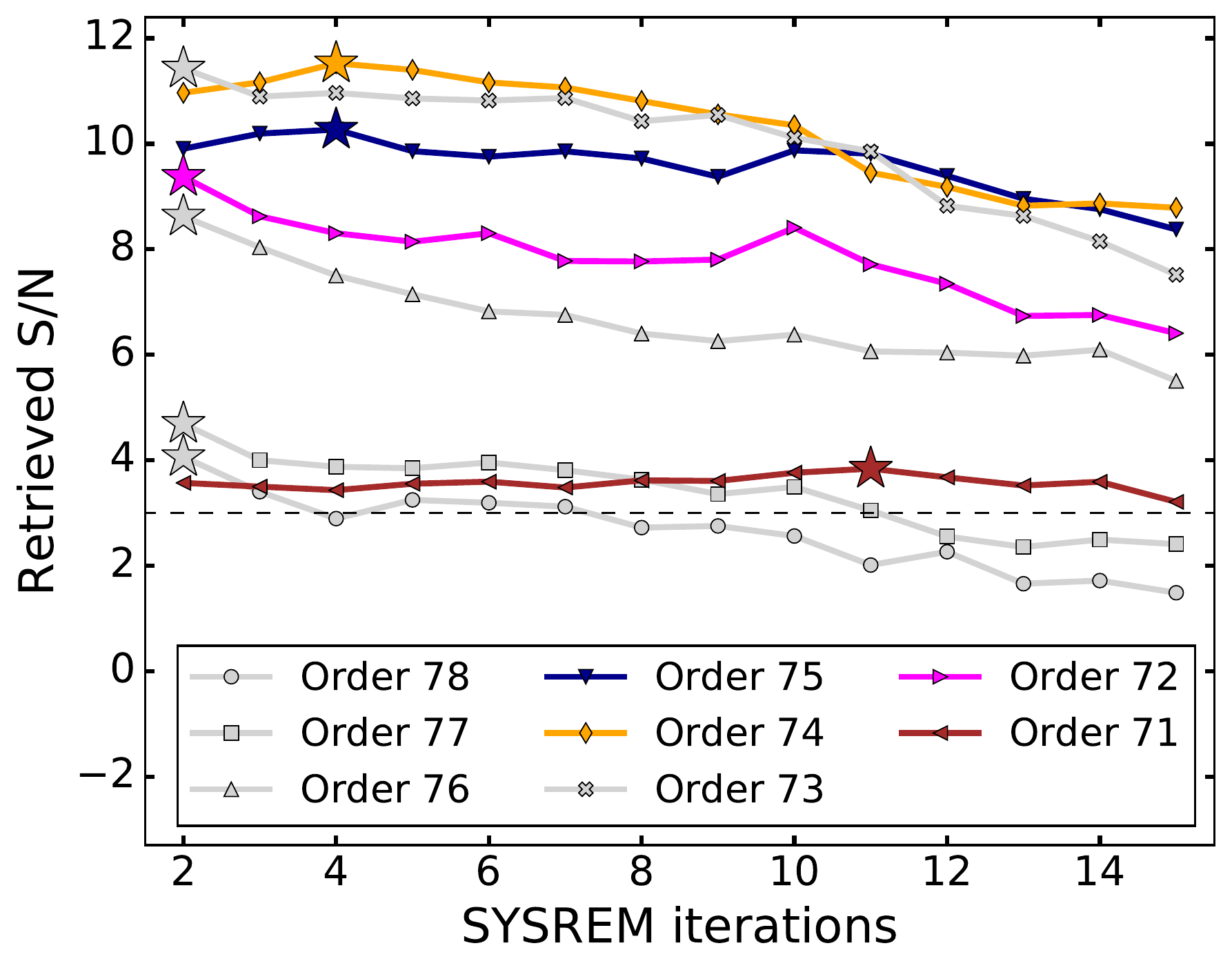}\includegraphics[angle=0, width=0.34\columnwidth]{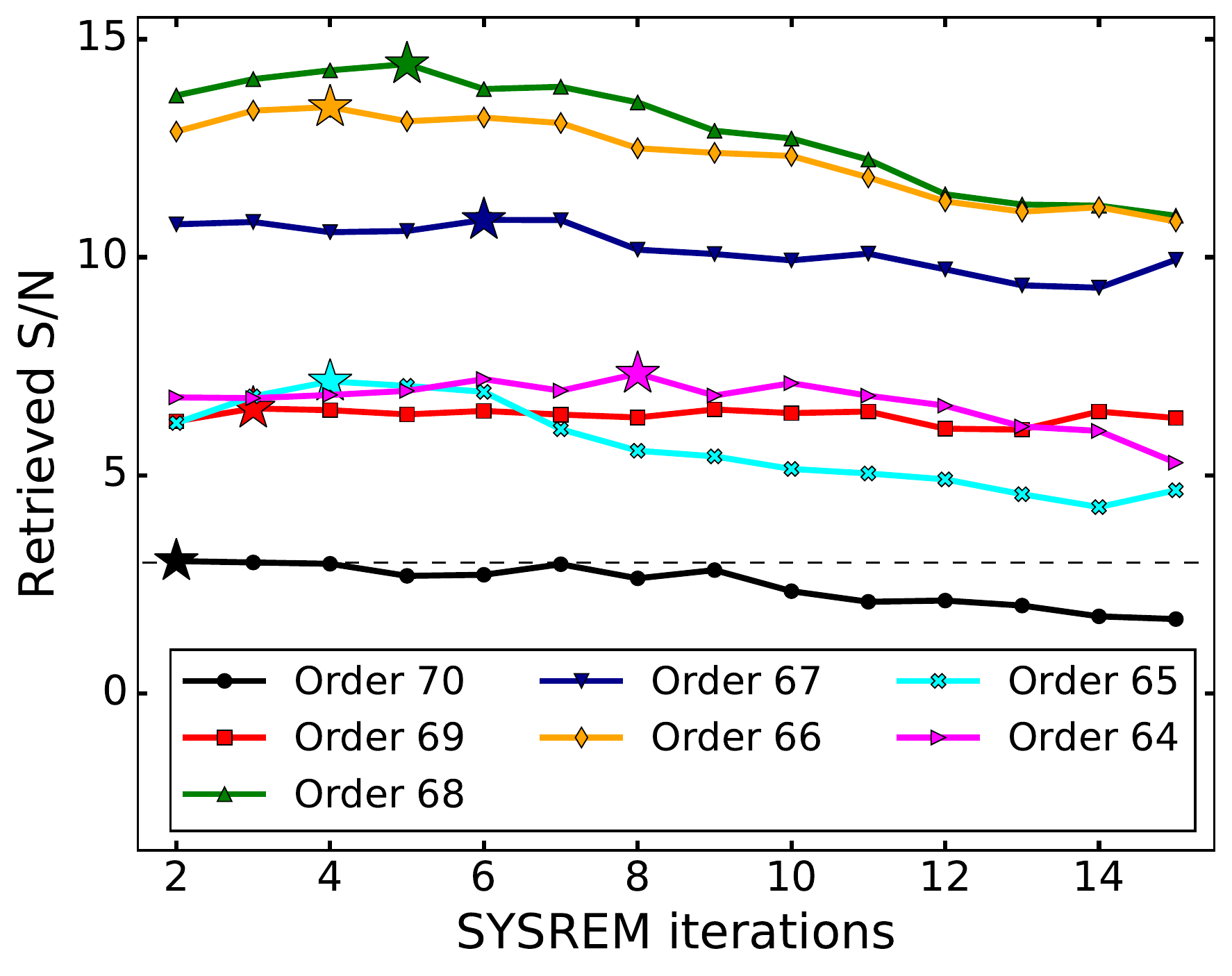}
\caption{Same as Fig.\,\ref{sn_evol_hd20}, but for \hdu18 on the nights of $N_{B,\,1}$ (first row), $N_{B,\,2}$ (second row), $N_{B,\,3}$ (third row), and $N_{B,\,4}$ (fourth row).} 
\label{sn_evol_hd18}
\end{figure*}

\clearpage

\begin{figure*}[ht]
\centering
\includegraphics[angle=0, width=0.30\columnwidth]{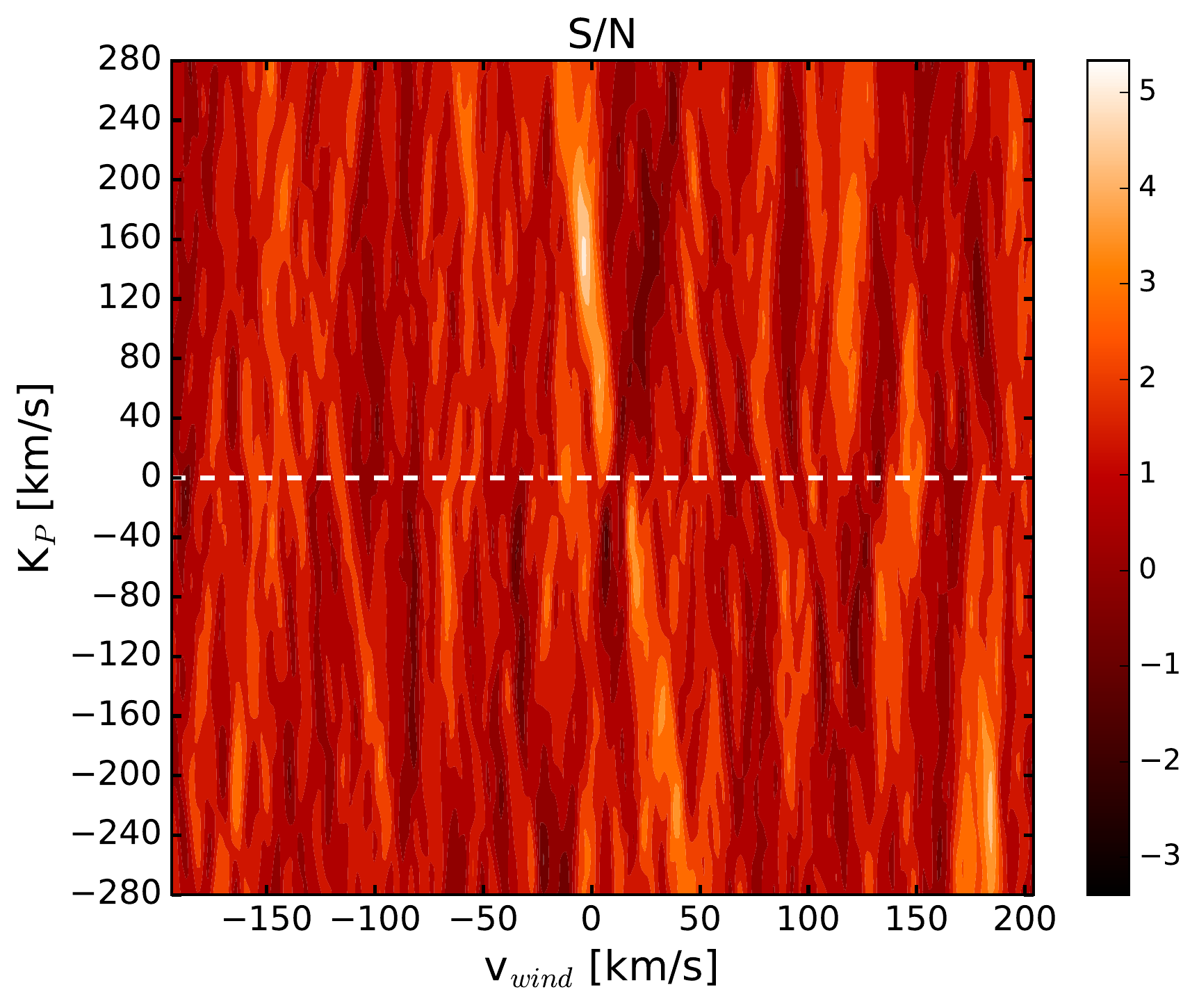}\includegraphics[angle=0, width=0.31\columnwidth]{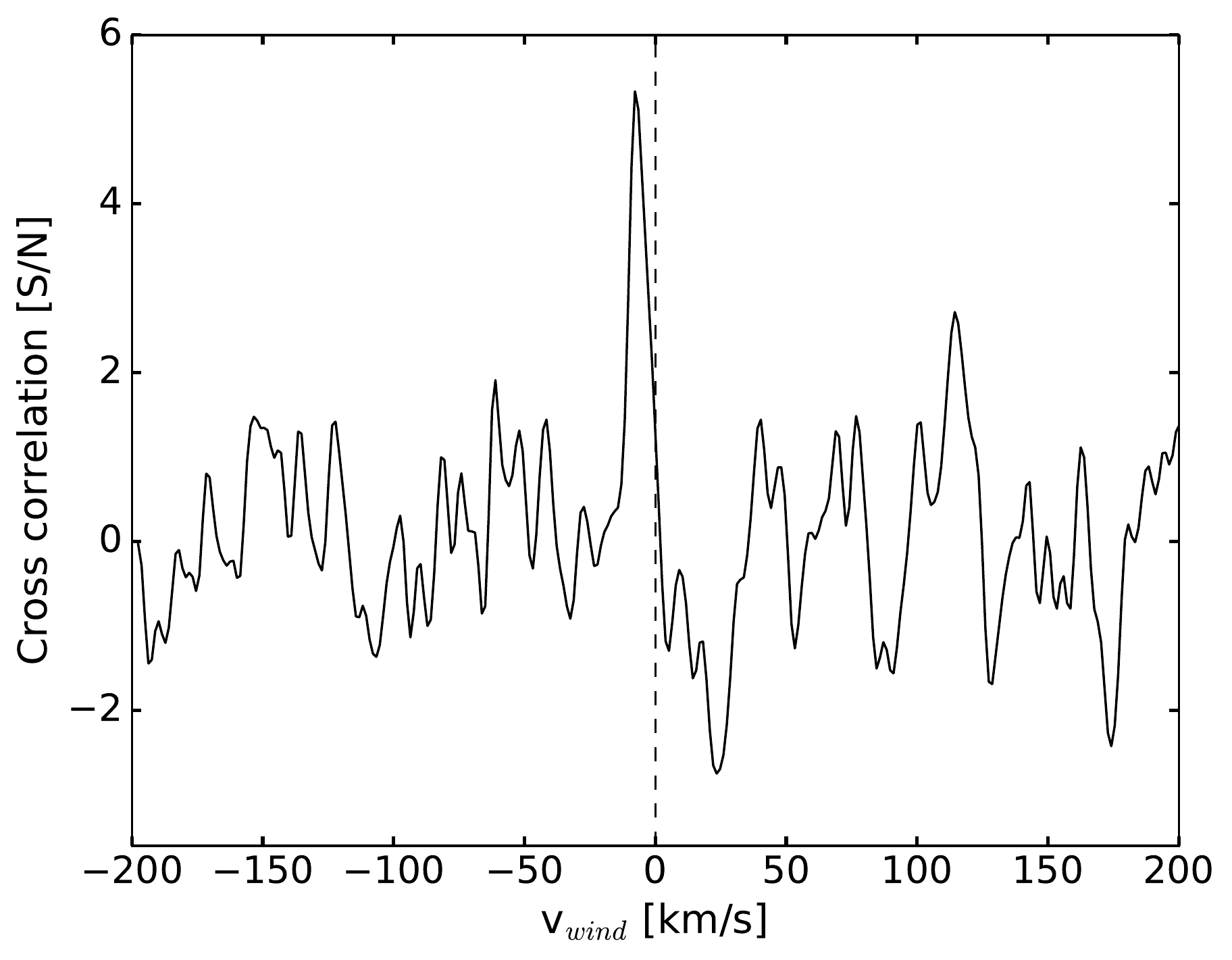}\includegraphics[angle=0, width=0.36\columnwidth]{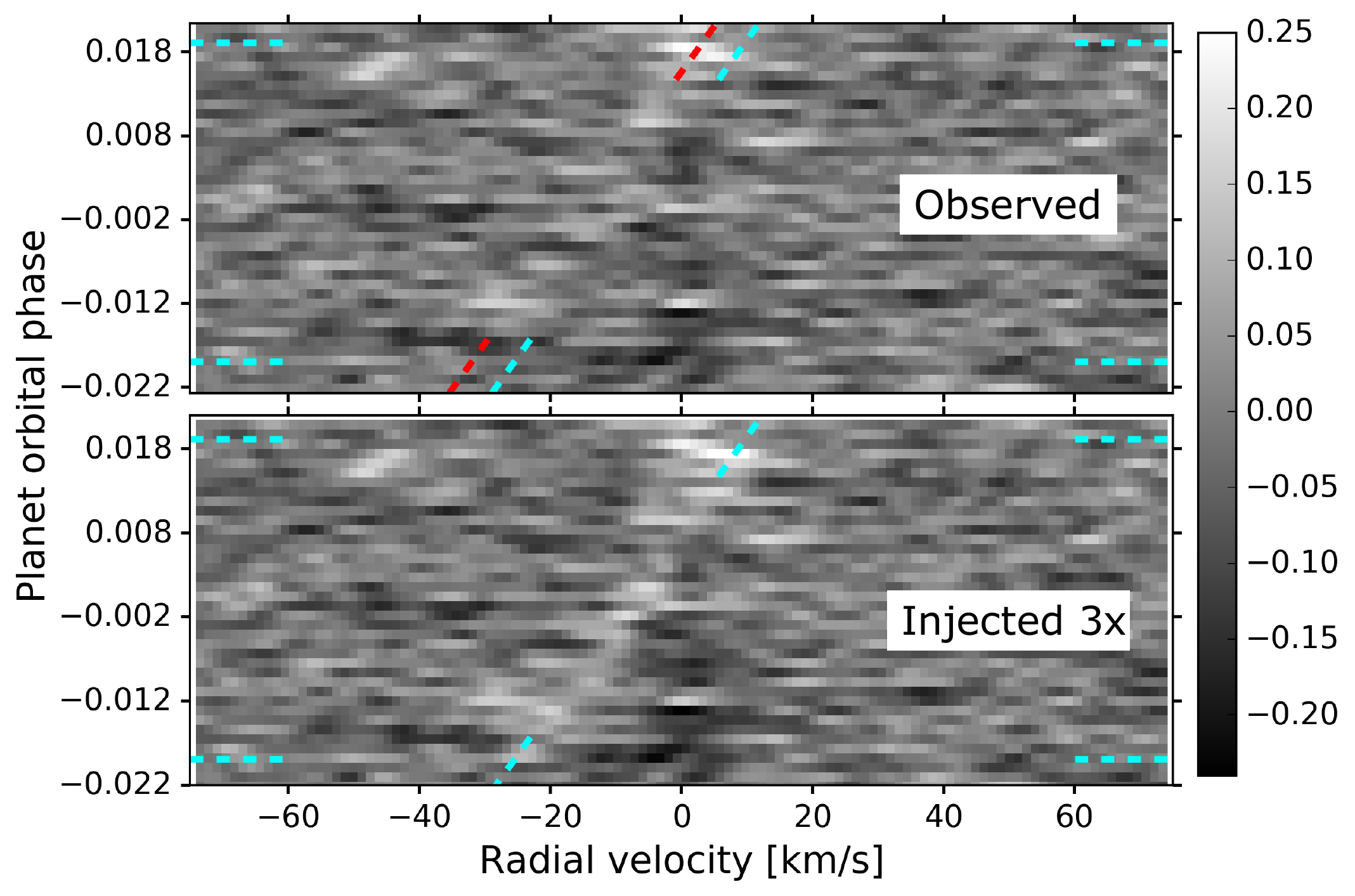}
\caption{Same as the top row of Fig.\,\ref{results}, but for \hd20 when using five {\tt SYSREM} iterations for all spectral orders.} 
\label{sn_map_5_its}
\end{figure*}

\begin{figure*}[ht]
\centering
\includegraphics[angle=0, width=0.4\columnwidth]{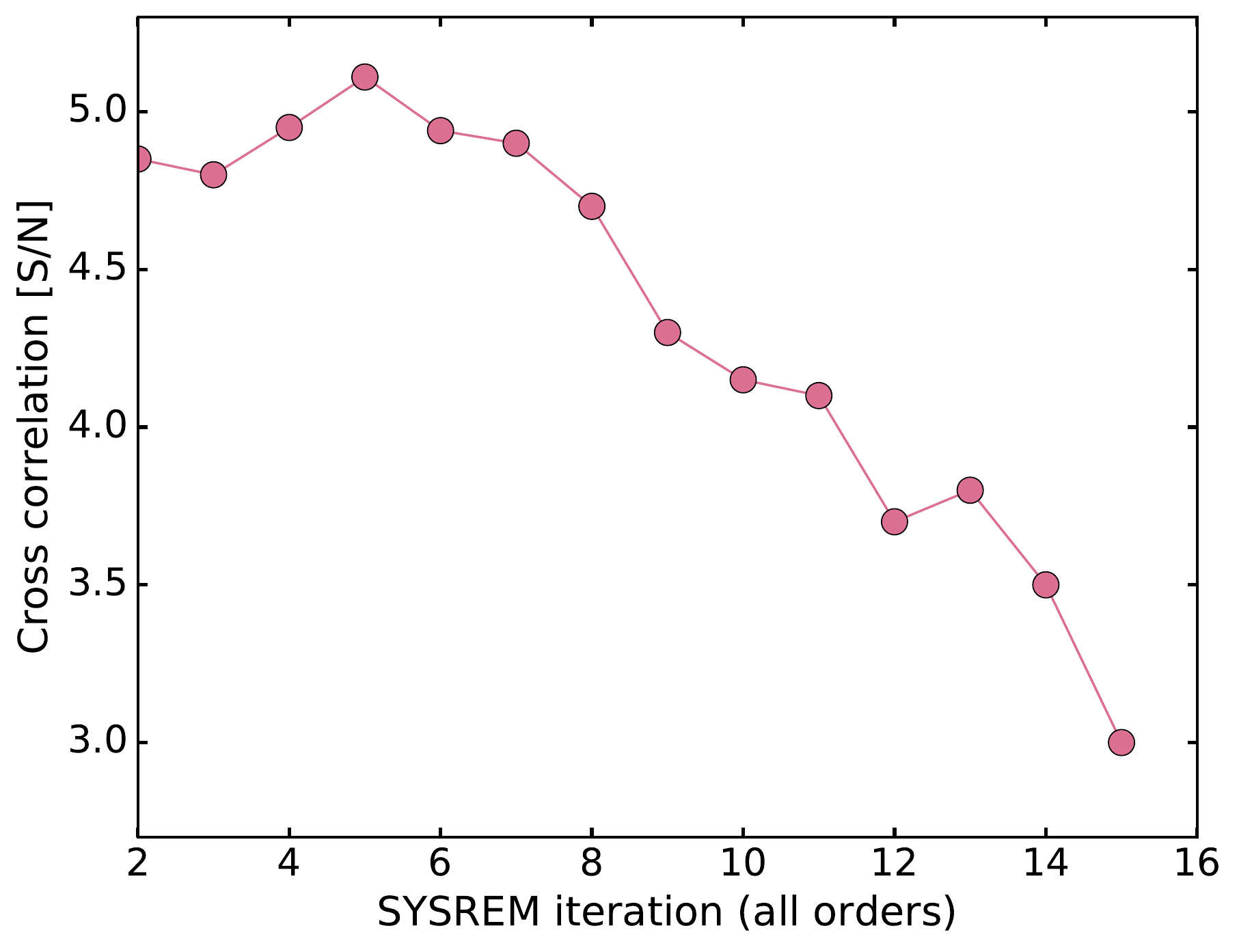}
\caption{S/N of the CCF peak obtained in $N_{A, 1}$ for \hd20 when cross-correlating the residual matrix after each {\tt SYSREM} iteration with the \h2o absorption model. The values correspond to the highest S/N at a $K_p$ in the range of 130 to 160\,km\,s$^{-1}$  and $\varv_{\text{wind}}$, ranging from $-$10 to $+$10\,km\,s$^{-1}$. The first iteration is omitted since it is mostly dominated by telluric residuals.} 
\label{sysrem_evol_freeze}
\end{figure*}

\begin{figure*}[ht]
\centering
\includegraphics[angle=0, width=0.32\columnwidth]{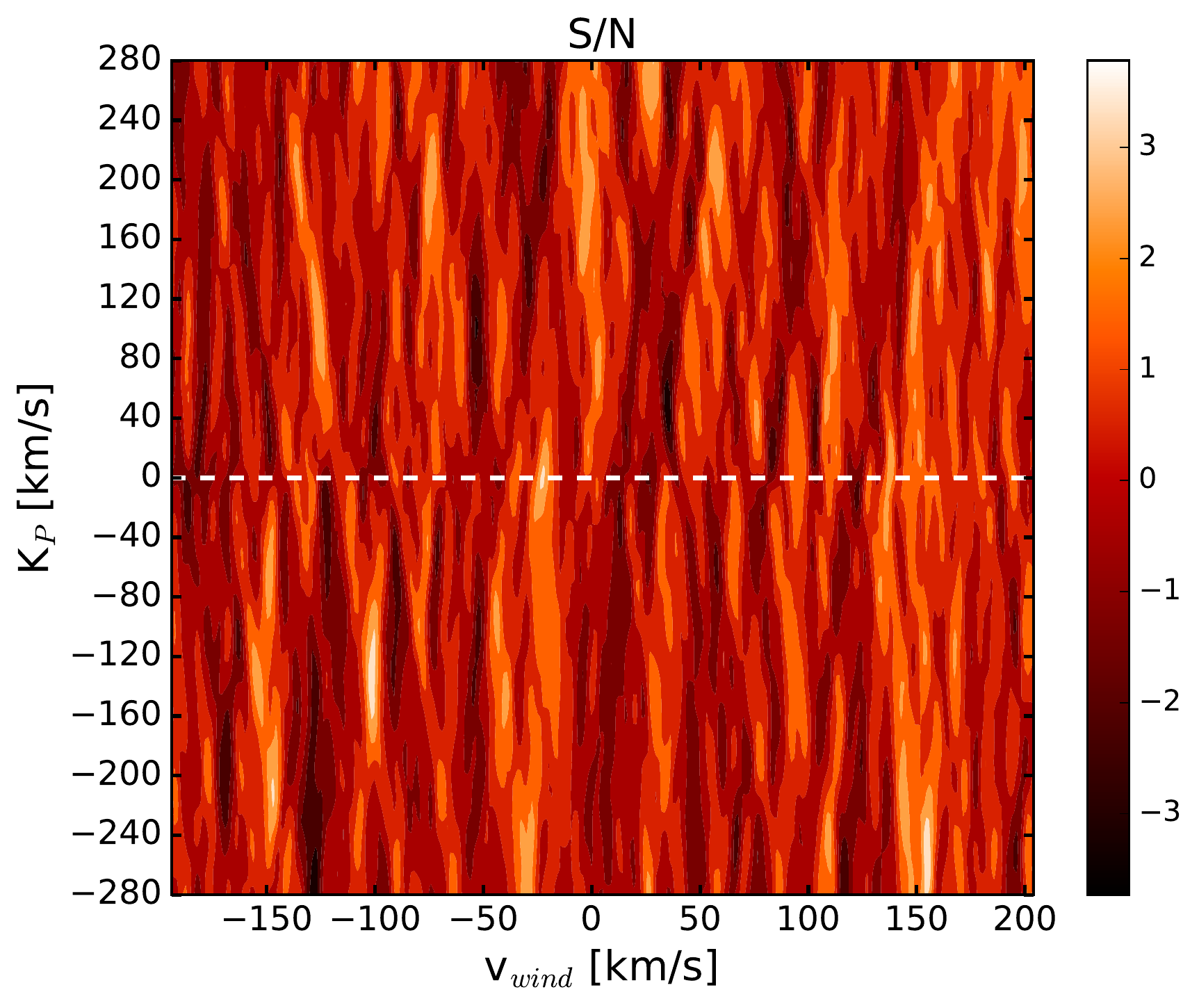}\includegraphics[angle=0, width=0.32\columnwidth]{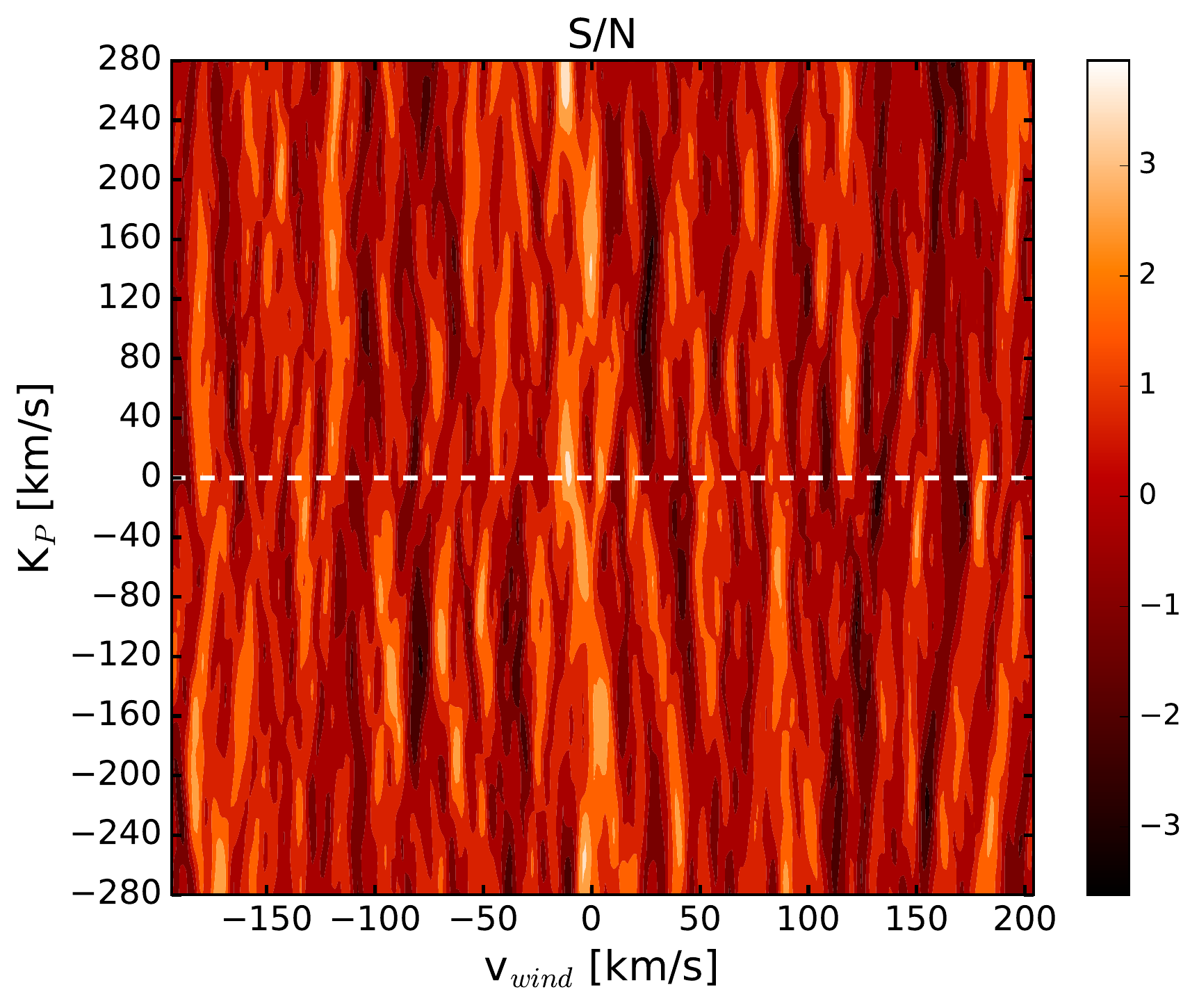}\includegraphics[angle=0, width=0.32\columnwidth]{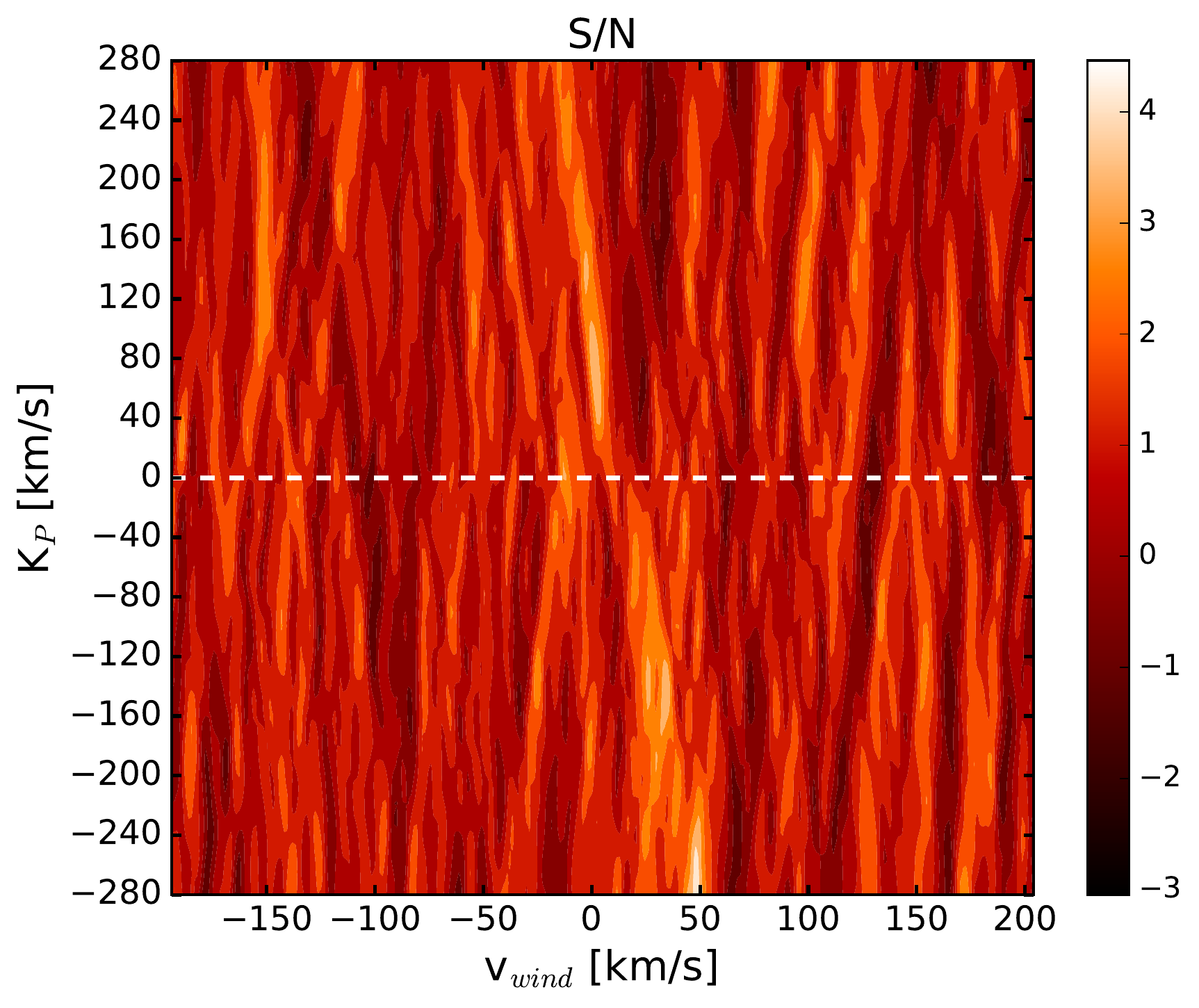}
\caption{S/N map for \hd20 for potential water signals with respect to the exoplanet rest-frame (horizontal axis) and K$_p$ (vertical axis) for the $\sim$0.72\,$\mu$m band (\textit{left panel}), the $\sim$0.82\,$\mu$m band (\textit{middle panel}), and the $\sim$0.95\,$\mu$m band (\textit{right panel}).} 
\label{multiband_hd20}
\end{figure*}

\begin{figure*}[htb!]
\centering
\includegraphics[angle=0, width=0.31\columnwidth]{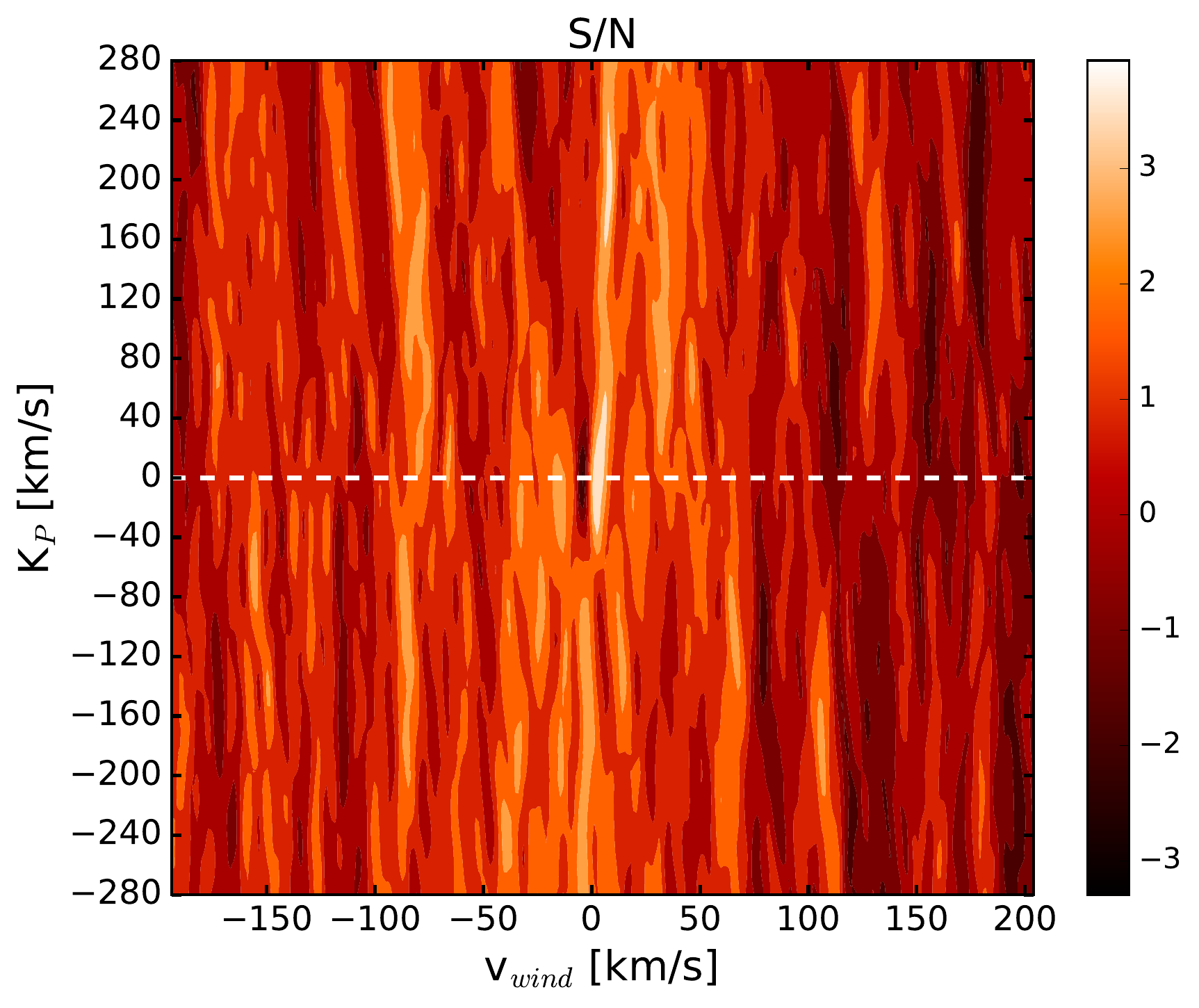}\includegraphics[angle=0, width=0.325\columnwidth]{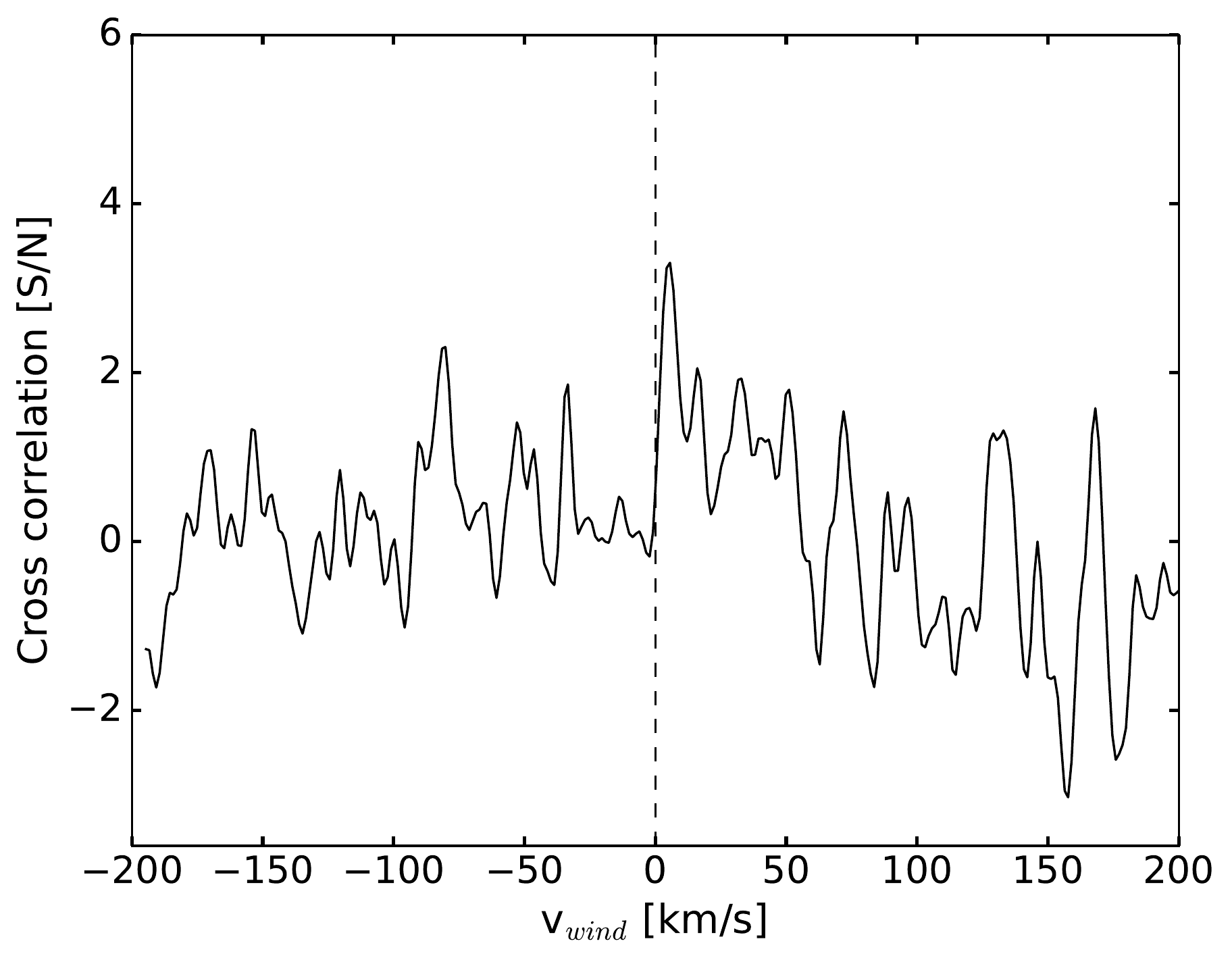}\includegraphics[angle=0, width=0.38\columnwidth]{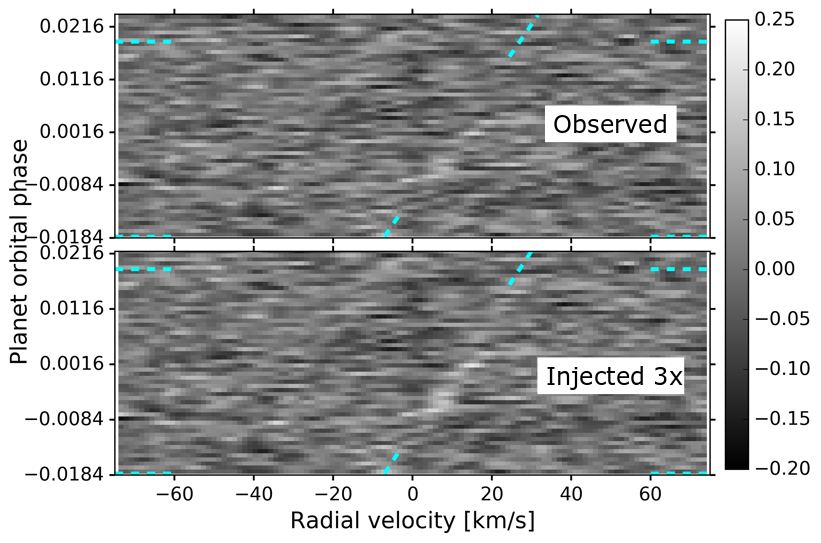}
\includegraphics[angle=0, width=0.31\columnwidth]{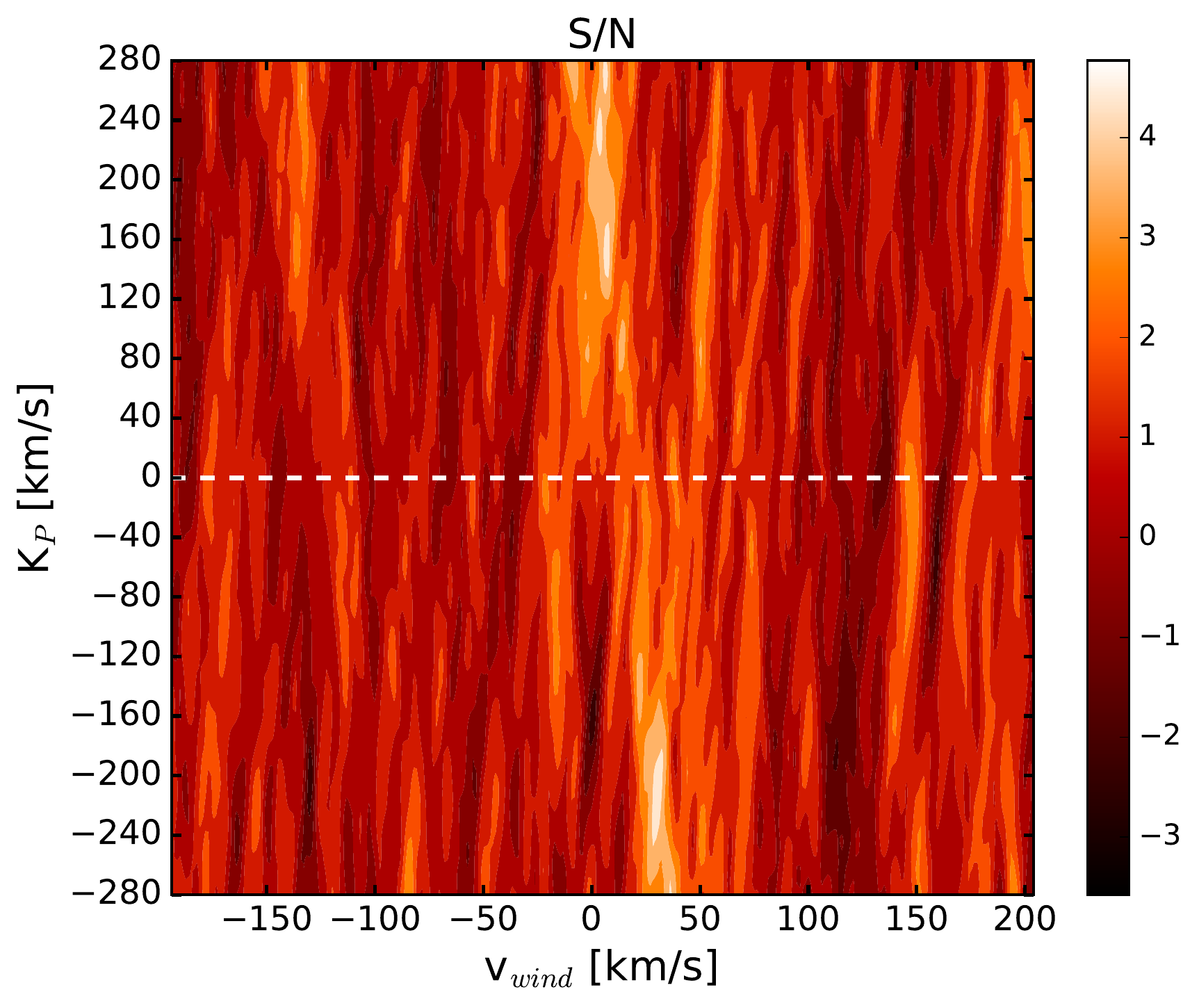}\includegraphics[angle=0, width=0.325\columnwidth]{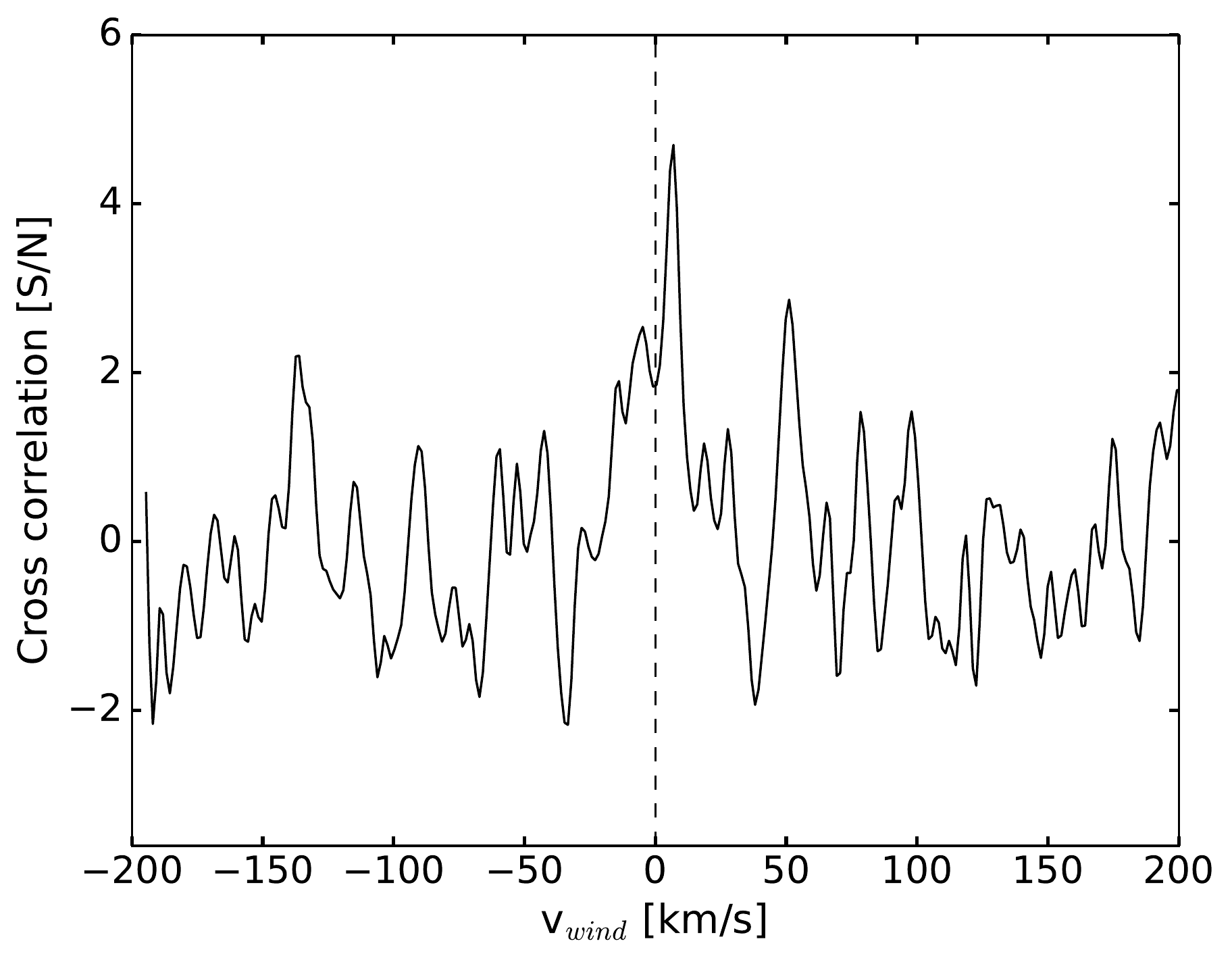}\includegraphics[angle=0, width=0.38\columnwidth]{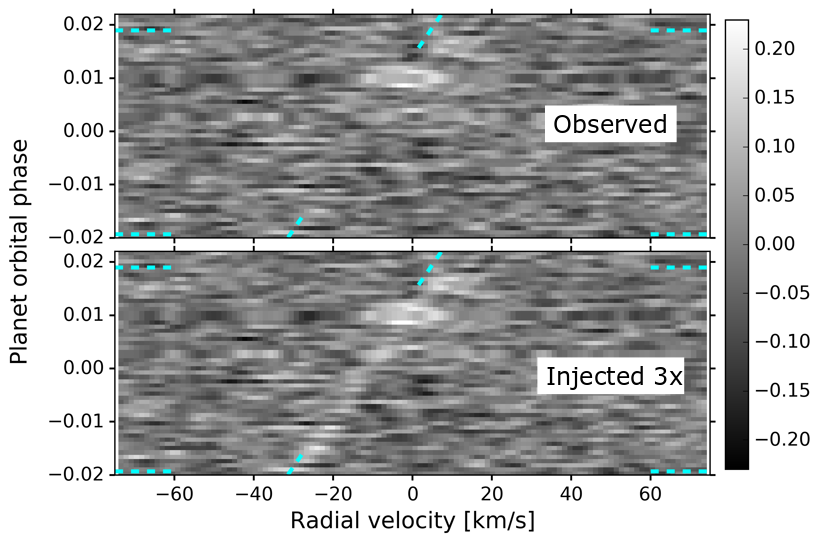}
\caption{Left column: S/N maps for \hd20 on the nights $N_{A,\,2}$ (top) and $N_{A,\,3}$ (bottom) for potential water signals with respect to the exoplanet rest-frame (horizontal axis) and K$_p$ (vertical axis). The horizontal dashed lines mark the $K_p$\,=\,0\,km\,s$^{-1}$ value. Middle column: Slice through the left panels at the expected $K_p$ for \hd20 (146\,km\,s$^{-1}$). Right column: Cross-correlation matrices in the Earth rest-frame. The transits occur between the horizontal dashed lines. The tilted dashed lines trace the expected planet velocities during the observations.} 
\label{sn_cc_hd20_mask_02}
\end{figure*}

\begin{figure*}[htb!]
\centering
\includegraphics[angle=0, width=0.31\columnwidth]{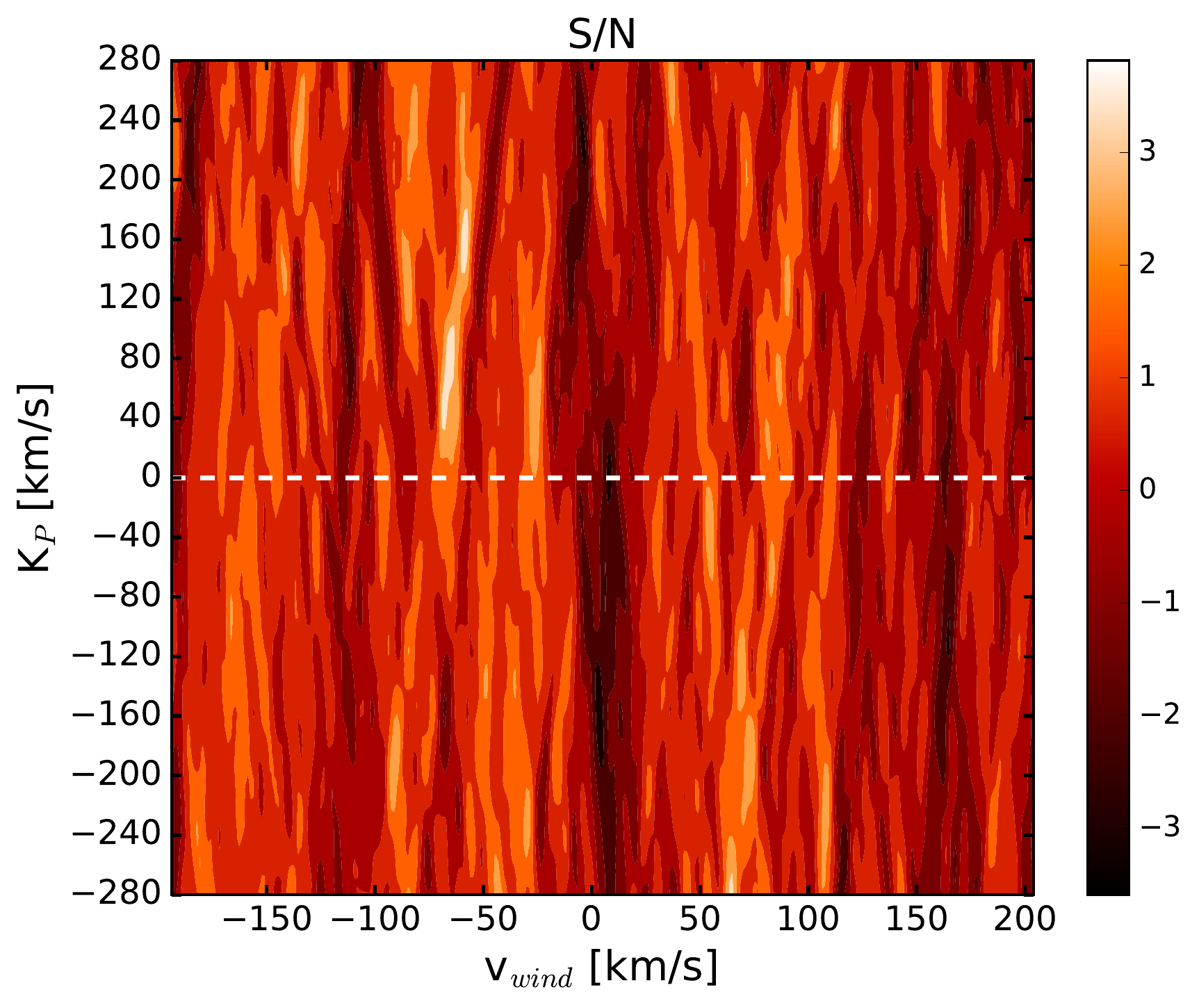}\includegraphics[angle=0, width=0.325\columnwidth]{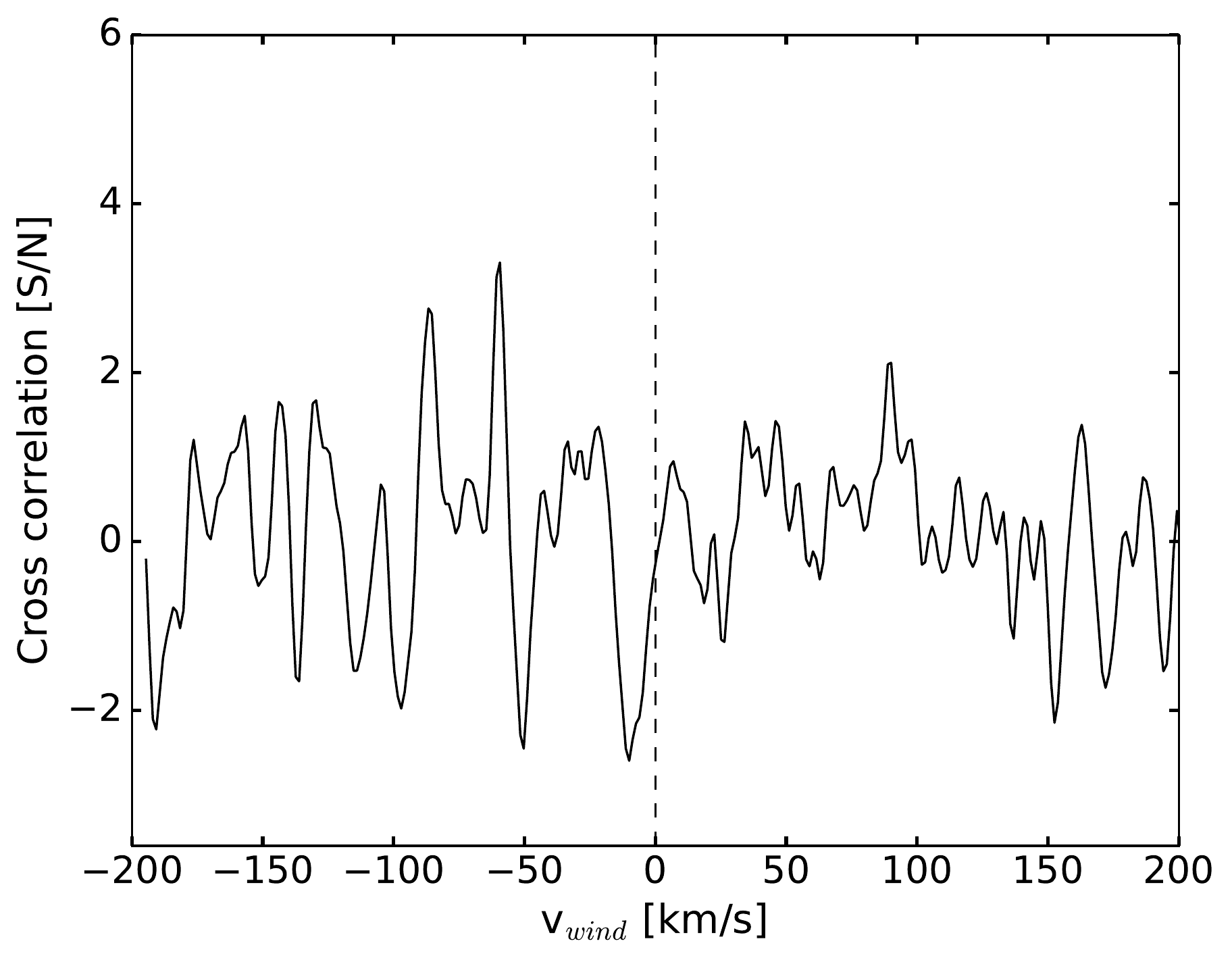}\includegraphics[angle=0, width=0.38\columnwidth]{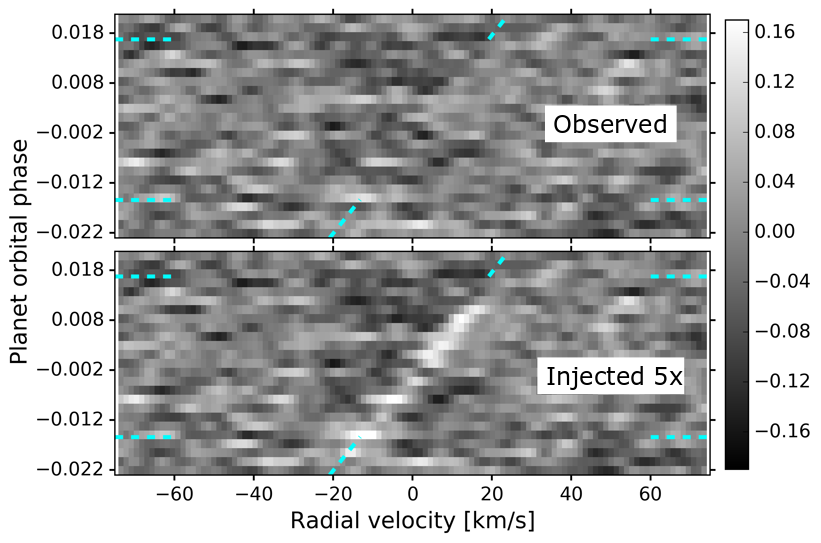}
\includegraphics[angle=0, width=0.31\columnwidth]{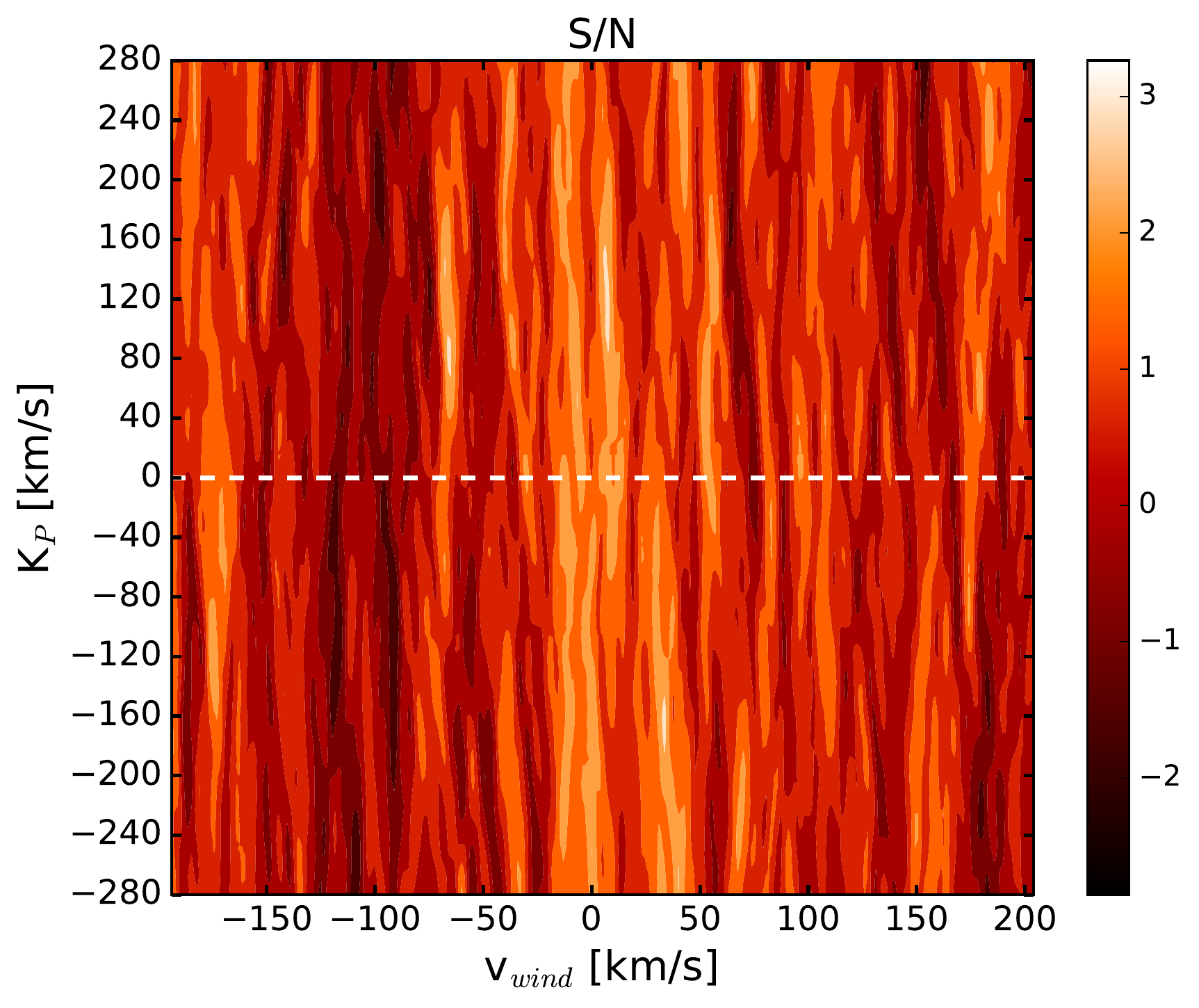}\includegraphics[angle=0, width=0.325\columnwidth]{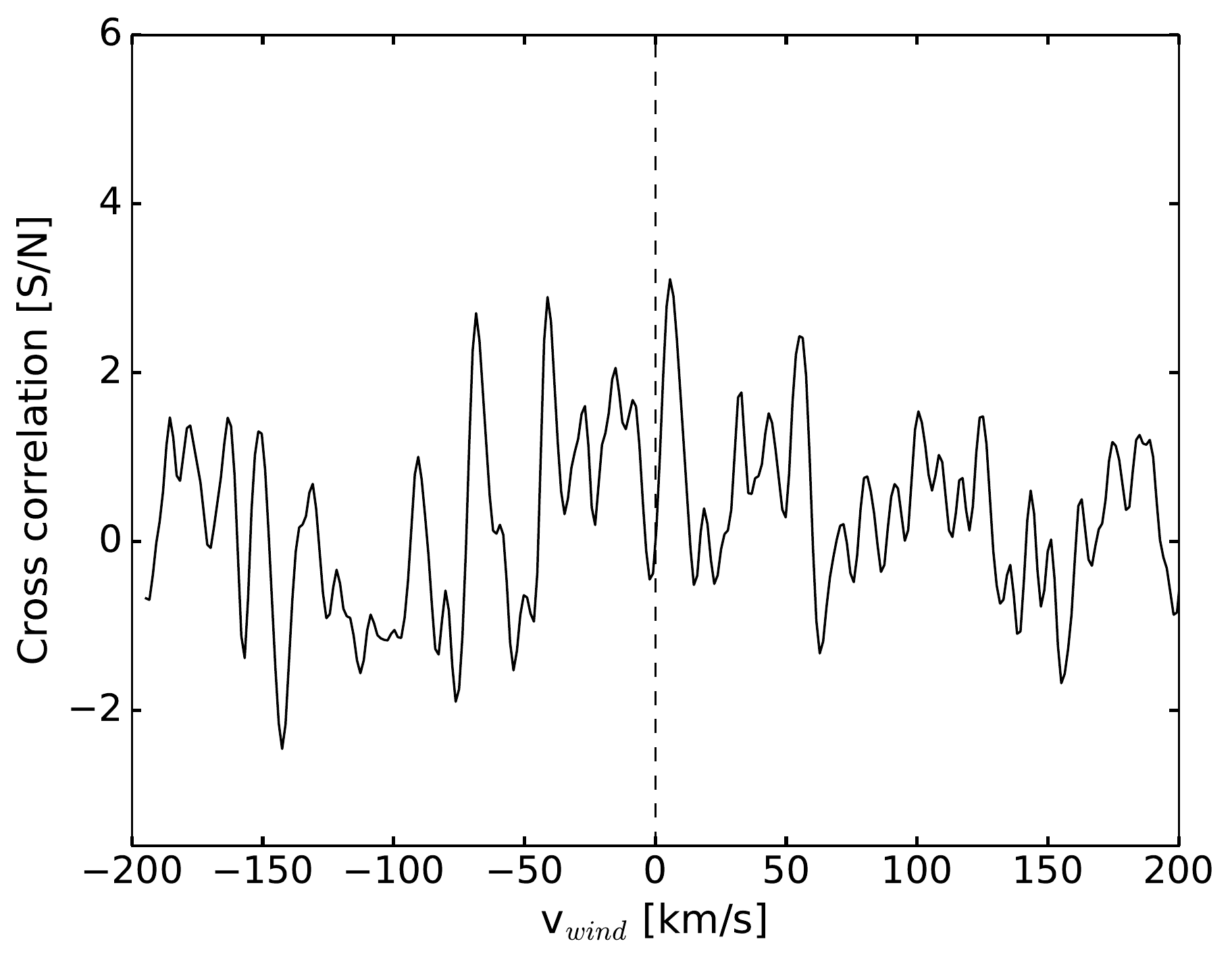}\includegraphics[angle=0, width=0.38\columnwidth]{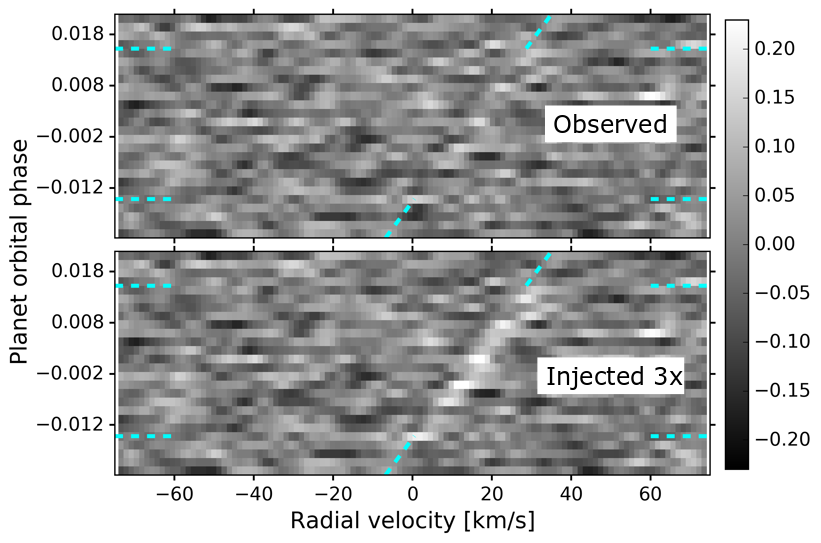}
\includegraphics[angle=0, width=0.31\columnwidth]{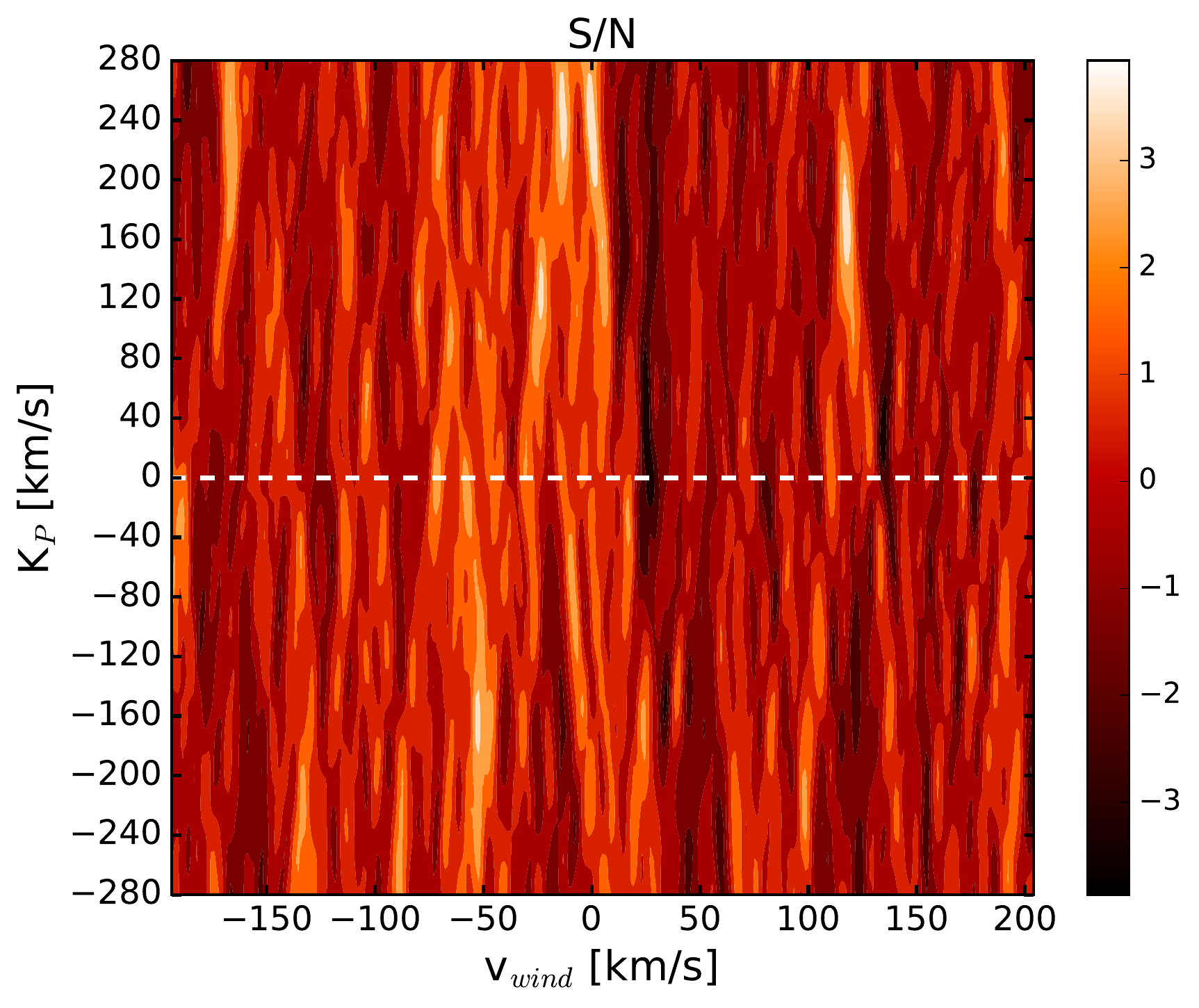}\includegraphics[angle=0, width=0.325\columnwidth]{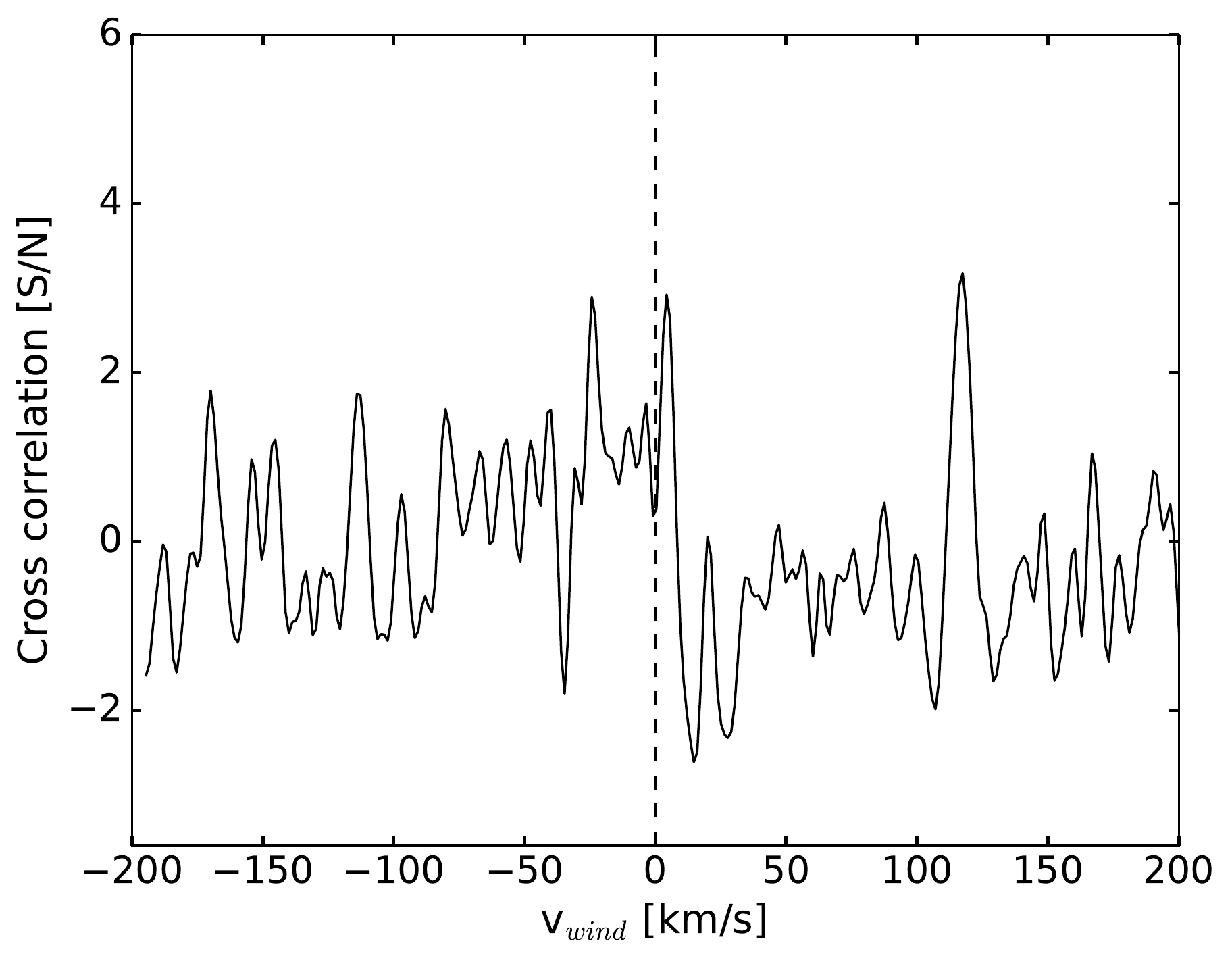}\includegraphics[angle=0, width=0.38\columnwidth]{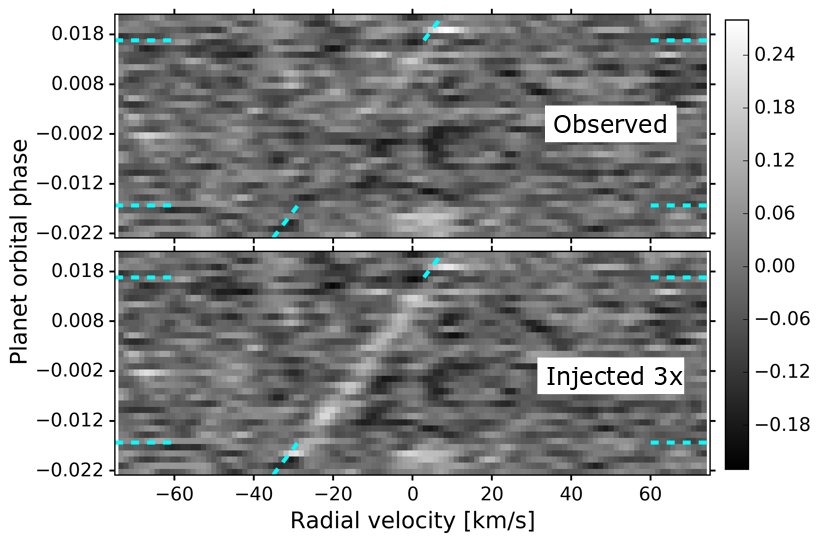}
\includegraphics[angle=0, width=0.31\columnwidth]{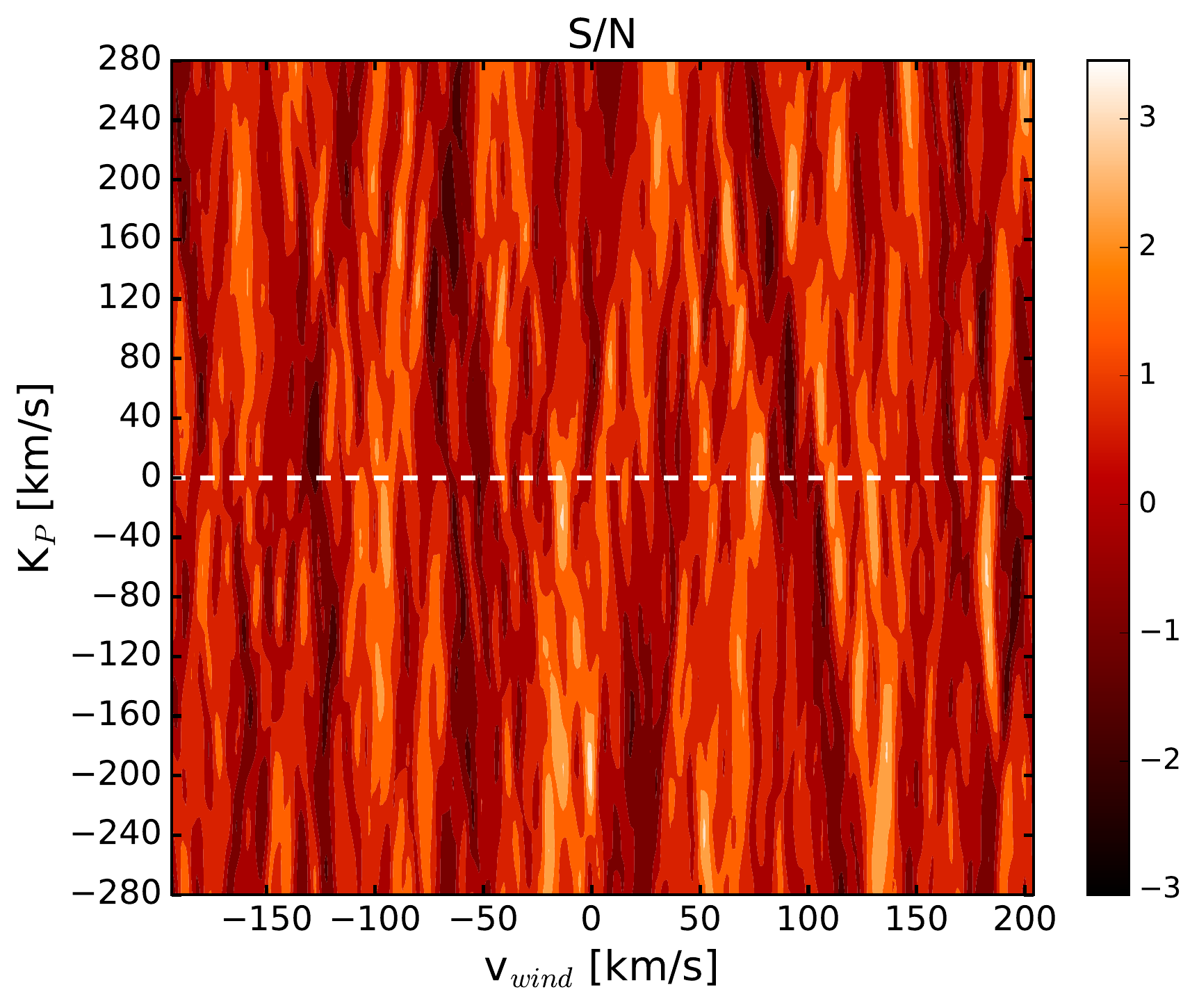}\includegraphics[angle=0, width=0.325\columnwidth]{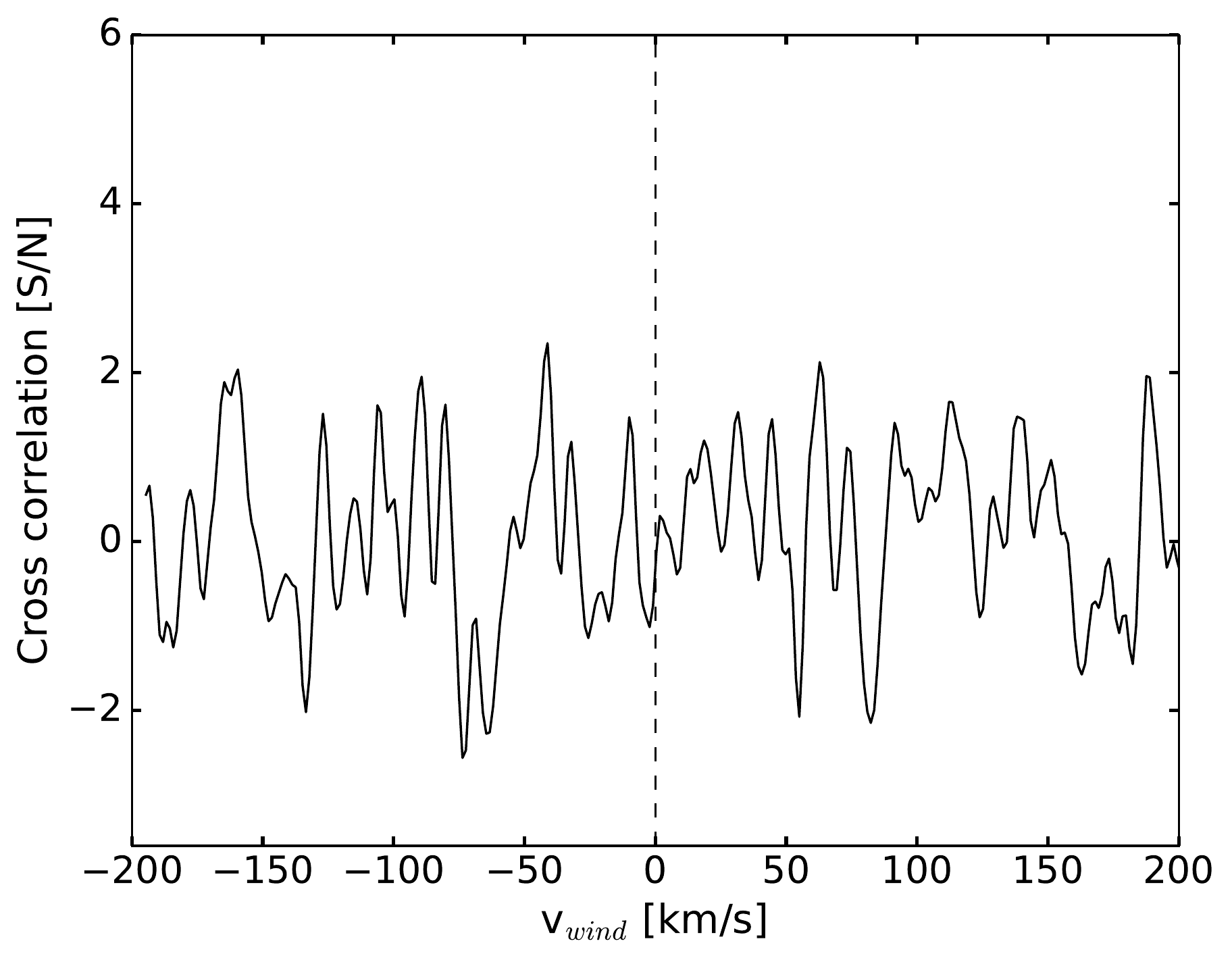}\includegraphics[angle=0, width=0.38\columnwidth]{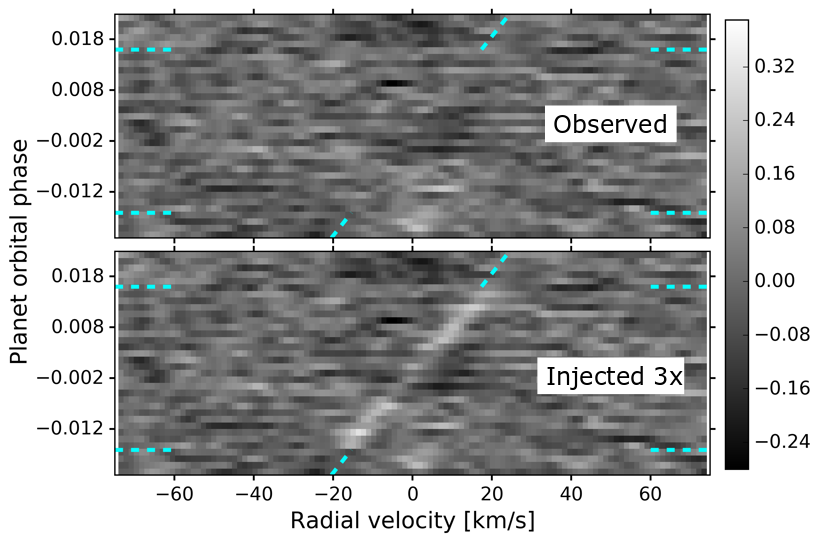}
\caption{Same as \ref{sn_cc_hd20_mask_02}, but for \hdu18 ($K_p = 152\,km\,s^{-1}$) on the nights $N_{B,\,1}$ (first row), $N_{B,\,2}$ (second row), $N_{B,\,3}$ (third row), and $N_{B,\,4}$ (fourth row).} 
\label{sn_cc_hd18_mask_02}
\end{figure*}

\clearpage

\end{document}